\documentclass[12pt]{article}
\usepackage{bm}
\usepackage[numbers]{natbib}
\usepackage[colorlinks,citecolor=blue,linkcolor=blue,urlcolor=blue]{hyperref}
\usepackage{latexsym, epsfig, epstopdf, amssymb, amsmath, amsthm, graphicx, mathrsfs, booktabs, mathabx}
\usepackage{rotating, tabularx, setspace,paralist}
\usepackage{color}
\usepackage[OT1]{fontenc}
\usepackage{lscape}
\usepackage{multirow,multicol}
\usepackage{anysize}

\usepackage{subcaption} 
\usepackage{amsfonts}
\usepackage{accents}
\usepackage{bbm} 
\usepackage[linesnumbered,ruled,vlined]{algorithm2e} 
\usepackage{algorithmic}
\usepackage{mathtools}

\usepackage{verbatim} 
\usepackage[utf8]{inputenc}
\usepackage[english]{babel}

\usepackage[e]{esvect}
\usepackage{hyperref}
\usepackage{multibib}

\newcites{SM}{Appendix References}

\newtheorem*{defn*}{Definition} 
\newtheorem{exmp}{Example} 
\newtheorem{condi}{Condition} 

\newtheorem{theorem}{Theorem}
\newtheorem{corollary}{Corollary}
\newtheorem{lemma}{Lemma}
\newtheorem*{lemma*}{Lemma}

\theoremstyle{plain}
\newtheorem{remark}{Remark}

\newtheorem*{claim*}{Claim}
\newtheorem*{rwr*}{Removed When Ready}

\usepackage{enumitem}
\newlist{steps}{enumerate}{1}
\setlist[steps, 1]{label = Step \arabic*:}

\usepackage{mathtools}

\SetKwInput{KwInput}{Input}                
\SetKwInput{KwOutput}{Output}              

\newcommand{\pkg}[1]{{\fontseries{b}\selectfont #1}}

\usepackage[ruled]{algorithm2e}

\renewcommand{\top}{^{T}}
\renewcommand{\dots}{\cdots}

\DeclareMathOperator*{\argmax}{arg\,max}

\usepackage[margin=0.8in,nohead]{geometry}

\begin{document}


	\title{\bf Analyze Additive and Interaction Effects via Collaborative Trees}

	\author{Chien-Ming Chi\thanks{
							Chien-Ming Chi is Assistant Research Fellow, Institute of Statistical Science, Academia Sinica, Taipei 11529, Taiwan (Email: \textit{xbbchi@stat.sinica.edu.tw}). %
				This work was supported by grant 111-2118-M-001-012-MY2 from the National Science and Technology Council, Taiwan.	}
			\hspace{.2cm}\\
			Academia Sinica}

\date{}
 \maketitle

	\begin{abstract}
	We present Collaborative Trees, a novel tree model designed for regression prediction, along with its bagging version, which aims to analyze complex statistical associations between features and uncover potential patterns inherent in the data. We decompose the mean decrease in impurity from the proposed tree model to analyze the additive and interaction effects of features on the response variable. Additionally, we introduce network diagrams to visually depict how each feature contributes additively to the response and how pairs of features contribute interaction effects. Through a detailed demonstration using an embryo growth dataset, we illustrate how the new statistical tools aid data analysis, both visually and numerically. Moreover, we delve into critical aspects of tree modeling, such as prediction performance, inference stability, and bias in feature importance measures, leveraging real datasets and simulation experiments for comprehensive discussions. On the theory side, we show that Collaborative Trees, built upon a ``sum of trees'' approach with our own innovative tree model regularization, exhibit characteristics akin to matching pursuit, under the assumption of  high-dimensional independent binary input features (or one-hot feature groups). This newfound link sheds light on the superior capability of our tree model in estimating additive effects of features, a crucial factor for accurate interaction effect estimation.
    
	\end{abstract}
	
	\noindent%

 \textit{Running title}: CTE
	

	\section{Introduction}

Insightful observations derived from statistical analyses of variable associations often lead to the formulation of good data-driven questions. For instance, when investigating how variables such as gender, college majors, and job categories influence the wage function of individual workers, analyzing the additive and interaction effects of these variables prompts the development of hypotheses for further exploration. Suppose we conduct an analysis and find that the influence of gender on wage rates is pronounced and primarily additive, suggesting  $\textnormal{Wage} = f_{1}(\textnormal{Other Variables}) + f_{2}(\textnormal{Gender})$, where $f_{1}$ and $f_{2}$ are some functions. When other variables fully characterize workers' capacity, this analysis raises the possibility of unequal pay for equal work, highlighting a social issue deserving attention. The example demonstrates how important subsequent analysis and inferences can stem from a thorough examination of variable effects, thereby motivating our efforts to establish a comprehensive analysis of the impact of variables on responses.

This paper introduces a new tree model called Collaborative Trees and its bagging version to estimate  effects of variables. Our estimation framework shares similarities with the Mean Decrease in Impurity (MDI) feature importance measure of Random Forests~\citep{Breiman2001}. Notably, we make a significant advancement by offering a more detailed analysis of both additive and interaction effects of input features, providing insights beyond the overall importance of each feature. Among various works~\citep{lou2007generalized, sobolprime1993sensitivity, apley2020visualizing, antoniadis2021random} concerning interactions, to name just a few, our work is closely aligned with Sobol' indices~\citep{sobolprime1993sensitivity, saltelli2010variance} for sensitivity analysis, which decompose the variance of response into the contribution shares of each feature and the interaction between all features. This decomposition is often referred to as the ANOVA decomposition~\citep{friedman2001greedy, antoniadis2021random}.
Unlike Sobol' indices, our analysis emphasizes the consistency of effect estimation and is attentive to applications involving high-dimensional and correlated input features. Additionally, to enhance the presentation of results from our analysis, which may yield a substantial number of estimated effects, we have introduced network diagrams for visually summarizing and screening the essential signals of interest. These diagrams aim to provide a comprehensive assessment of the statistical associations among all features, distinguishing them from plots of partial dependence~\citep{friedman2001greedy} and accumulated local effects~\citep{apley2020visualizing}, which typically focus on analysis of specific features.

\begin{figure}[t!]
    \centering    
\subfloat{
\centering
 \includegraphics[width=0.52\linewidth]{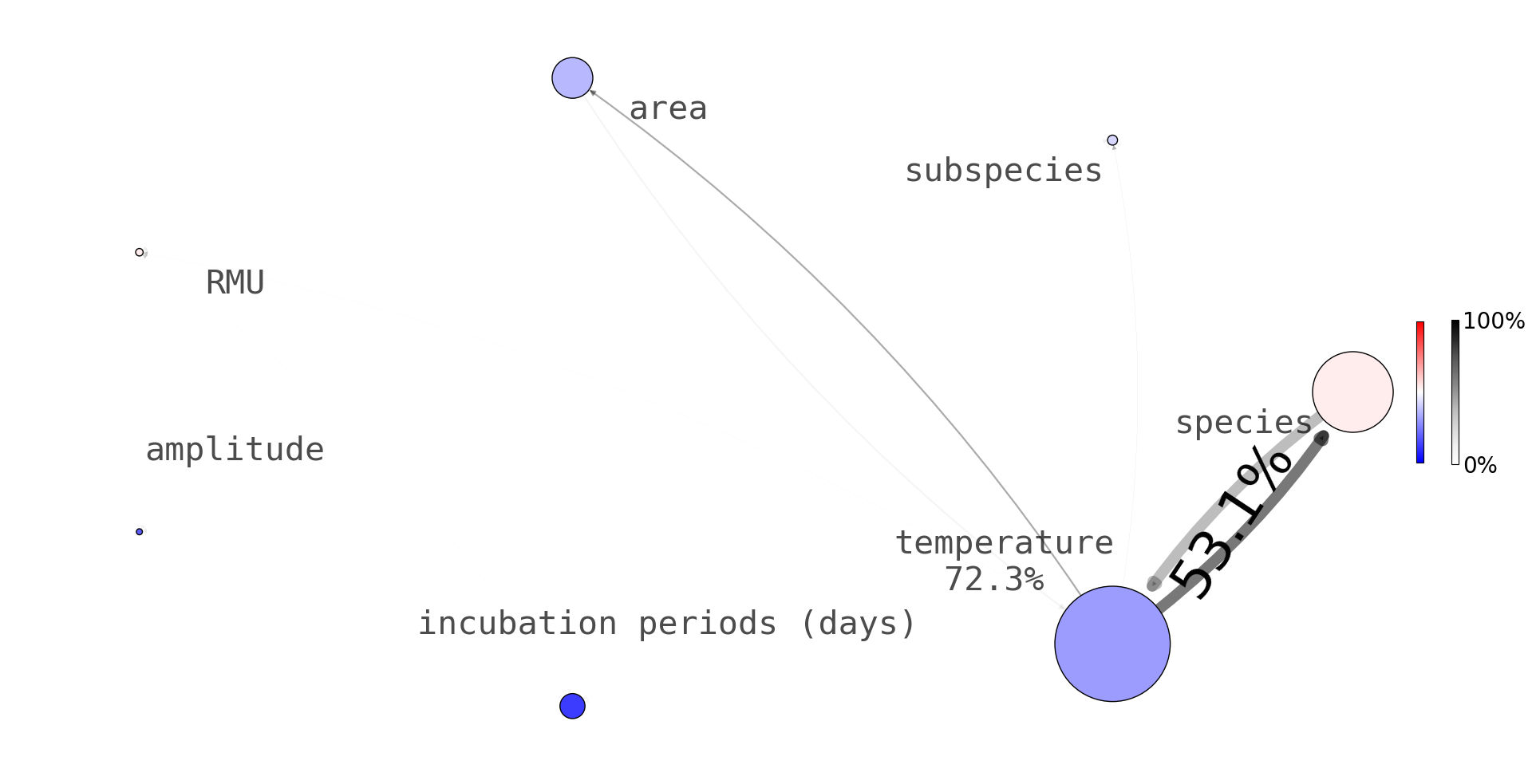}
}
\subfloat{
  \includegraphics[width=0.55\linewidth]{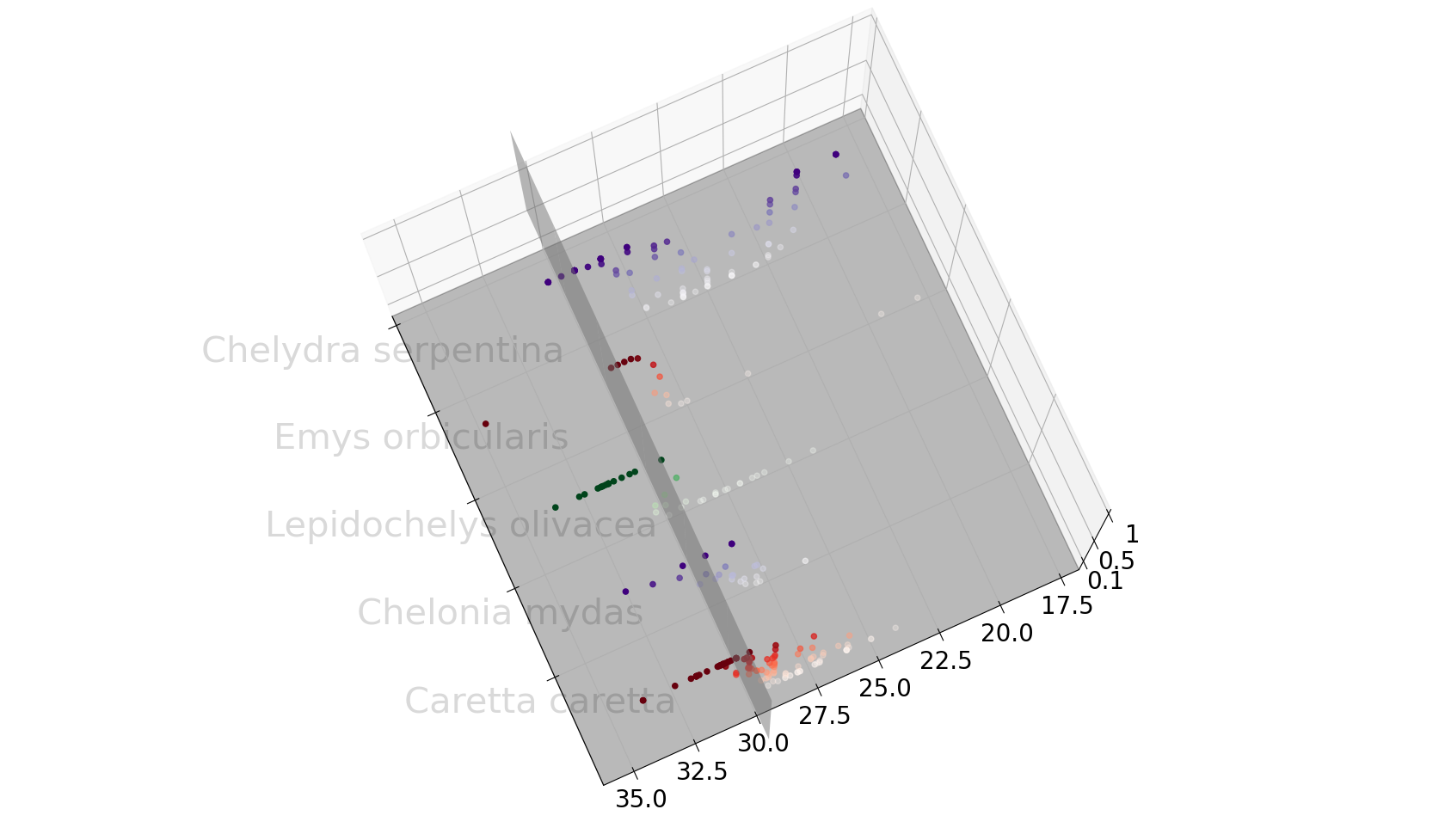}
}
    \caption{Embryo growth dataset (right panel) and the network diagram (left panel) for analyzing temperature-dependent sex determination. The z-axis, y-axis, and x-axis are respectively the sex ratio, species name, and incubation temperature.}
    \label{fig:relations}
\end{figure}

An example of a network diagram is provided in Figure~\ref{fig:relations} for the analysis of temperature-dependent sex determination in animals, which is crucial for conserving endangered species impacted by climate change~\citep{girondot2019embryogrowth, abreu2020recent}. In Figure~\ref{fig:relations}, the ratio of females to total births (z-axis) for newborns from five distinct species (y-axis), including various crocodile and turtle types, are shown. The x-axis indicates the incubation temperature. From the right panel of Figure~\ref{fig:relations}, most species exhibit higher ratios of females to total births at higher incubation temperatures. Particularly, for species Lepidochelys olivacea, Chelonia mydas, Caretta caretta, the sex ratios are well approximated by a logistic function with a single variable \texttt{temperature}. In fact, a majority of species in the data has a similar temperature-dependent sex determination, suggesting that additive effect may account for a significant proportion of the overall importance measure of \texttt{temperature} in determining sex ratios. On the other hand, we observe from the samples of Chelydra serpentina and Emys orbicularis that sex ratios varies across different species, which is clearly an interaction effect between \texttt{temperature} and \texttt{species} on the sex ratio.

Our network diagram summarizes the above observation. In the diagram of  Figure~\ref{fig:relations}, large circles represent features with high importance measures, thick edges between circles indicate strong interactions, and blue (red) circles highlight features with relatively strong additive (interaction) effects. Further diagram details are provided in Section~\ref{Sec3.3qq}. Our diagram suggests the observed pattern---the strong interaction between \texttt{species} and \texttt{temperature} for the 5 selected species---is consistent across all species in the dataset. Additionally, it illustrates the strong additive effect of \texttt{temperature} on the sex ratio. Moreover, the diagram prompts further analysis on the importance of \texttt{area} and \texttt{incubation periods (days)}, as discussed in Section~\ref{Sec6.1}. Notably, this new diagram can be used independently from Collaborative Trees, provided the required feature importance measures are available. However, a key advantage of using Collaborative Trees with the diagram is their ability to effectively distinguish truly important effects from unimportant ones. This capability is crucial for visually screening and highlighting relevant information in the diagram. This merit of Collaborative Trees is further discussed in Section~\ref{Sec3.3b} and Section~\ref{Sec5.1}.

The novel visualization tool arises from our in-depth analysis of feature association due to Collaborative Trees. Trees, known for their ability to reveal interactions between features, are utilized for analyzing both additive and interaction effects. However, trees are acknowledged to have limitations in capturing additive signals effectively~\citep{hastie2009elements, tan2022cautionary}. To augment Collaborative Trees' capability in handling additive components, several tree-based techniques have been proposed, including incorporating linear regression into tree nodes~\citep{friedberg2020local} and training a sum of additive trees. In this work, we leverage the "sum of trees" approach. Details of this enhancement are provided in the following section.

\subsection{Learning numerous additive components} \label{learn.additive.1}

\begin{figure}
\centering
 \includegraphics[width=0.7\linewidth]{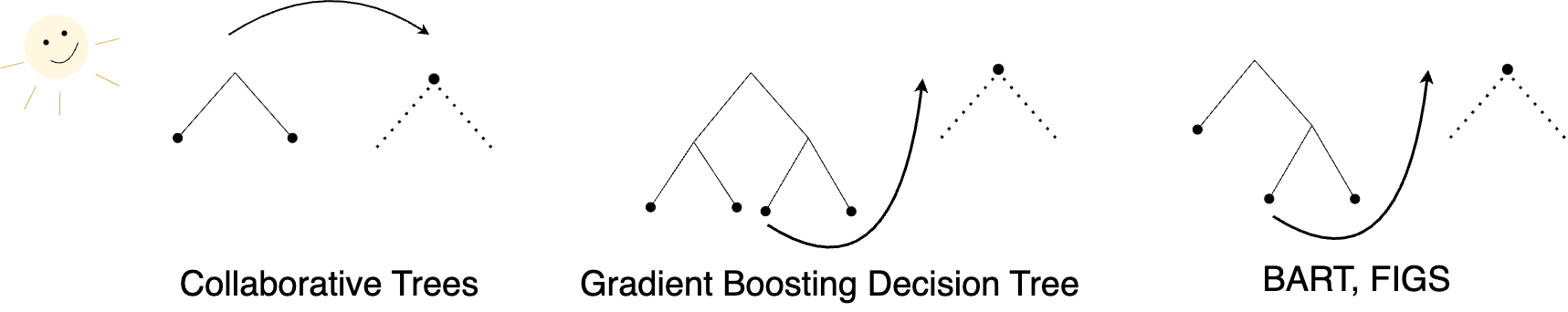}
 \caption{Collaborative Trees grow a new root node after splitting each root node, GBDT finishes the current tree before growing a new root node, while BART and FIGS rely on a sample-driven decision for each update, allowing them to potentially split on a new root node at any round. The number of potential root nodes is a hyperparameter. } 
 \label{boosting}
\end{figure}

A single decision tree is known to have limited capacity for fitting linear combinations of features~\citep{hastie2009elements}, which cannot be improved by merely applying methods like bagging or column subsampling~\cite{Breiman2001}. To overcome this problem, the proposed Collaborative Trees, adopts a specific tree-growing approach where new trees in the model are grown from the root nodes, utilizing residuals from the current round of updates. Intuitively, this tree-growing approach allows trees to extract the remaining additive signals in the residual by splitting on shallow nodes of new trees, which is often more efficient than learning additive signals by splitting deep nodes of the existing trees.
This idea of learning from residuals is widely used for building tree models, and the resulting trees are often referred to a sum of trees~\citep{friedman2003multiple, tan2022fast, chipman2010bart}. See, for example, BART~\citep{chipman2010bart}, FIGS~\citep{tan2022fast}, and Gradient Boosting Decision Trees (GBDT) such as XGBoost~\citep{chen2016xgboost}, lightGBM~\citep{ke2017lightgbm}, and MART~\citep{friedman2003multiple}. Among them, Collaborative Trees insist on growing a new root node after splitting each root node, GBDT insists on finishing the current tree before growing a new root node, while BART and FIGS rely on a sample-driven decision for each update, allowing them to potentially split on a new root node at any round; see Figure~\ref{boosting} for a graphical illustration. 

Our update priority and other model regularization techniques (see Section~\ref{sec.cte.1} for the detail) enhance Collaborative Trees as a specific variant of matching pursuit~\citep{mallat1993matching, davis1997adaptive}. These enhancements ensure the model's capability in capturing numerous additive components. Notably, our matching pursuit approach involves tree heredity and projections onto multiple feature groups to assess feature interactions, thereby differing from the standard matching pursuit methods~\citep{mallat1993matching} in the literature. We formally show that the sequential decisions of split coordinates of Collaborative Trees can be viewed as a feature selection path of our matching pursuit for applications with independent binary features or feature groups with one-hot indicators (e.g., resulting from binning continuous features). Additionally, our analysis shows that: 1) Extracting all additive signals (i.e., sure screening~\citep{fan2008sure}) is crucial for subsequently learning interaction effects. 2) To sure screen important components, the numbers of significant additive and interaction components are permitted to increase at some polynomial order with respect to the sample size. 3) The dimension of the input high-dimensional feature vector is permitted to increase at a rate on the order of the sample size raised to an arbitrarily large constant. To the best of our knowledge, we are among the first to formally establish this result, providing an interesting insight into the tree model building method ``learning from residuals,'' as depicted in Figure~\ref{boosting}. Moreover, our theory allows us to develop new feature importance measures for both additive effects and interaction effects of features. These measures serve as crucial components in constructing our network diagrams, offering a comprehensive visualization of the relationships and contributions of individual features in the model.

The rest of the paper is organized as follows. The Collaborative Trees algorithm is introduced in Section~\ref{Sec2}. The additive and interaction effects and their estimation are presented in Section~\ref{Sec3qq}. Theoretical results for ollaborative Trees and matching pursuit are in Section~\ref{Sec4qq}. Simulation experiments and real data studies are in Sections~\ref{Sec5qq}--\ref{Sec6}. 


\subsection{Notation}\label{notation}

Let $(\Omega, \mathcal{F}, \mathbb{P})$ be the probability triple, and $\boldsymbol{X} = (X_{1}, \dots, X_{p})^{\top}$ represent the $p$-dimensional random feature vector in the probability space. For $H\subset \{1, \dots, p\}$, $\boldsymbol{X}_{H} = (X_{j}, j \in H)^{\top}$ and $\boldsymbol{X}_{-H} =(X_{j}, j \not\in H)^{\top}$. Similar notation is used for their sample counterparts. Define $a \vee b = \max\{a, b\}$ for $(a, b)\in \mathbb{R}^2$. Let $\boldsymbol{1}_{A} = \boldsymbol{1}\{A\}$ be the indicator function, where $\boldsymbol{1}_{A} = 1$ if $A$ is true, and $\boldsymbol{1}_{A} = 0$ otherwise. Summation over an empty set is defined as zero. Sets contain unique elements, e.g., $\{j, j, k\} =\{j, k\}$. Integer parameters, like tree depth, are rounded to the nearest integer when necessary. For two real sequences $a_{n}$'s and $b_{n}$'s, define $a_{n} = O(b_{n})$ to mean that $\limsup_{n\rightarrow \infty} |a_{n} / b_{n}| <\infty$; $a_{n} = o(b_{n})$ implies that $\lim_{n\rightarrow \infty} |a_{n} / b_{n}| = 0$. Lastly, $\log a = \log_2 a$ for real $a>0$.

\section{Collaborative Trees}\label{Sec2}

The proposed tree model is termed as Collaborative Trees, with its bootstrap aggregation ensemble version called Collaborative Trees Ensemble (CTE). The name reflects the development of our tree model, which is briefly introduced here. 1.) Our tree model is a sum of predictive trees, which can be seen as collaboration between trees in making improved prediction. 2.) Child nodes with the same parent node collaborate, in the sense that these child nodes are updated together, as detailed in Section~\ref{sec.cte.1}. 3.) Features within the same group collaborate, meaning that Collaborative Trees split on feature groups rather than individual features when updating nodes. The concept of feature groups is introduced in Section~\ref{Sec2.1.b}. Our tree model regularization is introduced in detail in \eqref{gini2}.

\subsection{Feature groups of one-hot indicators}\label{Sec2.1.b}

Collaborative Trees apply to general features, but our  primary inference applications consider binary features and indicator feature groups, in which the feature groups may be formed by transforming  categorical features and binning continuous features into one-hot vectors. The use of feature groups significantly broadens the applications of our inference method. For example, in the embryo growth data set, the categorical variable \texttt{species} is represented by an one-hot vector, and the continuous variables incubation time and temperature are respectively binned into two one-hot vectors. See Section~\ref{Sec6.1} for details.

 Specifically, let set $\mathcal{X}(m)\subset\{1, \dots, p\}$ with $\texttt{\#}\mathcal{X}(m) >1 $ denote a group of  features of interest for each $m\in \{1, \dots, M_0\}$ for some $0\le M_{0}\le M$, where $M$ is the number of groups. Here,  $\mathcal{X}(l)\cap \mathcal{X}(k) = \emptyset$ for every $1\le l < k \le M_{0}$ and $\sum_{j\in\mathcal{X}(m)}\boldsymbol{1}_{X_{j} = 1} = 1$ for $1\le m \le M_{0}$. For notation completeness, each $\mathcal{X}(m)$ has a single binary feature (e.g., the gender variable) for $M_{0}<m\le M$ such that $\cup_{m=1}^M\mathcal{X}(m) = \{1, \dots, p\}$. To clarify, two special cases are $(M_{0}, M) = (0, p)$ and $M_{0} = M$, corresponding respectively to the scenario with only single features and the scenario without any single features. Throughout this paper, ``feature groups'' is used with awareness that they may include single features.

\subsection{Collaborative Trees}\label{sec.cte.1}

Collaborative Trees employ $K$ collaborative decision trees, denoted as $T_{k}:\mathbb{R}^p \mapsto \mathbb{R}$ for $k\in\{1, \dots, K\}$. These trees are trained on the sample $\{\boldsymbol{X}_{i} = (X_{i1}, \dots, X_{ip})^{\top}, Y_{i}\}_{i=1}^n$, where each $(\boldsymbol{X}_{i}, Y_{i})$ and $(\boldsymbol{X}, Y)$ are independently and identically distributed (i.i.d.). For any test point $\vv{x} \in \mathbb{R}^p$, Collaborative Trees generate predictions by summing up individual tree predictions: $ \sum_{k=1}^{K} T_{k}(\vv{x})$, which is often referred to a sum of trees~\citep{friedman2003multiple, tan2022fast, chipman2010bart}. In what follows, we will introduce the recursive updating algorithm for training the $K$ collaborative decision trees $T_{k}(\cdot)$.

\begin{figure}[h]
\centering
 \includegraphics[width=0.7\linewidth]{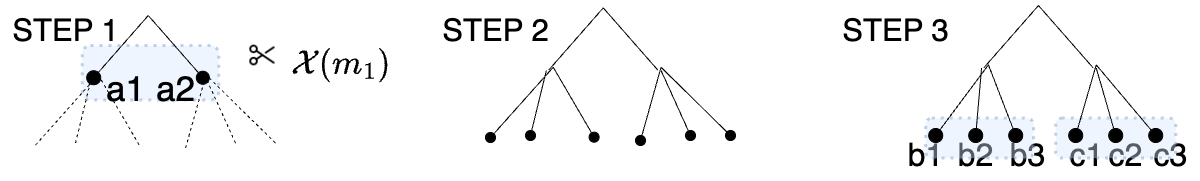}
 \caption{Here, $K=1$ and $W=\{A\}$ at STEP 1, where  $A= \{(a_1, 1), (a_2, 1)\}$; elements in $A$ are pairs of (node, index of tree having the node). An update on $A$ is made in three steps: At STEP 1, $\widehat{Q} = A$, and $\widehat{m} = m_{1}$ is decided by \eqref{gini2}, where $\texttt{\#}\mathcal{X}(m_{1}) = 3$.  At STEP 2, the tree is updated as in \textbf{Step 12} of Algorithm~\ref{algorithm1}, and $A$ is removed from $W$, leading to $W = \emptyset$. At STEP 3, two new sets of associated nodes, $B = \{(b_1, 1), (b_2, 1), (b_3, 1)\}$ and $C = \{(c_1, 1), (c_2, 1), (c_3, 1)\}$, are added to $W$, leading to $W = \{ B, C\}$.} 
 \label{step1}
\end{figure}

Let us consider a hyper rectangle $C_{0} \subset \mathbb{R}^{p}$ , also known as a node. When splitting $C_{0}$ on the $m$th feature group, the child nodes are denoted by $C_{1} = \{\vv{x} \in C_{0}: x_{j} > \widehat{c}\}$ and $C_{2} = \{\vv{x} \in C_{0}: x_{j} \le \widehat{c}\}$ if $\mathcal{X}(m) = \{j\}$, where the split point $\widehat{c}$ is decided by CART~\citep{Breiman2001}  as in \eqref{child.1} below. If $\texttt{\#}\mathcal{X}(m) > 1$, the child nodes are given by $C_{j} = \{\vv{x} \in C_{0}: x_{j} > 0\}$ for $j\in \mathcal{X}(m)$. In the latter case, the indices of child nodes $C_{j}$'s may not be consecutive integers. Child nodes with the same parent node are referred to as associated nodes. For example, in Figure~\ref{step1}, $\{a_{1}, a_{2}\}$, $\{b_{1}, b_{2}, b_{3}\}$, and $\{c_{1}, c_{2}, c_{3}\}$ are three sets of associated nodes. For completeness, each root node (the feature space) is also seen as a set of associated nodes. Collaborative Trees are trained using recursive splitting. During each iteration of updating Collaborative Trees, the algorithm decides a set of associated nodes to be updated, where each node in this set of associated nodes is split on some feature group. In addition, the feature group to be split should be the same across these associated nodes. For a graphical example, the set of associated nodes $a_1$ and $a_2$ are split on the same $m_1$th feature group at STEP 1 in Figure~\ref{step1}.

In this work, $W$ denotes the set of all eligible node sets for splitting at each round, along with the indices of their corresponding trees, which are needed for Algorithm~\ref{algorithm1}. For example, at the first round,
\begin{equation}
    \label{w0}
    W = \Big\{ \big\{(\textnormal{Root Node}, 1) \big\}, \dots , \big\{(\textnormal{Root Node},K) \big\} \Big\},
\end{equation}
where the $k$th element in $W$ represents a set containing only the root node of the $k$th tree, with the tree index indicated in parentheses. As the algorithm progresses, an element in $W$ may encompass multiple associated nodes; associated nodes in each element are updated at the same round. A graphical example illustrating how each element in $W$ is removed and new elements are included at each round is shown in Figure~\ref{step1}. During each round, given $W$ and the residuals $\{\widehat{r}_{i} = Y_{i} - \sum_{k=1}^{K}T_{k}(\boldsymbol{X}_{i})\}_{i=1}^n$, the optimization in \eqref{gini2} decides the best update. At this round, we obtain a set of  associated nodes $\widehat{Q}\in W$ to be updated and the $\widehat{m}$th feature group to be split as follows, with ties broken randomly.
\begin{equation}
    \begin{split}\label{gini2}
        (\widehat{Q}, \widehat{m}) & = \argmax_{Q \in W,\ m\in \{1, \dots, M\}} \Big[ \Big( \sum_{ (C,k) \in Q}  
        \textnormal{ SplitScore} (C, m)\Big)  - \lambda(W, Q)\Big],
    \end{split}
\end{equation}
where $\lambda(W, Q)$ is defined as follows: $\lambda(W, Q) = \infty$ if some $Q^{\dagger}\in W$ includes a root node (at depth zero) but $Q$ does not, or some $Q^{\dagger}\in W$ includes associated nodes at depth $1$ while $Q$ has deeper associated nodes. Otherwise,  $\lambda(W, Q) = 0$. The term $\lambda(W, Q)$ ensures the updating priority: 1. Root nodes. 2. Associated nodes at depth one. 3. Other associated nodes with depth deeper than one. The rationale behind our tree  regularization aligns with the idea of tree boosting~\citep{friedman2001greedy}, which also emphasizes learning from shallow nodes.

In addition, the split score, denoted as $\textnormal{SplitScore}(C_{0}, m)$, is the result of splitting a node $C_{0}$ using the $m$th feature group, and it is given by
\begin{equation}
    \label{gini1}
    \begin{cases}
    \max_{a \in N_{0}}\left[ \left(\sum_{i\in N_{0}} \widehat{r}_{i}^2\right) -  \sum_{l=1}^2 \sum_{i\in N_{l}(a)}  (\widehat{r}_{i} -  \frac{1}{1\vee\texttt{\#} N_{l}(a)} \sum_{i\in N_{l}(a) } \widehat{r}_{i}  )^2 \right] \ \  \textnormal{ if }   \texttt{\#}\mathcal{X}(m) = 1 , \\
        \left(\sum_{i\in N_{0}} \widehat{r}_{i}^2 \right)- \sum_{j\in \mathcal{X}(m)} \sum_{i\in N_{j}}  (\widehat{r}_{i} -\frac{1}{1\vee\texttt{\#} N_{j}} \sum_{i\in N_{j}} \widehat{r}_{i}  )^2  \hspace{2.8cm} \textnormal{ if } \texttt{\#}\mathcal{X}(m) > 1, 
    \end{cases}     
\end{equation}
where $N_{j} = \{i:\boldsymbol{X}_{i} \in C_{0}, X_{ij} > 0\}$ if $\texttt{\#}\mathcal{X}(m) > 1$,  $N_{1}(a) = \{i:\boldsymbol{X}_{i} \in C_{0}, X_{iq}> X_{aq}\}$ and $N_{2}(a) = \{i:\boldsymbol{X}_{i} \in C_{0}, X_{iq}\le X_{aq}\}$ if  $\mathcal{X}(m) = \{q\}$ for some $q\in \{1, \dots,  p\}$,  and $N_{0} = \{i:\boldsymbol{X}_{i} \in C_{0}\}$. See Algorithm~\ref{algorithm1} below for further details, where the validity of nodes for splitting and tree update is introduced in Section~\ref{Sec2.4}. 

For clarity in \textbf{Step 9} of Algorithm~\ref{algorithm1}, when splitting $C_{0}$ with respect to $\mathcal{X}(\widehat{m}) = \{q\}$ for some $ q\in \{1, \dots,  p\}$, we denote $\widehat{a}$ as the sample index in $N_{0}$ that maximizes \eqref{gini1} (note that $\texttt{\#}\mathcal{X}(\widehat{m}) = 1$). In this case, the two corresponding child nodes are given by
\begin{equation}
    \label{child.1}
    C_{1} = \{\vv{x} \in C_{0}: x_{q} > \widehat{c}\} \textnormal{ and }  C_{2} = \{\vv{x} \in C_{0}: x_{q} \leq \widehat{c} \}, \textnormal{ where } \widehat{c} = X_{\widehat{a}q}.
\end{equation}
\begin{spacing}{1.1}
{
    \begin{algorithm} 
		\SetAlgoLined
\KwIn{
Initial decision trees with $\sup_{\vv{x} \in \mathbb{R}^p}|T_{k}(\vv{x})| = 0$, and $W$ as in \eqref{w0}.
}%
\KwOut{$K$ collaborative decision trees $T_{1}(\cdot), \dots, T_{K}(\cdot)$.}

\While {$W\not = \emptyset$} 
{ 
Get  $\widehat{Q}$ and $\widehat{m}$ from  \eqref{gini2} with $\widehat{r}_{i} = Y_{i} - \sum_{k = 1}^{K}T_{k}( \boldsymbol{X}_{i})$ and $W$. 

Remove $\widehat{Q}$ from $W$. 

\For{$(C_{0},\widehat{k})\in \widehat{Q}$}{

\eIf{$\texttt{\#}\mathcal{X}(\widehat{m}) > 1$}{
Denote the $\texttt{\#}\mathcal{X}(\widehat{m})$ child nodes by $C_{j} = \{\vv{x} \in C_{0} : x_{j} > 0\}$ for $j\in \mathcal{X}(\widehat{m})$

Append $\{(C_{j}, \widehat{k}): j \in \mathcal{X}(\widehat{m}) \textnormal{ and } C_{j} \textnormal{ is a valid node for splitting} \}$ to $W$.

}{ 
 
Let $C_{1} $ and $C_{2} $ be child nodes obtained as in \eqref{child.1}. 

Append $\{(C_{l},\widehat{k}): l\in \{1,2\} \textnormal{ and } C_{l} \textnormal{ is a valid node for splitting}\}$ to $W$.

}
$T_{\widehat{k}}(\vv{x}) \gets T_{\widehat{k}} ( \vv{x}) + \sum_{j\in \{l: \textnormal{ valid child nodes } C_{l} \textnormal{ for tree update}\}} \frac{\boldsymbol{1} \{\vv{x} \in C_{j}\}  }{1\vee \texttt{\#} \{i: \boldsymbol{X}_{i} \in C_{j}\} } \sum_{i\in \{ \boldsymbol{X}_{i} \in C_{j} \}} \widehat{r}_{i}$.

}

}
\caption{Recursive Splitting for Growing $K$ Collaborative Trees}\label{algorithm1}
	\end{algorithm}} %
 \end{spacing}
 \vspace{-1em}
\subsection{Collaborative Trees Ensemble}\label{Sec2.4}

In this section, we introduce additional algorithm implementations in the following four points for Collaborative Trees Ensemble. 1.) To enhance prediction, we employ Collaborative Trees with bootstrap aggregating, or bagging.
2.) The validity of a node for splitting and/or tree updating depends on the number of subsamples in the node and its current depth. Our tree model employs hyperparameters \textsf{min\_samples\_split}, \textsf{min\_samples\_leaf}, and \textsf{max\_depth} for node validation.
3.) To optimize computation time, after the initial $2K$ updates, we only update a subset of nodes in the update node list $W$.
4.) In order to reduce estimation variance, Collaborative Trees make update decisions based on probability weights derived from split scores, which are related to the split sampling decision for BART~\citep{chipman2010bart}.

Furthermore, when required for inference applications, we group each continuous variable, such as the temperature variable in the embryo growth dataset, into \textsf{n\_bins} bins of equal sizes, creating a feature group with one-hot indicators representing data in each bin. The detail of hyperparameter tuning and an algorithm runtime discussion are given in the Supplementary Material. All source codes for this paper are available upon request.

\section{Extended mean decrease in impurity}\label{Sec3qq}

In this section, we introduce how to calculate the extended mean decrease in impurity (XMDI) for each Collaborative Trees model grown in Algorithm~\ref{algorithm1}, assuming a centered response. To calculate the XMDI for Collaborative Trees Ensemble (see Section~\ref{Sec2.4}), the values of XMDIs are the respective averages over all bagging tree models.

\begin{figure}
\centering
 \includegraphics[width=0.7\linewidth]{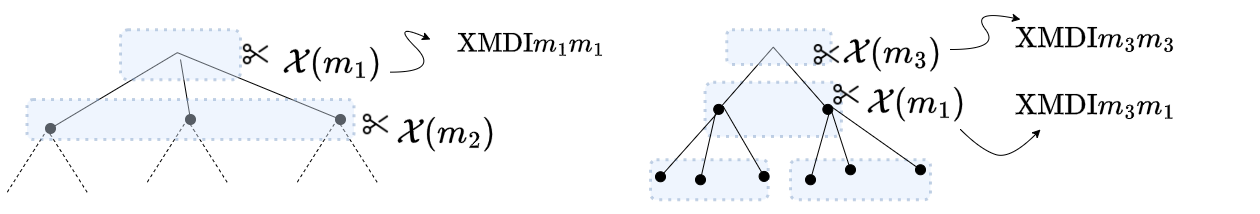}
 \caption{In this graphical example  with $2$ collaborative decision trees, each set of associated nodes is highlighted in blue. 
 At the $1$st round, the root node of the left tree is split on group $\mathcal{X}(m_{1})$ with $\texttt{\#}\mathcal{X}(m_{1}) = 3$, resulting in three child nodes. At the $4$th round, a set of associated nodes on the left tree is about to be updated on some group $\mathcal{X}(m_{2})$ with $\texttt{\#}\mathcal{X}(m_{2}) = 2$. In addition,  after each update, the split score is attributed to the corresponding group feature importance, as illustrated in this graphical example.} 
 \label{fig_cte_growing2}
\end{figure}

At the end of the $s$th round of update (i.e., \textbf{Step 13} of Algorithm~\ref{algorithm1}) with $s>0$, let $\widehat{U}_{s}(\vv{x}) = \sum_{k=1}^K T_{k}(\vv{x})$ for each $\vv{x} \in \mathbb{R}^p$, with $\widehat{U}_{0}(\vv{x}) = 0$, and let $\widehat{Q}_{s}$ denote the set of associated nodes to be updated. These associated nodes will be split on the $\widehat{m}_{s}$th feature group for some $1\le \widehat{m}_{s} \le M$. Let $1\le \widehat{e}_{s}\le s-1$ denote the index of round when the parent node of $\widehat{Q}_{s}$ is updated, in which $\widehat{e}_{s} = \emptyset$ if $\widehat{Q}_{s}$ consists of a root node. In addition, for each $s>0$, define $\widehat{u}_{s}$ such that $\widehat{u}_{s} = \widehat{m}_{s}$ if $\widehat{e}_{s} = \emptyset$, $\widehat{u}_{s} = \widehat{u}_{\widehat{e}_{s}}$ if $\widehat{m}_{s} = \widehat{m}_{\widehat{e}_{s}}$, and $\widehat{u}_{s} = \widehat{m}_{\widehat{e}_{s}}$ if $\widehat{m}_{s} \not= \widehat{m}_{\widehat{e}_{s}}$. Then, the XMDI is defined to be such that for every $1\le l \le k \le M$,

{\small \begin{equation*}
    \begin{split}
\textnormal{XMDI}_{lk} &= \sum_{s\in \{q:(\widehat{m}_{q}, \widehat{u}_{q}) = (l, k) \textnormal{ or }(\widehat{m}_{q}, \widehat{u}_{q}) = (k, l)\} } \left[\left(\sum_{i=1}^n (Y_{i} - \widehat{U}_{s-1}(\boldsymbol{X}_{i}))^2 \right)- \sum_{i=1}^n (Y_{i} - \widehat{U}_{s}(\boldsymbol{X}_{i}))^2\right]\frac{1}{n}.
    \end{split}
\end{equation*}}%
The sample responses are centered before calculating the XMDI in practice. Notably, the definition of XMDI also accounts for applications with continuous and categorical variables, where consecutive splits on the same feature are possible. Meanwhile, the overall feature importance is given by 
\begin{equation}
    \label{xmdi.overall}
    \textnormal{XMDI}_{m} = \sum_{1\le l\le M} \textnormal{XMDI}_{lm}
\end{equation} 
for  $m\in \{1, \dots, M\}$. 
Let us demonstrate the calculation of the XMDI with Example~\ref{xmda.calculation}.
\begin{exmp}\label{xmda.calculation}
   In an application involving binary features and feature groups, child nodes are not split on the same feature group as their parent nodes to avoid trivial splits yielding zero split scores. Therefore, the calculation of XMDI involves two key steps: 1.) $\textnormal{XMDI}_{mm}$ aggregates all split scores resulting from splitting the $m$th feature group at the root nodes. 2.) $\textnormal{XMDI}_{lk}$ is the summation of split scores from splitting the $k$th (or $l$th) feature group on certain nodes, provided their parent nodes have been split on the $l$th (or $k$th) group. For a visual demonstration of computing XMDIs in this context, refer to Figure~\ref{fig_cte_growing2}.
\end{exmp}



We now introduce the additive effects (or first-order effects or main effects) and interaction effects at the population level. The overall prediction power of $\boldsymbol{X}_{\mathcal{X}(m)}$ can be measured by the variance of the residual $f(\boldsymbol{X})  - \mathbb{E} (f(\boldsymbol{X})| \boldsymbol{X}_{-\mathcal{X}(m)})$, where we recall $\boldsymbol{X}_{-\mathcal{X}(m)}$ is a random vector consisting of all features but those in $\mathcal{X}(m)$. 
The additive effect of $\boldsymbol{X}_{\mathcal{X}(m)}$ is measured by $\textnormal{Var}(g_{m}(\boldsymbol{X}))$, while the two-way interaction effect of $\boldsymbol{X}_{\mathcal{X}(l)}$ and $\boldsymbol{X}_{\mathcal{X}(k)}$ is measured by $\textnormal{Var}(g_{lk}(\boldsymbol{X}) - g_{l}(\boldsymbol{X}) - g_{k}(\boldsymbol{X}))$. Here, 
\begin{equation}
\begin{split}
    \label{gj1}
    g_{J}(\boldsymbol{X}) &= \mathbb{E}(f(\boldsymbol{X})  - \mathbb{E} (f(\boldsymbol{X})| \boldsymbol{X}_{-\mathcal{X}(J)}) |\boldsymbol{X}_{\mathcal{X}(J)}),
    \end{split}
\end{equation}
for each $J \subset \{1, \dots, M\}$, with $\mathcal{X}(J) = \cup_{m\in J}\mathcal{X}(m)$. We subtract $g_l(\boldsymbol{X})$ and $g_k(\boldsymbol{X})$ from $g_{lk}(\boldsymbol{X})$ to better focus on estimating the two-way interaction effect between $\boldsymbol{X}_{\mathcal{X}(l)}$ and $\boldsymbol{X}_{\mathcal{X}(k)}$.
 For simplicity, we write $g_{J}(\boldsymbol{X})$ as $g_{m}(\boldsymbol{X})$ for univariate $J = \{m\}$, and $g_{lk}(\boldsymbol{X})$ as $g_{J}(\boldsymbol{X})$ for $J=\{l, k\}$.

 An example of $f(\boldsymbol{X})$ consisting of linear and XOR interaction components is given in Example~\ref{example.xor} of Section~\ref{Sec4.1qq}. In Theorem~\ref{theorem2}, we demonstrate that XMDI$_{mm}$ and XMDI$_{lk}$ = XMDI$_{kl}$ are respectively  consistent estimators of  $\textnormal{Var}(g_{m}(\boldsymbol{X}) )$ 
 and $\textnormal{Var}(g_{lk}(\boldsymbol{X}) - g_{l}(\boldsymbol{X}) - g_{k}(\boldsymbol{X}))$ when feature groups are independent and regularity conditions hold. Therefore, XMDI$_{m}$ defined as in \eqref{xmdi.overall} evaluates the sum of the additive effect of the $m$th feature group and its interactions with the other $M-1$ feature groups.

\subsection{Comparison with Sobol' indices and conventional two-way interaction models} \label{Sec3.3b}

To illustrate with an example, let us consider $f(\boldsymbol{X}) = \beta X_{1}X_{2} + X_{1} + X_{2}$ with $\mathbb{E}(f(\boldsymbol{X})) = 0$. In this example, the conventional two-way interaction inference~\citep{bien2013lasso} aims at testing whether $\beta = 0$. Meanwhile, Sobol' indices measure the additive effect of $X_{j}$ by $\textnormal{Var}[\mathbb{E}(f(\boldsymbol{X}) | X_{j})]$ for $j\in \{1, 2\}$, while the interaction effect between $X_{1}$ and $X_{2}$ is measured by $\textnormal{Var}[ \mathbb{E}(f(\boldsymbol{X}) | X_1, X_{2})] - \textnormal{Var}[[\mathbb{E}(f(\boldsymbol{X}) | X_{1})] - \textnormal{Var}[ \mathbb{E}(f(\boldsymbol{X}) | X_{2}) ]$. As a result, when features are independent, there is no interaction effect if and only if $\beta = 0$, regardless of which of the three approaches mentioned here is used. Among these approaches, a major restriction of the conventional one is the assumption of a parametric model, compared to Sobol' indices and XMDIs.

On the other hand, the additive and interaction Sobol' indices often exhibit a bias towards irrelevant features that are correlated with relevant ones, and hence may not be apt for statistical applications. For example, given $f(\boldsymbol{X}) = \gamma X_{1} + X_{3}$ with a sufficiently large $\gamma > 0$, it can be shown that $\textnormal{Var}[ \mathbb{E}(f(\boldsymbol{X}) | X_{2}) ] > \textnormal{Var}[ \mathbb{E}(f(\boldsymbol{X}) | X_{3}) ]$ if $X_{1}$ and $X_{2}$ are highly correlated, and $X_{3}$ is independent of $(X_{1}, X_{2})$. In contrast, the population XMDI correctly identifies $X_{2}$ as an irrelevant feature, yielding $\textnormal{Var}(g_{2}(\boldsymbol{X})) = 0$ in this example, regardless of the arbitrary correlation between $X_{1}$ and $X_{2}$.

In fact, XMDIs are closely related to the total Sobol' index~\cite{benard2022mean}, which quantifies the importance of $X_{\mathcal{X}(m)}$ using $\mathbb{E}[\textnormal{Var}(f(\boldsymbol{X}|\boldsymbol{X}_{-\mathcal{X}(m)})] = \mathbb{E}\{[f(\boldsymbol{X}) - \mathbb{E}(f(\boldsymbol{X})| \boldsymbol{X}_{-\mathcal{X}(m)})]^2\}$, addressing the bias concern by considering the variance of the residual function $f(\boldsymbol{X}) - \mathbb{E}(f(\boldsymbol{X})| \boldsymbol{X}_{-\mathcal{X}(m)})$. By way of comparison, $\textnormal{Var}(g_{m}(\boldsymbol{X}))$ given in \eqref{gj1} quantifies the variance of the residual function $f(\boldsymbol{X}) - \mathbb{E}(f(\boldsymbol{X})| \boldsymbol{X}_{-\mathcal{X}(m)})$ projected onto the $m$th feature group. Similarly, our measure of two-way interaction is also based on decompositions of the residual functions. Therefore, XMDIs may offer enhanced reliability in the presence of correlated features compared to Sobol' indices~\citep{sobolprime1993sensitivity}.

Our theory in Section~\ref{Sec4.2qq} provides a solid guarantee that positive XMDI values are exclusive to significant components asymptotically when feature groups are independent. Meanwhile, our empirical analysis in Section~\ref{Sec5.1} demonstrates the advantages of XMDIs over Sobol' indices in distinguishing significant components from insignificant ones, particularly when features are highly correlated. It is noteworthy that XMDIs may benefit from incorporating debiasing techniques, as in~\citep{li2019debiased, zhou2021unbiased, benard2022mean, strobl2007bias}, although the details are beyond the scope of this paper.

\subsection{Network diagrams}\label{Sec3.3qq}

Let us introduce the four key elements in network diagrams (see Figure~\ref{fig:relations}) as follows: 1.) The sizes of the circles represent the overall feature importance $\textnormal{XMDI}_{i}$'s. 2.) The thickness of each edge depends on $\textnormal{XMDI}_{ij}$. 3.) The colors (blue or red) of the features depend on the ratio $\frac{\textnormal{XMDI}_{ii}}{\textnormal{XMDI}_{i}}$. A feature with high value of this ratio is represented in a deep blue shade, indicating that it is more likely to be an additive component in the model. Otherwise, this feature tends to be an interaction components, since $\textnormal{XMDI}_{i} = \sum_{1\le l\le M} \textnormal{XMDI}_{il}$. 4.) The grayness of edges and the direction of arrows are determined by $\frac{\textnormal{XMDI}_{ij}}{\textnormal{XMDI}_{i}}$, reflecting the strength of interaction between $X_{i}$ and $X_{j}$ based on the overall importance of $X_{i}$. The arrow points towards the feature in the denominator.
In each diagram, we are interested in features represented by larger circles in a deep blue shade, and interactions depicted with greater thickness and darker edges. We also present the maximum values of $\frac{\textnormal{XMDI}_{ij}}{\textnormal{XMDI}_{i}}$ and $\textnormal{XMDI}_{i} \times \left[\textnormal{Sample variance of the response}\right]$. These values, expressed as percentages, are displayed atop the corresponding measures in the diagrams.

\section{Asymptotic analysis of Collaborative Trees} \label{Sec4qq}

In this section, we show the sure screening property~\cite{fan2008sure} of Collaborative Trees in Algorithm~\ref{algorithm1} and consistent XMDI estimation under proper model assumptions, allowing the number of important additive and interaction components to increase polynomially with the sample size. In contrast, it is well-known that decision trees cannot effectively capture smooth signals in additive models~\citep{hastie2009elements, tan2022cautionary} when the number of significant components is of polynomial order in the sample size.

A key observation here is that our tree model is a type of matching pursuit~\citep{mallat1993matching} (see \eqref{mp.3} in Section~\ref{Sec4.1qq}) within the first $2K$ updates. Therefore, from a theoretical viewpoint, we are interested in some important topics of asymptotic analysis for trees and/or matching pursuit, including sure screening property~\citep{fan2008sure}, model consistency and convergence rates for matching pursuit~\citep{barron2008approximation, livshitz2003two, klusowski2023sharp} and for trees~\citep{chi2022asymptotic, klusowski2021universal}, and estimation consistency of feature importance measures~\citep{scornet2023trees, louppe2013understanding}. In this paper, we formally analyze the sure screening property of Collaborative Trees in Theorem~\ref{theorem3} and the estimation consistency of XMDIs in Theorem~\ref{theorem2}, while leaving the tree model consistency for future work.

Let us summarize the analysis challenges for our tree model. We will demonstrate that the first $K$ updates of Collaborative Trees are a standard matching pursuit~\citep{mallat1993matching}, while the subsequent $K$ updates involve tree heredity (see Condition~\ref{heredity.1}) and projections onto multiple feature groups. Hence, the existing model consistency results~\citep{mallat1993matching, barron2008approximation, klusowski2021universal, livshitz2003two} developed for the standard matching pursuit are not directly applicable to our tree model. In addition, the updates after the first $2K$ steps in Collaborative Trees are more akin to sums of trees~\citep{chipman2010bart} than matching pursuit, making the analysis more involved and may further require analysis techniques for high-dimensional trees~\citep{chi2022asymptotic, klusowski2021universal}. Therefore, we shall focus on analyzing our tree inference (the sure screening property and XMDIs consistency) and postpone the consistency analysis of Collaborative Trees to future work.

On the other hand, sure screening at the early updates within the first $2K$ updates ensures reliable assessment of important features. The sure screening property has never been formally established for trees, to the best of our knowledge. Our matching pursuit with the non-standard projection scheme under tree heredity makes a rigorous analysis of sure screening a challenging task. Furthermore, the XMDI estimation of $\textnormal{Var}(g_{J}(\boldsymbol{X}))$'s (see \eqref{gj1}) involves the analysis of the prediction function $\mathbb{E}(f(\boldsymbol{X})| \boldsymbol{X}_{-\mathcal{X}(J)})$, making our technical analysis of the XMDI estimation consistency especially difficult when considering correlation between feature groups. We overcome these technical challenges by controlling the distributional dependence between feature groups (see Condition~\ref{signal.strength.1} in Section~\ref{Sec4.1qq}) and assuming that the feature selection path has no repeated elements (see Theorem~\ref{theorem3}). Despite having theoretical limitations, our theory is among the first formal results that establish equivalence between trees and matching pursuit, which has been discussed and inspected by many authors~\citep{klusowski2023sharp, tan2022fast, chipman2010bart}.

\subsection{Sure screening additive and interaction components} \label{Sec4.1qq}

In this section, we show that Collaborative Trees introduced in Section~\ref{sec.cte.1} can identify all significant components in the data generating models.  Since individual binary feature can be seen as a group of a single feature, our analysis is established without loss of generality for feature groups $\mathcal{X}(1), \dots, \mathcal{X}(M)$ defined in Section~\ref{Sec2.1.b}. The set of significant additive components is given by 
$S_{1}^* = \{m: \textnormal{Var}( g_{m}(\boldsymbol{X})  )>0, 1\le m\le M \}$, while the set of significant interaction components is defined to be $S_{2}^* = \{\{l,k\} : \textnormal{Var}(g_{lk}(\boldsymbol{X}) - g_{l}(\boldsymbol{X}) - g_{k}(\boldsymbol{X})   ) > 0, 1\le l<k\le M\}$. The function $g_{J}(\boldsymbol{X})$'s are defined in \eqref{gj1}.

 Our asymptotic analysis is based on a crucial observation: When features are either binary or groups of one-hot indicators, Collaborative Trees exhibit asymptotic similarity to the matching pursuit method. To explain this connection, we begin with a recap of Algorithm~\ref{algorithm1} from Section~\ref{sec.cte.1}, and provide some definitions of notation. In each round of tree update,   a set of associated nodes is updated together w.r.t. some group $\mathcal{X}(m)$. Consider the $s$th round of update on some set of associated nodes in the $k$th tree. Let $\widehat{J}_{s}\subset \{1, \dots, M\}$ denote the set of group indices representing the update history from the root node of the $k$th tree to the current set of associated nodes. Particularly, the index of the feature group to be split at the $s$th round of update is also in $\widehat{J}_{s}$. For example, $\widehat{J}_{1} = \{m_{1}\}, \widehat{J}_{2} = \{m_{3}\}, \widehat{J}_{3} = \{m_{3}, m_{1}\}$, and $\widehat{J}_{4} = \{m_{1}, m_{2}\}$ in Figure~\ref{fig_cte_growing2}.

 To gain insights into the asymptotic equivalence between Collaborative Trees and matching pursuit, we consider functions $R_{l}^{\star}:\mathbb{R}^p\mapsto\mathbb{R}$ defined recursively for $0 < l \le 2K$:
\begin{equation}
    \label{mp.3}
    R_{l}^{\star}(\boldsymbol{X})  = R_{l-1}^{\star} (\boldsymbol{X})+   \mathbb{E}( f(\boldsymbol{X}) - R_{l-1}^{\star}(\boldsymbol{X}) | \boldsymbol{X}_{\mathcal{X}(\widehat{J}_{l})}),
\end{equation} 
where $R_{0}^{\star}(\vv{x}) = 0$ and $\mathcal{X}(J) = \cup_{m\in J}\mathcal{X}(m)$ for $J\subset\{1, \dots, M\}$. In our technical proofs, we show that setting $J = \widehat{J}_{s}$ maximizes the sample counterpart of the function $L(R_{s-1}^{\star}, J) =   \textnormal{Var}\left[\mathbb{E}( f(\boldsymbol{X}) - R_{s-1}^{\star}(\boldsymbol{X})|\boldsymbol{X}_{\mathcal{X}(J)}) \right]$, subject to $J \in \Theta_{s}$ for some constraints $\Theta_{s}$'s. These constraints are such that $\Theta_{1} = \dots = \Theta_{K} = \{\{1\}, \dots, \{M\}\}$, and $\Theta_{K+1}, \dots, \Theta_{2K}$ are related to tree heredity. Furthermore, $L(R_{s-1}^{\star}, \widehat{J}_{s})$ is asymptotically equivalent to the sample split score at the $s$th split. In essence, given our tree model regularization \eqref{gini2}, the first 
$K$ updates of Collaborative Trees follow a standard matching pursuit~\citep{mallat1993matching}, while the subsequent 
$K$ updates involve tree heredity and projections onto multiple feature groups. In Theorem~\ref{theorem3},  our analysis relies on this connection to establish results ensuring the sure screening property of $\widehat{J}_{1}, \dots, \widehat{J}_{2K}$, as stated in the same theorem. The following technical conditions are required for Theorem~\ref{theorem3}.
\begin{condi}\label{model.1}
Assume $\boldsymbol{X}$ is a binary random vector. If  $\texttt{\#}\mathcal{X}(m)>1$ for some $1\le m \le M$, then $\sum_{j\in\mathcal{X}(m)}\boldsymbol{1}_{X_{j} = 1} = 1$  and $ \texttt{\#} \mathcal{X}(m)\le M_{\mathcal{X}}$ for some  constant $2\le  M_{\mathcal{X}}<\infty$. Let $Y= f(\boldsymbol{X}) + \varepsilon$ with $\mathbb{E}(f(\boldsymbol{X})) = 0$,  $\max_{\vv{c} \in \{0, 1\}^{p}}|f(\vv{c})|\le M_{f}$ for some constant $1< M_{f}<\infty$, and zero mean model error $\mathbb{E}(\varepsilon) = 0$ with $\mathbb{E}(\varepsilon \boldsymbol{1}\{|\varepsilon| \le  c \} |\boldsymbol{X}) = 0$ for each $c>0$ and $\mathbb{E}|\varepsilon|^{q_{0}} < M_{\varepsilon}$  for some constants $q_{0} >0$ and  $M_{\varepsilon} <\infty$. 
\end{condi}

\begin{condi} \label{signal.strength.1}
 Assume $\textnormal{Var}(g_{m}(\boldsymbol{X})) > 2d_{n} +  42\delta_{0} p_{\min}^{-2}  \mathbb{E}[(f(\boldsymbol{X}))^2] $ and \sloppy $\textnormal{Var}( g_{J}(\boldsymbol{X}) - (\sum_{j\in J} g_{j}(\boldsymbol{X}))) > 2d_{n} + 7616  \delta_{0} (\texttt{\#}S_{2}^*)^4  p_{\min}^{-2} \mathbb{E}[(f(\boldsymbol{X}))^2]$ for some $d_{n} >0$,  each $m \in S_{1}^*$, and each $J \in S_{2}^*$, where $p_{\min}  = \min_{ 1\le l<k\le M , i \in \mathcal{X}(l) , j \in \mathcal{X}(k) , (a, b)\in \{0, 1\}^2 }\mathbb{P}(X_{i} = a, X_{j} = b)$, and
 \begin{equation*}
    \begin{split}        
        \delta_{0} &=  \max_{1\le m\le M}\inf \left\{\delta\ge 0 : \mathbb{P} \left( \max_{i\in\mathcal{X}(m)}| \mathbb{P}(X_{i} = 1|\boldsymbol{X}_{-\mathcal{X}(m)})  - \mathbb{P}(X_{i} = 1) | \le \delta \right) = 1 \right\}.
    \end{split}
\end{equation*}
\end{condi}

\begin{condi}[Tree heredity] \label{heredity.1}
    
    For every $S^{\dagger}\subset S_{1}^*$ and $J\in S_{2}^*$ such that each element in $S^{\dagger}$ belongs to a distinct set in $(S_{2}^*\backslash J)$, it holds that $J \cap (S_{1}^*\backslash S^{\dagger}) \not=\emptyset$. 
\end{condi}


The regularity conditions on $f(\boldsymbol{X})$ and model error in Condition~\ref{model.1} are assumed to simplify the application of standard concentration inequalities. No specific heteroskedastic assumption on $\textnormal{Var}(\varepsilon | \boldsymbol{X})$ is required.
 The condition $p_{\min} \ge p_{01} > 0$ assumed in Theorem~\ref{theorem3} holds if feature groups are independent  and that  $ \mathbb{P}(X_{j} = 1)$ is bounded away from one and zero for all binary features.  Additionally, when all components in $S_1^*$ are selected, Condition~\ref{heredity.1} ensures tree heredity, allowing for the selection of all components in $S_2^*$. For example, the case where $S_1^* = \{2, 4\}$ and $S_2^* = \{\{2, 3\}, \{2, 4\}\}$ is not permitted, whereas the case where $S_1^* = \{3, 4\}$ and $S_2^* = \{\{2, 3\}, \{2, 4\}\}$ satisfies Condition~\ref{heredity.1}. Hence, Condition~\ref{heredity.1} is more restrictive than weak heredity~\citep{bien2013lasso}, but it is natural for sum of trees.

Condition~\ref{signal.strength.1} requires the signal strength of significant components to surpass the sum of $2d_{n}$, representing estimation variance, and an additional term capturing bias due to feature group distributional dependence, measured by $\delta_{0}$. This condition ensures that all important components are sufficiently strong and can be effectively screened by Collaborative Trees, as stated in Theorem~\ref{theorem3}. The technical bias terms may benefit from improvement through a more careful analysis. 

In Theorem~\ref{theorem3}, we assume that all nodes in the initial $2K$ updates are valid  in Algorithm~\ref{algorithm1}, and define $\mathbb{E}(h(\boldsymbol{X})|\boldsymbol{X} = \vv{c}) = 0$ when $\mathbb{P}(\boldsymbol{X} = \vv{c}) = 0$ for every $\vv{c} \in \{0, 1\}^{p}$ and every measurable function $h:\mathbb{R}^p\mapsto\mathbb{R}$. Recall $R_{l}^{\star}(\boldsymbol{X})$ and $\widehat{J}_{l}$ from equation \eqref{mp.3}. The technical proofs are provided in the Supplementary Material.

\begin{theorem}\label{theorem3}
Assume $\widehat{J}_{l} \not= \widehat{J}_{k}$ for each $\{l, k\}\subset \{1, \dots, 2K\}$, Condition~\ref{model.1}, and that  \sloppy$(\boldsymbol{X}_{1}, Y_{1}), \dots, (\boldsymbol{X}_{n}, Y_{n}), (\boldsymbol{X}, Y)$ are i.i.d. Assume Conditions~\ref{signal.strength.1}--\ref{heredity.1} hold with $p_{\min}\ge p_{01} \vee \frac{2\log{(n)}}{\sqrt{n}}$, and $d_{n} = L_{1}  n^{-\frac{1}{2}} (\log{n}) (n^{ \frac{1}{q_{1}}}\sqrt{2K} + \sqrt{\sum_{l=1}^{2K} \iota_{l-1}^2} ) (\iota_{2K-1}^2 + \iota_{2K-1} + 3)$, in which $0<p_{01}<1$ is a constant, $L_{1}> 0$ is a sufficiently large constant, $q_{1}$ is an arbitrary constant with $q_{0} > q_{1} > 0$, and that $\iota_{-1} = \iota_{0}  = 0$ and
\begin{equation}
    \label{iota.1}
    \begin{split}
    \iota_{s} &= \sum_{l=1}^{s}  \max_{\vv{c} \in \{0, 1\}^{\texttt{\#} \mathcal{X}(\widehat{J}_{l}) }} \left| \mathbb{E}[(f(\boldsymbol{X}) - R_{l-1}^{\star}(\boldsymbol{X})) |\boldsymbol{X}_{\mathcal{X}(\widehat{J}_{l})} = \vv{c} ] \right| \qquad \textnormal{ for } s > 0.
    \end{split}
\end{equation}
Assume $M\ge K \ge \texttt{\#}S_{1}^* \vee \texttt{\#}S_{2}^*$, $d_{n} \le 1$, and that $M = O(n^{K_{01}})$ for some arbitrary constant $K_{01}>0$. Then, $\mathbb{P}(\{S_{1}^* \not\subset \cup_{l=1}^{K} \widehat{J}_{l}\}\cup ( S_{2}^* \not\subset \{\widehat{J}_{K+1}, \dots, \widehat{J}_{K+\texttt{\#}S_{2}^*}\} ) ) = o(1)$.

\end{theorem}

Theorem~\ref{theorem3} investigates Collaborative Trees' capability to select numerous significant components from high-dimensional input features; note that the tree model is not necessarily consistent after sure screening important components. According to the current theory, a necessary condition for the estimation error $d_{n}$ to diminish asymptotically is $K\le n^{\frac{1}{7}}$, which is derived from Supplementary Material, where we show that $\iota_{s} \le s c_{0}$ for some constant $c_{0}>0$. Corollary~\ref{corollary.1} below relaxes the requirement on $K$ by assuming independent features, although the results still do not achieve optimality in the number of significant components allowed in the model. For example, the Lasso allows
$o(\sqrt{n / \log{p}})$ important features given proper regularity conditions~\citep{buhlmann2011statistics}; this rate may be further improved under stronger conditions. Despite technical limitations that prevent optimal results, Theorem~\ref{theorem3} demonstrates Collaborative Trees' ability to handle well beyond $\log{(n)}$ significant components, which is advantageous than decision trees, as have introduced at the beginning of Section~\ref{Sec4qq}.
Such a result is further illustrated in Corollary~\ref{corollary.1} below, where notation follows that in Theorem~\ref{theorem3}, while $S_{1}\subset\{1, \dots, M\}$ and $S_{2}$ takes the form $\{\{m_{1}, m_{2}\}, \{m_{3}, m_{4}\}, \dots \}$.


\begin{corollary}\label{corollary.1}
Let $M_{f}>1$ be an arbitrary constant, and $S_{1}$ and $S_{2}$ be two sets satisfying Condition~\ref{heredity.1} with $\texttt{\#}S_{1} \vee \texttt{\#}S_{2}  \le K \le (\log{n})^{-1}n^{\frac{1}{5} - \frac{2}{5q_{1}}}$.  Let $f(\boldsymbol{X}) = \sum_{m\in S_{1}} h_{m}(\boldsymbol{X}_{\mathcal{X}(m)}) + \sum_{J\in S_{2}} h_{J}(\boldsymbol{X}_{\mathcal{X}(J)})$ with $\delta_{0} = 0$ (i.e., independent feature groups), in which $\boldsymbol{X}$ satisfies Condition~\ref{model.1} with $p_{\min} \ge p_{01}$, $\mathbb{E}(h_{m}(\boldsymbol{X}_{\mathcal{X}(m)})) = 0$ and $\mathbb{E}(h_{J}(\boldsymbol{X}_{\mathcal{X}(J)})|\boldsymbol{X}_{\mathcal{X}(l)}) = 0$ for each $m\in S_{1}$ and each $l\in J\in S_{2}$, and that $\textnormal{Var}(h_{J}(\boldsymbol{X}_{\mathcal{X}(J)})) = 16L_{1} n^{-\frac{1}{2}} (\log{n}) n^{ \frac{1}{q_{1}}}\sqrt{2K} \eqqcolon w_{n}$ for each $J \in \{\{m\}: m\in S_{1} \} \cup S_{2}$. Then, 
 $o(1) = w_{n} \ge 2d_{n}$ and $\max_{\vv{c}\in\{0, 1\}^{p} } |f(\vv{c})| < M_{f}$ for all large $n$. In addition, if other required conditions for Theorem~\ref{theorem3} are assumed, then $\mathbb{P}(\{S_{1} \not\subset \cup_{l=1}^{K} \widehat{J}_{l}\}\cup ( S_{2} \not\subset \{\widehat{J}_{K+1}, \dots, \widehat{J}_{K+\texttt{\#}S_{2}}\} ) ) = o(1)$.
\end{corollary}
Given the assumption of a bounded regression function, the variance of each component $w_{n}$ has to decrease asymptotically to accommodate an increasing (polynomially with respect to 
$n$) $\texttt{\#}S_{1} \vee \texttt{\#}S_{2}$, as stated in Corollary~\ref{corollary.1}. Example~\ref{example.xor} below is an instance of  $f(\boldsymbol{X})$ in Corollary~\ref{corollary.1}, where $S_{1}$ and $S_{2}$  respectively consist of linear  components and XOR interaction components. 
\begin{exmp}
    \label{example.xor}
    Assume each feature group has a single feature with $M=p$. Let 
    $f(\boldsymbol{X}) = \big[\sum_{m\in S_{1}} \beta_{m}(X_{m} - \mathbb{E}(X_{m}))    \big]
    + \sum_{\{l, k\}\in S_{2}} [\beta_{lk}\boldsymbol{1}_{(X_{l} - \mathbb{E}(X_{l}) ) (X_{k} - \mathbb{E}(X_{k})) >0 } - \beta_{lk}\boldsymbol{1}_{(X_{l} - \mathbb{E}(X_{l})) (X_{k} - \mathbb{E}(X_{k})) <0 }]$, in which  $\beta_{m}$'s and $\beta_{lk}$'s are real coefficients, $X_{j}$'s are i.i.d. Bernoulli random variables.
\end{exmp}
When dealing with non-identically distributed features, groups with more than one feature, and a small $\delta_{0}>0$, one cannot anticipate a clean model setting as in Example~\ref{example.xor}. Nevertheless, Theorem~\ref{theorem3} guarantees Collaborative Trees identifying all important components, whose count potentially grows polynomially with sample size given regularity conditions and small $\delta_{0}$, in high-dimensional nonlinear settings.



\subsection{Analysis of additive and interaction effects} \label{Sec4.2qq}

In this section, we analyze the XMDIs defined in Section~\ref{Sec3qq}. Recall $\sum_{k=1}^K T_{k}(\boldsymbol{X})$ from Algorithm~\ref{algorithm1}, $R_{l}^{\star}(\boldsymbol{X})$ from \eqref{mp.3}, and $g_{J}(\boldsymbol{X})$ from Section~\ref{Sec3qq}. The technical proof of Theorem~\ref{theorem2} is provided in Section~\ref{technical.proof} of the Supplementary Material.

\begin{theorem}\label{theorem2}
Assume all conditions in Theorem~\ref{theorem3}. Then, the following results hold with probability approaching one. For each feature group $m\in \{1,\dots, M\}$,
{\begin{equation*}
    \begin{split}
         |\textnormal{XMDI}_{mm} - \textnormal{Var}(g_{m}(\boldsymbol{X}))| & \le d_{n} + L_{3}\delta_{0},\\
   \sum_{k\not= m} |\textnormal{XMDI}_{mk} - \textnormal{Var}(g_{mk}(\boldsymbol{X}) - g_{m}(\boldsymbol{X}) - g_{k}(\boldsymbol{X}))|  &\le \Delta +  \widehat{\Gamma} + (\texttt{\#}S_{2}^* + 1)d_{n} + L_{3}\delta_{0}  (\texttt{\#}S_{2}^*)^5,
    \end{split}
 \end{equation*}}%
where $d_{n}$ and $\delta_{0}$ are given in Theorem~\ref{theorem3}, $L_{3}>0$ is a sufficiently large constant, $\Delta = \textnormal{Var}(f(\boldsymbol{X}) - R_{K + \texttt{\#}S_{2}^*}^{\star}(\boldsymbol{X}))$, and $\widehat{\Gamma}  = (n^{-1}\sum_{i=1}^n \varepsilon_{i}^2)  - n^{-1}\sum_{i=1}^n (Y_{i} - \sum_{k=1}^K T_{k}(\boldsymbol{X}_{i}) )^2$.
\end{theorem}


In the context of modern MDI feature importance analysis~\citep{scornet2023trees, louppe2013understanding, li2019debiased}, Theorem~\ref{theorem2} is the first to formally separate feature importance of trees into additive and interaction effects. 
Our results ensure the reliable inference when feature dependence is mild and the number of significant interaction components is limited. In Theorem~\ref{theorem2}, the second upper bound involves $\texttt{\#}S_{2}^*\times d_{n}$ since we sum up all estimation errors resulting from interaction estimation. The term $(\texttt{\#}S_{2}^*)^5$ may be improved through a more careful analysis.

\begin{remark}
As mentioned at the beginning of Section~\ref{Sec4qq}, the analysis of $\Delta$ and $\widehat{\Gamma}$, which is related to tree model consistency, could be too involved for the current paper. Instead, we empirically assess the consistency quality of our tree model in Section~\ref{Sec6.2}. On the other hand, it is well-documented that impurity-based measures exhibit bias towards irrelevant features correlated with relevant ones~\citep{louppe2013understanding, strobl2008conditional}. Despite ongoing efforts to mitigate this bias~\citep{li2019debiased, loecher2020unbiased, zhou2021unbiased, benard2022mean, chi2022fact}, a formal analysis of bias in feature importance derived from sum-of-trees models is still lacking. While our Theorem~\ref{theorem2} is limited in assessing the bias issue due to the stringent technical requirements on $\delta_{0}$, we provide empirical experiments in Section~\ref{Sec5.1} demonstrating that the XMDIs are bias-resistant.

\end{remark}

\section{Simulation experiments}\label{Sec5qq}


 This section demonstrates XMDIs' numerical stability and resistance to bias from irrelevant correlated features. We also investigate key hyperparameters' impact on Collaborative Trees Ensemble's inferences using artificial datasets generated from Model~\eqref{Y1}, where each feature group contains a single feature. Groups are therefore ignored for simplicity.
\begin{align}
        Y & = 5X_{1} + 20 (X_{3} - 0.5)^{2} + 15X_{5} + 2X_{9} + 10\sin{(\pi (X_{9} - 0.5)(X_{10}-0.5))} + \varepsilon, \label{Y1}
\end{align}
    where $\varepsilon$ is an independent standard Gaussian model error. The covariate distribution of $(X_{1}, \dots, X_{p})^T$ is chosen as a Gaussian copula, ensuring that $0\le X_{j}\le 1$ almost surely, with an AR(1) covariance matrix $\Sigma = (\lambda^{|k-l|})$ for each $\{k,l\}\subset \{1,\dots, p\}$, in which $0\le \lambda<1$ and $p\in \{10, 100\}$. Specifically, $(X_1,\cdots, X_p)$ has a joint CDF function $\Phi_{\Sigma}(\Phi^{-1}(x_1),\cdots, \Phi^{-1}(x_p))$, where $\Phi_{\Sigma}$ is the CDF of multivariate Gaussian with mean zero and covariance matrix $\Sigma$, and $\Phi^{-1}$ is the inverse of the univariate standard Gaussian CDF. Model~\eqref{Y1} is a variant of the Friedman regression function~\citep{friedman1991multivariate}, which is  used for testing the selection power for the linear main effect, quadratic effect, and interaction effect. A weak additive component $2X_{9}$ is included in the model to ensure the tree heredity and stabilize the numerical results.


The hyperparameter tuning of Collaborative Trees Ensemble (see Section \ref{Sec2.4}) is performed as introduced in the Supplementary Material, with $20$ runs of optimization. Some important tuned hyperparameters are $K =12$ and $\textsf{n\_bins} \in\{ None, 5\}$ for the simulation experiments in Section~\ref{Sec5qq}. Here, the original features $X_{j}$'s are used when $\textsf{n\_bins} = None$. Otherwise, every continuous feature is binned into \textsf{n\_bins} bins, and becomes a feature group. We set the sample size $n = 500$ for all simulation experiments.

\subsection{XMDIs are bias-resistant and numerically stable}\label{Sec5.1}

In this section, we empirically demonstrate that the XMDIs 
 of Collaborative Trees Ensemble can effectively separate significant and insignificant components in learning environments with high correlation between features. To illustrate the bias resistance of XMDIs, we conduct simulations based on Model~\eqref{Y1} with $\lambda \in\{0.1, 0.8\}$ and $p=100$. The results, presented in Table~\ref{tab:xmdi_stable}, indicate the rates at which a threshold exists to distinguish the importance measures of significant and insignificant components in Model~\eqref{Y1}. The significant features are defined within the index set $S = \{1, 3, 5, 9, 10\}$, where four additive components are respectively based on $(X_{1}, X_{3}, X_{5}, X_{9})$ and an interaction component is based on $(X_{9}, X_{10})$. Meanwhile, the noisy features are in $S^c \coloneqq \{1, \dots, 100\}\backslash S$. It is important to note that our focus here is not on estimating the separation threshold. We compare the separation rates obtained from 100 independent and repeated simulation experiments with those obtained using Random Forests MDI, Sobol' indices~\citep{sobolprime1993sensitivity, saltelli2010variance}, and conditional permutation importance (CPI~\citep{strobl2008conditional}). Among many methods~\citep{benard2022mean, li2019debiased, loecher2020unbiased, zhou2021unbiased, chi2022fact}, CPI is a well-established and representative measure proposed for addressing the bias issue of Random Forests MDI. The total (overall), first-order (additive), and second-order (interaction) Sobol' indices~\citep{sobolprime1993sensitivity, saltelli2010variance} are calculated with Random Forests as the prediction model. See the Supplementary Material for additional details of Sobol' indices and hyperparameter tuning.

In column (I) of Table~\ref{tab:xmdi_stable}, XMDI and CPI consistently distinguish between important features and noisy variables across 100 experiments, showcasing their numerical stability and robustness in the presence of highly correlated features ($\lambda=0.8$) and numerous noisy variables ($p=100$). The column (III) shows that the estimated additive effects of $X_{1}$, $X_{3}$, and $X_{5}$ based on the XMDI indeed consistently dominate those from insignificant features in $S^c$. However, in addition to results in Table~\ref{tab:xmdi_stable}, we report here that the additive effect of $X_{9}$ is too weak so the separation rates of $X_{9}$ to $S^c$ based on the XMDI are roughly 80\% for $\lambda\in \{0.1, 0.8\}$; more samples and a stronger signal can increase these rates. Moving forward to column (II),  the interaction effect between $X_{9}$ and $X_{10}$ is accurately identified by the XMDI, further underscoring its precision in capturing meaningful interactions amidst complex data scenarios. On the other hand, even with mild correlation $\lambda=0.1$, Sobol' indices struggle to effectively distinguish the interaction effect between $X_{9}$ and $X_{10}$. The scenario with $\lambda=0.8$ illustrates that Sobol' indices exhibit some resistance to bias, except when it comes to selecting interaction effects. It is worth noting that the inferior performance of MDI in the high correlation scenario underscores the considerable challenges posed by the learning environment for the methods under investigation, thereby emphasizing the bias-resistant nature of XMDI and CPI, despite the fact that CPI does not apply in cases (II) and (III). These results reinforce the reliability and informativeness of our network diagrams for drawing insightful inferences.


\begin{table}
		\begin{center}
			{
				\begin{tabular}[t]{ |c|c cc|c cc|}
                    \hline
                       \multirow{2}{*}{Method}  & \multicolumn{3}{c|}{$\lambda = 0.1$} & \multicolumn{3}{c|}{$\lambda = 0.8$} \\
                        & (I) & (II) & (III) & (I) & (II) & (III) \\
					 \hline {  XMDI of Collaborative Trees Ensemble} & 1.00 & 1.00 & 1.00 & 1.00 & 1.00 & 1.00\\ 
      \hline {  Sobol' indices~\citep{sobolprime1993sensitivity} \hspace{3.55cm} } & 0.93 & 0.10 & 0.83 & 0.85 & 0.00 & 0.70 \\ 
      \hline {MDI~\citep{Breiman2001} of Random Forests \hspace{4.5em}} & 1.00 & \multicolumn{2}{c|}{\footnotesize Not Applicable} & 0.37 & \multicolumn{2}{c|}{\footnotesize Not Applicable} \\ 
      \hline { CPI~\citep{strobl2008conditional} of Random Forests \hspace{4.6em}} & 1.00 & \multicolumn{2}{c|}{\footnotesize Not Applicable}& 0.98 & \multicolumn{2}{c|}{\footnotesize Not Applicable}\\ 
                    \hline  
                    
			\end{tabular} }
			\caption{(I) Success rates of distinguishing significant and insignificant features in Model~\eqref{Y1} based on overall feature importance measures. (II) Rates of the interaction effect between $X_{9}$ and $X_{10}$ being the maximum among all pairs. (III) Rates of the additive effects of $(X_{1}, X_{3}, X_{5})$ dominating those from insignificant features in $S^c$. All rates are obtained from 100 simulation experiments with $p=100$ features and $n=500$. MDI and CPI do not infer feature interaction or additive effects. } \label{tab:xmdi_stable}
		\end{center}
	\end{table}

\subsection{Number of trees $K$, data binning, and feature correlation}

\begin{figure}
 \captionsetup[subfigure]{width=0.49\textwidth}

    \subfloat[CTE with   $\textsf{n\_bin} = 5$.]{
\centering
 \includegraphics[width=0.50\linewidth]{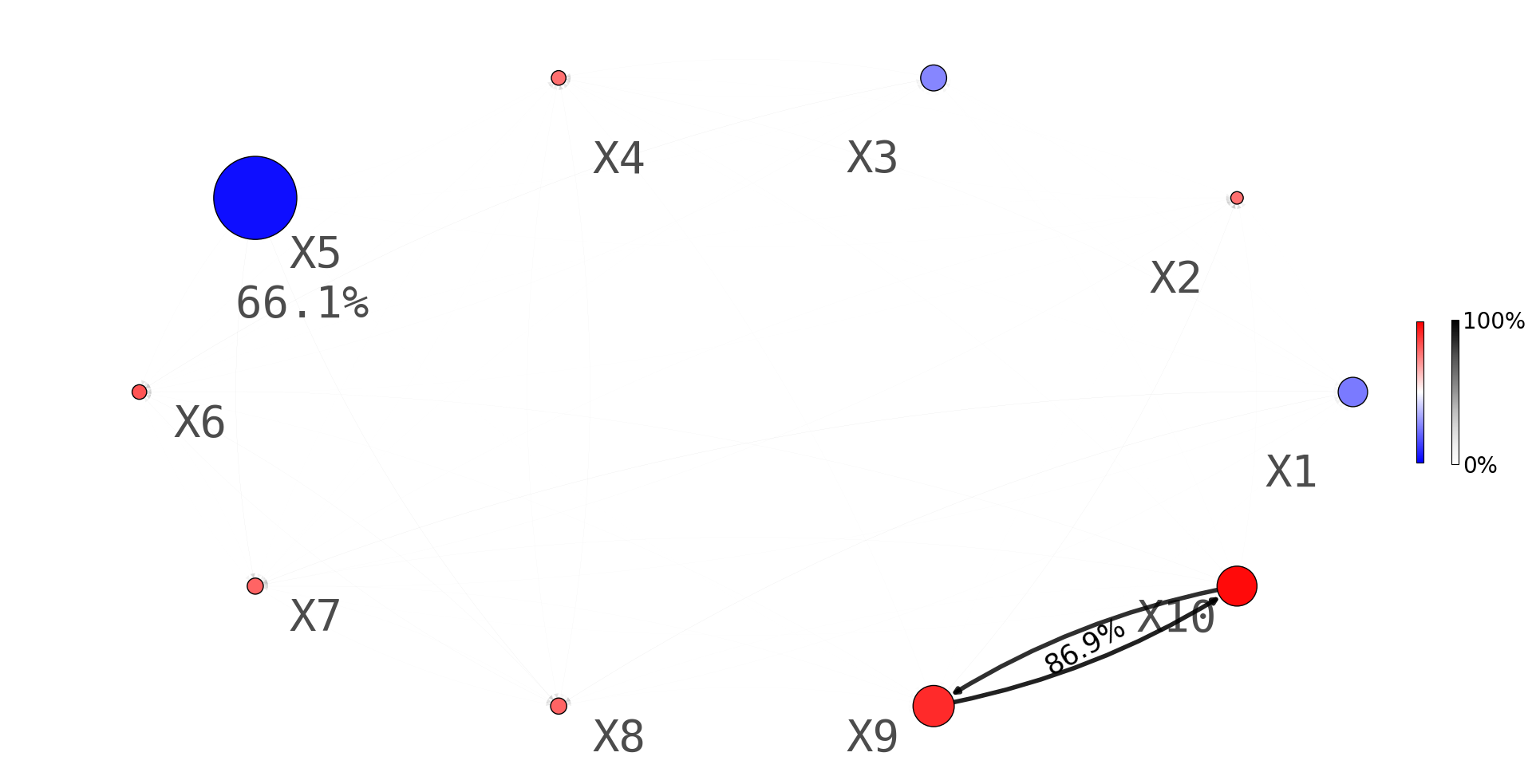}
}
\subfloat[CTE uses the original continuous features ($\textsf{n\_bin} = None$).]{
\centering
  \includegraphics[width=0.5\linewidth]{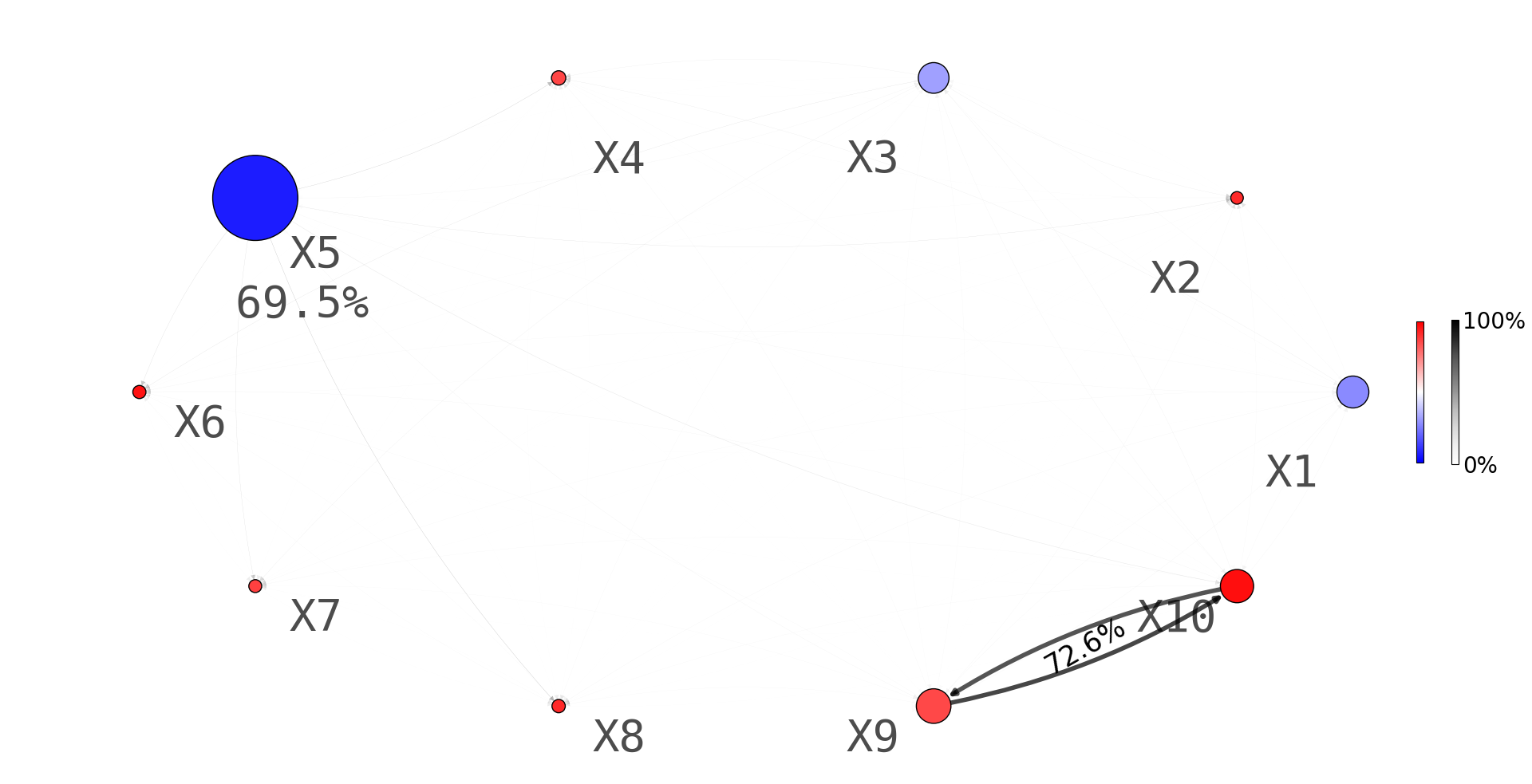}
}

 \subfloat[CTE with high feature correlation $\lambda = 0.8$.]{
\centering
 \includegraphics[width=0.5\linewidth]{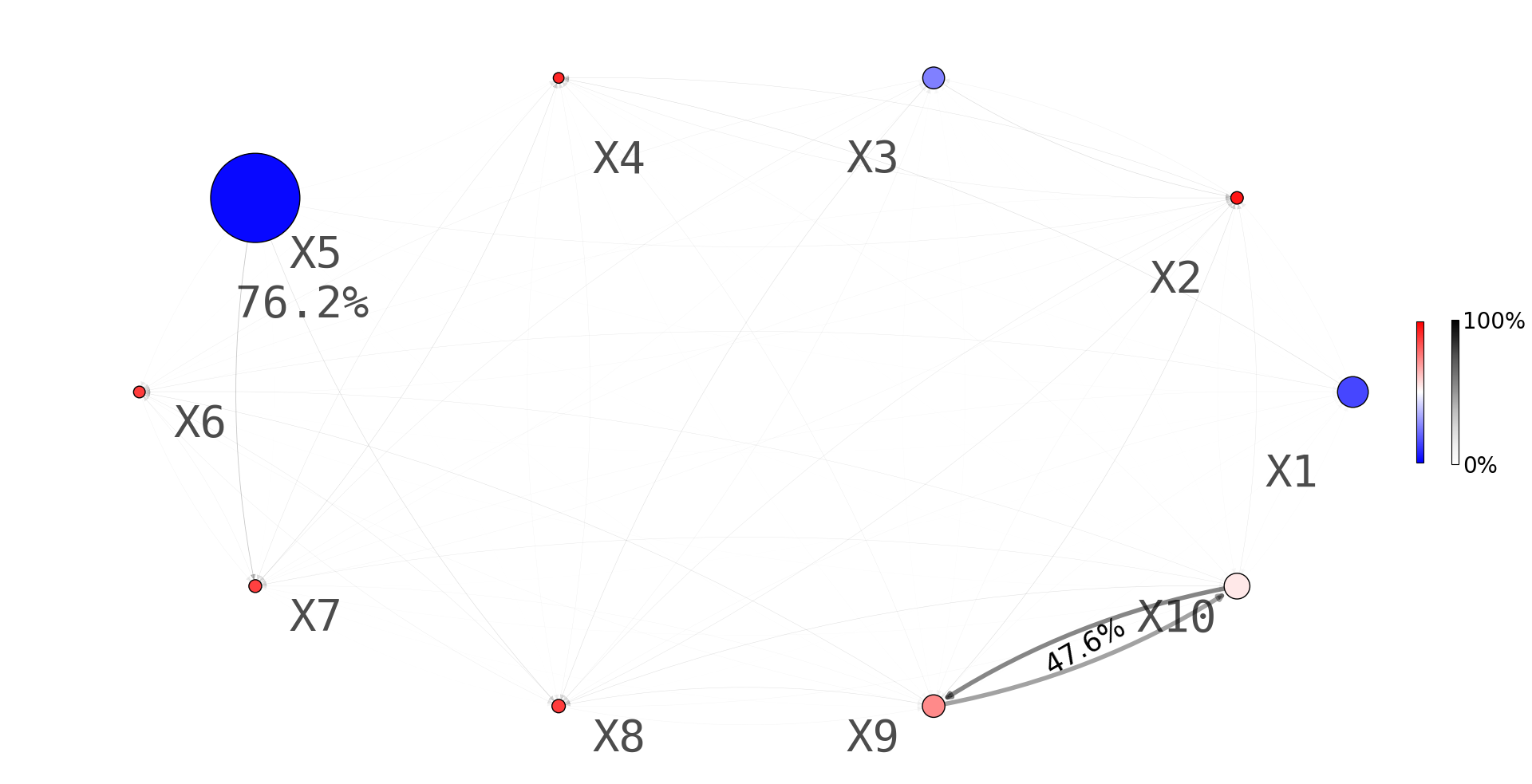}
}
\subfloat[CTE with only a single tree $K = 1$.]{
\centering
  \includegraphics[width=0.5\linewidth]{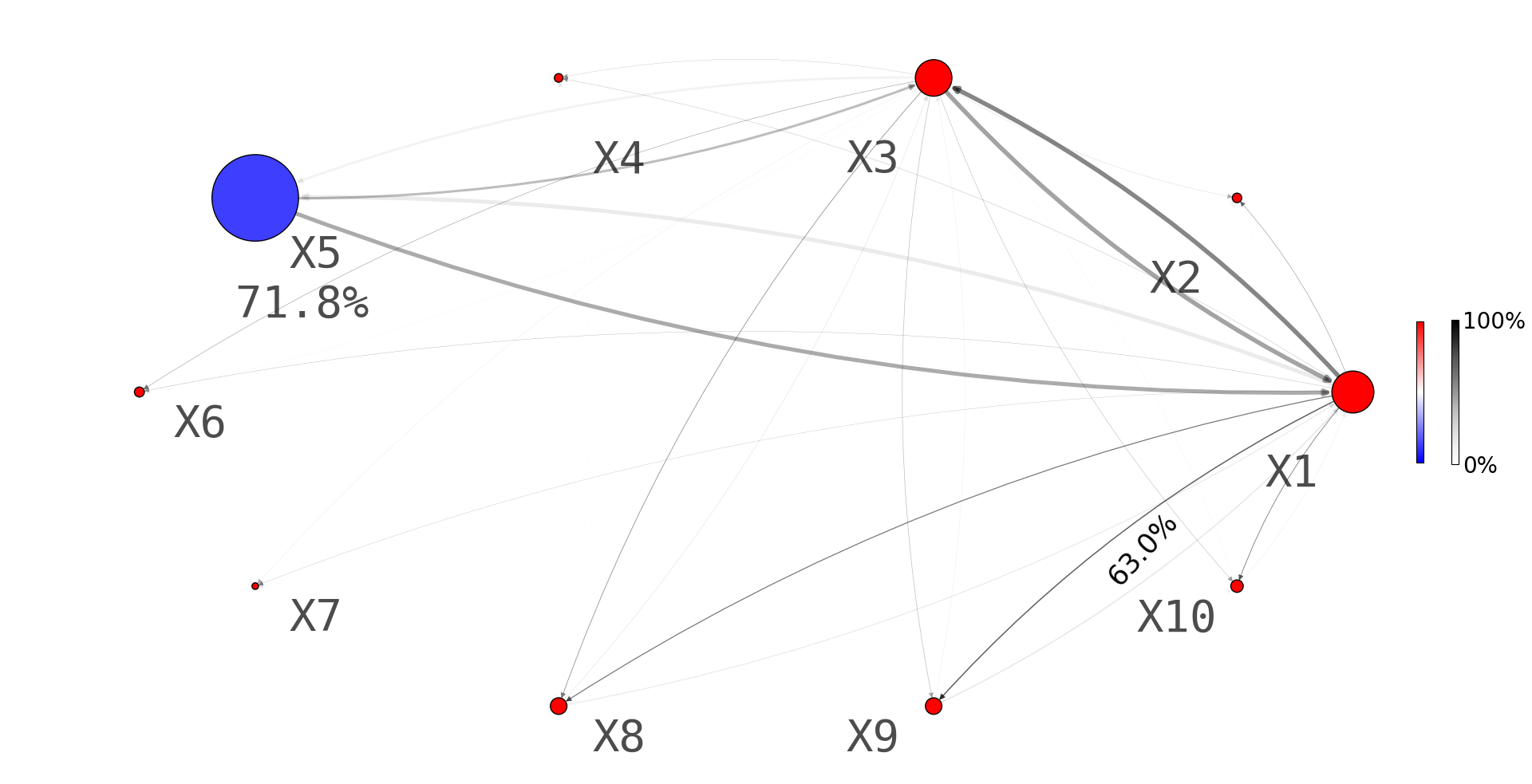}
}
\caption{Network diagrams for data simulated from Model~\eqref{Y1}  with $p=10$, $n= 500$, and other varying parameters indicated in the corresponding panels. }\label{fig:networks}
\end{figure}

Figure~\ref{fig:networks} presents network diagrams (see Section~\ref{Sec3.3qq}) for data simulated from Model~\eqref{Y1} with $p=10$, in which other varying parameters are indicated in the corresponding panels. From Figure~\ref{fig:networks}, XMDI successfully distinguishes significant features from noisy ones in Model~\eqref{Y1} across all cases, except for panel (d) with $K=1$. In this particular case, the importance of $X_{9}$ and $X_{10}$ might be underestimated, and $X_{3}$ and $X_{1}$ are incorrectly identified as interaction components. The results in panel (d) underscore the crucial role of $K$ in our tree model for making accurate inferences, consistent with our theoretical expectations. Meanwhile, the major distinction of the other three panels lies in the inference about the interaction components based on $X_{9}$ and $X_{10}$. A comparison between panels (a) and (b) shows that whether to use binned features has only a mild impact on the inferences of our tree models in this experiment. This result suggests the inferences of the XMDI may still be informative without binning.

The adverse effects of feature correlation on our inferences are clearly seen in panel (c). Notably, the importance of features $X_{9}$ and $X_{10}$ appears slightly underestimated in this scenario compared to panels (a) and (b), despite the excellent separation rates reported in Table~\ref{tab:xmdi_stable} (the same setup but with $p=10$ features here). Interestingly, the interaction effect account for roughly only $47\%$ for their overall feature importance, which is far lower than panels (a) and (b). The high feature correlation between $X_{9}$ and $X_{10}$ weakens the contribution of their interaction effects to their overall importance. This observation highlights the nuanced nature of studying interaction effects, emphasizing the importance of understanding both the regression function and feature correlation, namely, the joint distribution of $(Y, \boldsymbol{X})$. Collaborative Trees represent our attempt to address these complexities. Our finding echoes the empirical observations made by~\citep{chipman2010bart} regarding the importance of a sum of trees for BART. However,  their simulation study does not investigate complex feature effects or the potential bias of effect measures.

These simulation experiments enhance our understanding of the proposed tree model, and raise intriguing research questions. Key areas for further exploration include developing more efficient methods for binning features, exploring the feasibility of making valid inferences from general continuous features without binning, and advancing the XMDI to better resist bias by incorporating debiasing techniques, such as CPI~\citep{strobl2008conditional} and related works~\citep{benard2022mean, li2019debiased}.

\section{Real data inferences and prediction evaluation}\label{Sec6}

\subsection{Embryo growth data study}\label{Sec6.1}

Temperature-dependent sex determination, or TSD, is a common occurrence among vertebrates, including reptiles such as crocodilians, turtles, fish, and some lizards. TSD is crucial for species resilience to climate change impacts. Our analysis focuses on TSD using the``2022-03-29'' version of the embryo growth dataset comprising $874$ samples (with missing values in sex-ratio removed) from the R package \pkg{embryogrowth}~\cite{girondot2019embryogrowth}, covering data from 60 species, primarily reptiles.

Each hatch of eggs comprises an observation in the sample. For every hatch of eggs incubated in temperature-regulated chambers, the following information has been recorded: the sex ratio of females to total births in this hatch (response), species category (\texttt{species} and \texttt{subspecies}), the location where the eggs come from (\texttt{area} and \texttt{RMU}, short for Regional Management Unit), incubation period (\texttt{incubation periods (days)}), incubation temperature (\texttt{temperature}), and incubation temperature amplitude (\texttt{amplitude}). The temperature amplitude is the maximum variation of the incubation temperature, while temperatures are recorded in Celsius. Other details of the dataset can be found in~\citep{girondot2019embryogrowth, abreu2020recent}. There are $60$ species, $10$ subspecies, $81$ areas, and $14$ RMUs, which are all grouped into feature groups with one-hot indicators. We apply Collaborative Trees Ensemble based on these features for analyzing the sex ratio. The resulting network diagram is shown in the left panel of Figure~\ref{fig:relations}. Hyperparameter tuning involves 100 optimization runs, yielding $K = 11$ and $\textsf{n\_bins} = 40$, where continuous features are binned into \textsf{n\_bins} bins.

Different species are known to exhibit various patterns of TSD~\citep{lockley2021effects}, implying an interaction between \texttt{species} and \texttt{temperature} (each TSD pattern naturally depends on \texttt{temperature}). Our XMDI  inferences correctly identify the interaction between \texttt{species} and \texttt{temperature}, as illustrated in the left diagram of Figure~\ref{fig:relations}. Additionally, the diagram reveals a distinct additive effect associated with \texttt{temperature}, suggesting the prevalence of a specific TSD pattern in the dataset. This observation aligns with our experience, as the logistic-shaped TSD pattern, depicted in the right panel of Figure~\ref{fig:relations}, is frequently observed in the dataset. Now, let us investigate the importance of the variable \texttt{area}. Considering that same species may inhabit same areas, a plausible explanation for the clear interaction effect (the second strongest) between \texttt{area} and \textsf{temperature} is the potential synonymy of \texttt{area} with \texttt{species}. The distance correlation~\citep{szekely2009brownian} between \texttt{area} and \texttt{species} is as high as $0.67$, providing support for our explanation. Despite this, the overall importance of \texttt{area} and its interaction between \textsf{temperature} are weaker compared to \texttt{species}. This could be attributed to the presence of multiple distinct species in the same region, diminishing the importance of \texttt{area} relative to \texttt{species}.

On the other hand, the diagram in Figure~\ref{fig:relations} suggests that the effects of \texttt{incubation periods (days)} are not related to \texttt{species}, as the statistical association pattern for \texttt{incubation periods (days)} is markedly different from that of \texttt{species}. While we acknowledge that we are not experts in embryo growth analysis, the variable \texttt{incubation periods (days)} has less often been considered an important factor in TSD studies, making our results potentially interesting for further investigation.
Now, let us consider two potential explanations: 1.) Uncontrolled environmental factors, such as humidity, might influence \texttt{incubation periods (days)} and, consequently, the sex ratio. A more detailed examination of the samples is essential for a thorough understanding. 2.) The detected signal of \texttt{incubation periods (days)} in Figure~\ref{fig:relations} may be considered noise. Significance tests are required for further analysis.


Using the embryo growth dataset, we have demonstrated how to leverage Collaborative Trees Ensemble for advanced data analysis. We remark that the technical tree heredity requirement Condition~\ref{heredity.1} does not prevent the XMDI from correctly accounting for multiple interactions stemming from the same feature group in practice. A discussion, along with a larger version of Figure~\ref{fig:relations} and detailed numerical results, is provided in the Supplementary Material to save space.

\subsection{Prediction accuracy evaluation}\label{Sec6.2}

We evaluate the predictive performance of Collaborative Trees Ensemble and compare it with Random Forests~\citep{Breiman2001} and XGBoost~\citep{chen2016xgboost}. These models are known for their competitiveness when compared to modern deep learning models, as noted in~\citep{grinsztajn2022tree}. The benchmark comparison follows the methodology outlined in~\citep{grinsztajn2022tree}, a well-established work that provides a state-of-the-art benchmark procedure and simulation environments.

Following the approach outlined in~\citep{grinsztajn2022tree}, we utilize 19 selective datasets from OpenML, each containing no more than $10,000$ samples, for prediction evaluation based on the goodness-of-fit measure R$^2$ value. The calculation of each R$^2$ value involves splitting the full sample into training, validation, and test sets, constituting 48\%, 32\%, and 20\% of the full dataset, respectively. This division is employed for hyperparameter tuning based on the training and validation samples, as well as prediction loss calculation based on the test sample. We perform 100 optimization runs for the hyperparameter tuning procedure, without binning any features (i.e., \textsf{n\_bins} = $None$). To ensure result stability, we conducted each experiment on the 19 datasets independently 10 times. 

To summarize the comparison, we calculated the average (adjusted) win rates~\citep{grinsztajn2022tree} for Collaborative Trees Ensemble, XGBoost, and Random Forests, which are \textbf{0.8069, 0.5986, 0.2439}, respectively.  While XGBoost has been recognized for its strong performance against modern deep learning models in tabular regression~\citep{grinsztajn2022tree}, our tree model surpassed its performance, highlighting the competitive predictive capabilities of our proposed approach. For the detailed R$^2$ scores for each of the 19 datasets, along with other experiment specifics, please refer to the Supplementary Material.



\section{Discussion}
We introduce Collaborative Trees, a novel tree model designed for regression prediction, along with its bagging version, to analyze complex statistical associations between features and uncover potential patterns in the data. In conclusion, we highlight three challenges related to Collaborative Trees. Firstly, the study of bias in XMDI and rigorous significance tests are relevant for making more stable inferences. Moving forward, applying our analysis tools to scenarios with lots of features, like in genetic data analysis, poses a significant challenge. Lastly, investigating the consistency of Collaborative Trees and its bagging version is crucial for the applications of Collaborative Trees. 


	\bibliographystyle{chicago}
	\bibliography{references}

\clearpage

 \appendix

\setcounter{page}{1}
\setcounter{section}{0}
\renewcommand{\theequation}{A.\arabic{equation}}
\renewcommand{\thesubsection}{A.\arabic{subsection}}
\setcounter{equation}{0}
	
\begin{center}{\bf \Large Supplementary Material to ``Analyze Additive and Interaction Effects via Collaborative Trees''}
		
\bigskip
		
 Chien-Ming Chi
\end{center}
	
\noindent  The Supplementary Material provides supplementary content related to hyperparameter tuning (Section~\ref{hyper.tuning.1}), algorithm implementation (Section~\ref{SecA.2}), a runtime analysis (Section~\ref{computation.time}), details of the Sobol' indices (Section~\ref{sobol.appendix}), a larger version of Figure~\ref{fig:relations} and detailed numerical results for Figure~\ref{fig:relations} (Section~\ref{Sec6.1.d}), and simulation experiments (Section~\ref{Sec6.1.c} and Section~\ref{Sec6.2b}). Additionally, Section~\ref{technical.proof} includes the proofs of Theorems \ref{theorem3}--\ref{theorem2}, Corollary~\ref{corollary.1}, and several technical lemmas. The notation used throughout remains consistent with that of the main paper. All Python codes used in the paper, including the implementation of Collaborative Trees Ensemble, are available upon request.

\section{Supplementary Material}

 \subsection{Hyperparameter tuning}\label{hyper.tuning.1}

Table~\ref{tab:validation_sapce} displays all hyperparameters of Collaborative Trees Ensemble introduced in Section~\ref{Sec2.4}, where the number of aggregated predictors is denoted by \textsf{n\_estimators}. In Table~\ref{tab:validation_sapce}, $\textsf{n\_bins}$ is required when transforming certain continuous or categorical variables into feature groups. Additionally, when $\textsf{random\_update} = 0$, Collaborative Trees randomly update one node sampled from the update node list $W$ at each round.

We use the Python package \pkg{hyperopt} from \url{https://hyperopt.github.io/hyperopt/} for tuning all predictive models. The tuning procedure is introduced as follows. The full dataset is split into three subsamples, representing the training, validation, and test sets, which constitute 48\%, 32\%, and 20\% of the full dataset, respectively. To determine the optimal hyperparameters for our predictive models, we assess $B$ distinct sets. Utilizing the Python package \pkg{hyperopt}, we perform hyperparameter sampling in each round. The training set is used to train tree models based on the sampled hyperparameters. Subsequently, the model's performance is evaluated on the validation set to score each set of hyperparameters. After conducting this process for $B$ rounds, we select the set with the highest evaluation scores as the best hyperparameters. A final tree model is then trained using this optimal set, taking into account both the training and validation samples, which represent 80\% of the full dataset. Finally, we compute a R$^2$ score on the test set for assessing the model's predictive performance. When prediction evaluation is unnecessary, we split the sample into 60\% for training and 40\% for validation.

The hyperparameter spaces of Collaborative Trees, XGBoost, and Random Forests are respectively given in Tables~\ref{tab:validation_sapce}--\ref{tab:validation_sapce_rf}. In Table~\ref{tab:validation_sapce}, \textsf{n\_estimators} is the number of boostrap aggregated trees, \textsf{n\_trees} is the number of collaborative trees $K$, \textsf{alpha} is the probability weights ciefficients used in the softmax function, \textsf{n\_bins} is the number of bins for binning continuous and categorical features, and that \textsf{random\_update}, \textsf{min\_samples\_split}, \textsf{min\_samples\_leaf}, \textsf{max\_depth} have been introduced in Section~\ref{Sec2.4}.  The choice of the \textsf{n\_trees} parameter depends on the number of significant components in the data generating models. Based on our experience in the data study, we recommend an upper limit of $12$ for \textsf{n\_trees}. Meanwhile, Tables~\ref{tab:validation_sapce_xgb}--\ref{tab:validation_sapce_rf} respectively present the hyperparameters of XGBoost and Random Forests. Additional details about these hyperparameters can be found on the respective package websites.

\begin{table}[t]
		\begin{center}{
				\begin{tabular}[t]{ |ccc|}
    \hline
    Parameter name & Parameter type & Parameter space \\
                         \hline \textsf{n\_estimators} (\texttt{\#} baggings) & integer & 100  \\
                        \textsf{n\_trees} ($K$ in Algorithm~\ref{algorithm1}) & integer & $\{6, \dots, 12\}$ \\
                        \textsf{alpha} & non-negative number & $\{0.001, 1, 10, 100, 10000, \infty\}$ \\
                        \textsf{min\_samples\_split} & integer & $\{5, 10, 15, 20, 30\}$\\ 
                        \textsf{min\_samples\_leaf} & integer& $\{0, 5, 10, 15, 20, 30\}$ \\
                        \textsf{n\_bins} & integer &  $\{5, 7, 10, 15, 20, 40\}$\\
                        \textsf{random\_update} & non-negative number & $\{0, 0.0001, 0.001, 0.01, 0.1, 1\}$ \\
       \textsf{max\_depth} & integer & $\{3, 5, 10, 20, 30, \infty\}$\\
					 \hline 
			\end{tabular} }
			\caption{Hyperparameter space for tuning Collaborative Trees Ensemble. Note that even when \textsf{random\_update} is less than 1, the first $2K$ updates of Collaborative Trees still adhere to our tree model regularization~\eqref{gini2}} \label{tab:validation_sapce}
		\end{center}
	\end{table}

\begin{table}[t]
		\begin{center}{
				\begin{tabular}[t]{ |ccc|}
    \hline
    Parameter name & Parameter type & Parameter space \\
                         \hline \textsf{n\_estimators} & integer & 1000  \\
                        \textsf{gamma} &  non-negative real number & $\exp{(Z)}$, $Z\sim  \textnormal{Uniform}[-8\log_{e}(10), \log_{e}(7)]$ \\
                        \textsf{reg\_alpha} & non-negative real number & $\exp{(Z)}$, $Z\sim  \textnormal{Uniform}[-8\log_{e}(10), \log_{e}(100)]$ \\
                        \textsf{reg\_lambda} & non-negative real number & $\exp{(Z)}$, $Z\sim  \textnormal{Uniform}[\log_{e}(0.8), \log_{e}(4)]$ \\ 
                        \textsf{learning\_rate} & non-negative real number & $\exp{(Z)}$, $Z\sim  \textnormal{Uniform}[-5\log_{e}(10), \log_{e}(0.7)]$ \\
                        \textsf{subsample} & positive real number &  $ \textnormal{Uniform}[0.5, 1]$ \\
                        \textsf{colsample\_bytree} & positive real number &  $ \textnormal{Uniform}[0.5, 1]$ \\
                        \textsf{colsample\_bylevel} & positive real number &  $ \textnormal{Uniform}[0.5, 1]$ \\
                        \textsf{min\_child\_weight} & non-negative integer &  Uniform$\{0, \dots, 20\}$\\
       \textsf{max\_depth} & non-negative integer &  Uniform$\{2, \dots, 15\}$\\
					 \hline 
			\end{tabular} }
			\caption{Hyperparameter space for tuning XGBoost. Details of these parameters can be found on the website \url{https://xgboost.readthedocs.io/en/stable/parameter.html}. } \label{tab:validation_sapce_xgb}
		\end{center}
	\end{table}

\begin{table}[t]
		\begin{center}{
				\begin{tabular}[t]{ |ccc|}
    \hline
    Parameter name & Parameter type & Parameter space \\
                         \hline \textsf{n\_estimators} & integer & 1000  \\
                        \textsf{gamma} &  non-negative real number & $\textnormal{Uniform}[0, 1]$ \\
                        \textsf{min\_samples\_split} & integer & $\{1, \dots, 20\}$\\ 
                        \textsf{min\_samples\_leaf} & integer& $\{2, \dots, 20\}$ \\
                        \textsf{min\_impurity\_decrease} & non-negative real number &  $ \textnormal{Uniform}\{0, 0.01, 0.02,0.05\}$ \\
                        \textsf{max\_depth} & non-negative integer &  Uniform$\{5, 10, 20, \infty\}$\\
       \textsf{criterion} & string &  $[\textnormal{squared\_error}, \textnormal{absolute\_error}]$ \\
					 \hline 
			\end{tabular} }
			\caption{Hyperparameter space for tuning Random Forests. Details of these parameters can be found on the website \url{https://scikit-learn.org/stable/modules/generated/sklearn.ensemble.RandomForestClassifier.html}} \label{tab:validation_sapce_rf}
		\end{center}
	\end{table}

\subsection{Collaborative Trees Ensemble implementation}\label{SecA.2}

Collaborative Trees Ensemble includes the following implementations in addition to Algorithm~\ref{algorithm1}, with
all hyperparameters given in Table~\ref{tab:validation_sapce}. 1.) To enhance prediction, we build Collaborative Trees with bootstrap aggregating, or bagging.  2.) A node of depth less than \textsf{max\_depth} is valid for splitting if its subsample size is larger than \sloppy$\max\{\textsf{min\_samples\_split}, \textsf{min\_samples\_leaf}\}$.
If a node has at least two child nodes whose subsample sizes are larger than \textsf{min\_samples\_leaf}, then these child nodes are valid for tree update.
3.) To reduce computation time, after the initial $2K$ updates, we update nodes in $W^{\dagger}$, randomly sampled from $W$ with $\texttt{\#}W^{\dagger} = \max\{\textsf{random\_update} \times \texttt{\#}W, 1\}$.
See Section~\ref{sec.cte.1} for details of the update node list $W$.
4.) To reduce estimation variance, Collaborative Trees make update decisions based on probability weights derived from split scores.
At each round of update with some update node list $W$, we compute a vector of split scores $\big( \sum_{ (C,k) \in Q} \textnormal{SplitScore} (C, m)\big) - \lambda(W, Q)$ as in \eqref{gini2} for each $Q\in W$, denoted by $\vv{SS}$.
We update the split sampled from the probability vector $\textnormal{softmax}(\textsf{alpha}\times \vv{SS})$, where $\textnormal{softmax}(\vv{z}) = \big[\sum_{j =1}^k \exp{(z_{j})}\big]^{-1} \times (\exp{(z_{1})}, \dots, \exp{(z_{k})})$ for $\vv{z}\in\mathbb{R}^k$. The case with $\textsf{alpha} = \infty$ corresponds to maximization as in \eqref{gini2}. This split sampling approach draws inspiration from BART~\citep{chipman2010bart}, known to enhance tree prediction accuracy in certain applications. Here, to streamline the sampling process, we adopt a softmax transform on the split scores.

\subsection{Computational time analysis}\label{computation.time}

We extracted a subsample of size $n=10,000$ from the ``superconduct'' dataset in \citep{grinsztajn2022tree}, sourced from OpenML (\url{https://www.openml.org/}). This dataset comprises 79 features. This sample is also used in Table~\ref{tab:r_square}. The runtime evaluation for Collaborative Trees Ensemble with a single bagging tree is conducted with the following hyperparameters: $K\in \{1, 3, 5, 7, 9, 11\}$, \textsf{min\_samples\_split} = 5, \textsf{min\_samples\_leaf} = 5, $\textsf{random\_update} \in \{0, 0.001, 0.01, 0.1, 1\}$, $\textsf{alpha} = \infty$, $\textsf{max\_depth} =20$, $\textsf{n\_bins} = None$. The specifics of the hyperparameters for Collaborative Trees Ensemble are outlined in Section~\ref{Sec2.4} and Section~\ref{SecA.2}.

The evaluation was performed on a Mac laptop with a 2.6 GHz 6-Core Intel Core i7 processor and macOS Sonoma Version 14.1.2. The environment details are as follows: Spyder version 5.4.3 (conda), Python version 3.10.12 64-bit, Qt version 5.15.2, PyQt5 version 5.15.7, Operating System: Darwin 23.1.0.

Although the parameters related to the tree structure (\textsf{min\_samples\_split}, \textsf{min\_samples\_leaf}, \textsf{max\_depth}) can significantly impact the model's training runtime, the evaluation focuses on the number of collaborative trees $K$ and \textsf{random\_update}, which determine the number of associated nodes considered at each update round. The results of runtime are reported in Table~\ref{tab:cptime}.

From Table~\ref{tab:cptime}, it takes approximately 5.5 hours ($200 * \frac{1}{60} * \frac{1}{60} * 100 \approx 5.5$) to train a Collaborative Trees Ensemble model with full updates (i.e., \textsf{random\_update} = 1), $K = 11$, and $100$ bagging trees on a sample of size $10,000$ with $79$ features. It is important to note that while the current implementation of Collaborative Trees Ensemble in Python is not yet optimized for computation (codes are written using the Python \pkg{numpy} package), we have implemented parallel training to expedite processing.

Additionally, it is worth mentioning that the number of bagging trees could potentially be much smaller than our default setting of 100. Furthermore, setting $\textsf{random\_update} = 1$ may not be necessary across the entire hyperparameter space in Table~\ref{tab:validation_sapce}, which could significantly reduce computational time, as indicated in Table~\ref{tab:cptime}.



\begin{table}[t]
		\begin{center}{
				\begin{tabular}[t]{ |cc|cccc|}
    \hline
  $K$ $\Big\backslash$ & \textsf{random\_update} & 0 &  0.01 &  0.1 & 1\\\hline
           1 & &  2 &  2 & 4 & 14 \\
           3 & & 5  &  5 & 11 & 43 \\
           5 & & 8 & 9 & 19 & 93\\
           7 & & 13  & 12 & 30 & 115 \\
           9 & &  15 & 16 & 35 & 142 \\
           11 & &  21 & 22 & 51 & 203 \\
					 \hline 
			\end{tabular} }
			\caption{Table showing the computational time evaluation for Collaborative Trees Ensemble with a single bagging tree (i.e., the number of bagging trees = 1). The table presents the computational time for various hyperparameter configurations, including the number of collaborative trees ($K$) and \textsf{random\_update} settings. Time measurements are displayed in seconds, rounded to the nearest second. Each entry represents the average runtime over 5 repetitions. } \label{tab:cptime}
		\end{center}
	\end{table}

\subsection{Sobol' indices in Section~\ref{Sec5.1}}\label{sobol.appendix}

We implemented the Sobol' indices using the R package \pkg{sensitivity}. This package requires a trained predictive model for calculating various importance measures. Similar to our approach for MDI and CPI, we trained a Random Forests model for the Sobol' indices. It is worth noting that there are several implementations available for calculating the Sobol' indices with a predictive model. We opted for the `sobolSalt` function in the package due to its user-friendly interface for calculating up to second-order importance measures.

For the Sobol' indices, we estimated the total Sobol' indices~\citep{sobolprime1993sensitivity, benard2022mean}, first-order indices, which are similar to XMDI$_{jj}$'s, and the second-order indices for each pair of features.

It is important to mention that the Random Forests algorithm in R differs slightly from its Python counterpart. We followed a similar hyperparameter tuning procedure as in Python as in Section~\ref{hyper.tuning.1}, but for simplicity, we do not provide detailed reporting here.

\subsection{Simulation experiment details for Section~\ref{Sec6.1}}

\subsubsection{Numerical analysis results for Section~\ref{Sec6.1}}\label{Sec6.1.d}

\begin{figure}
 \captionsetup[subfigure]{width=1.0\textwidth}
\centering
 \includegraphics[width=1.0\linewidth]{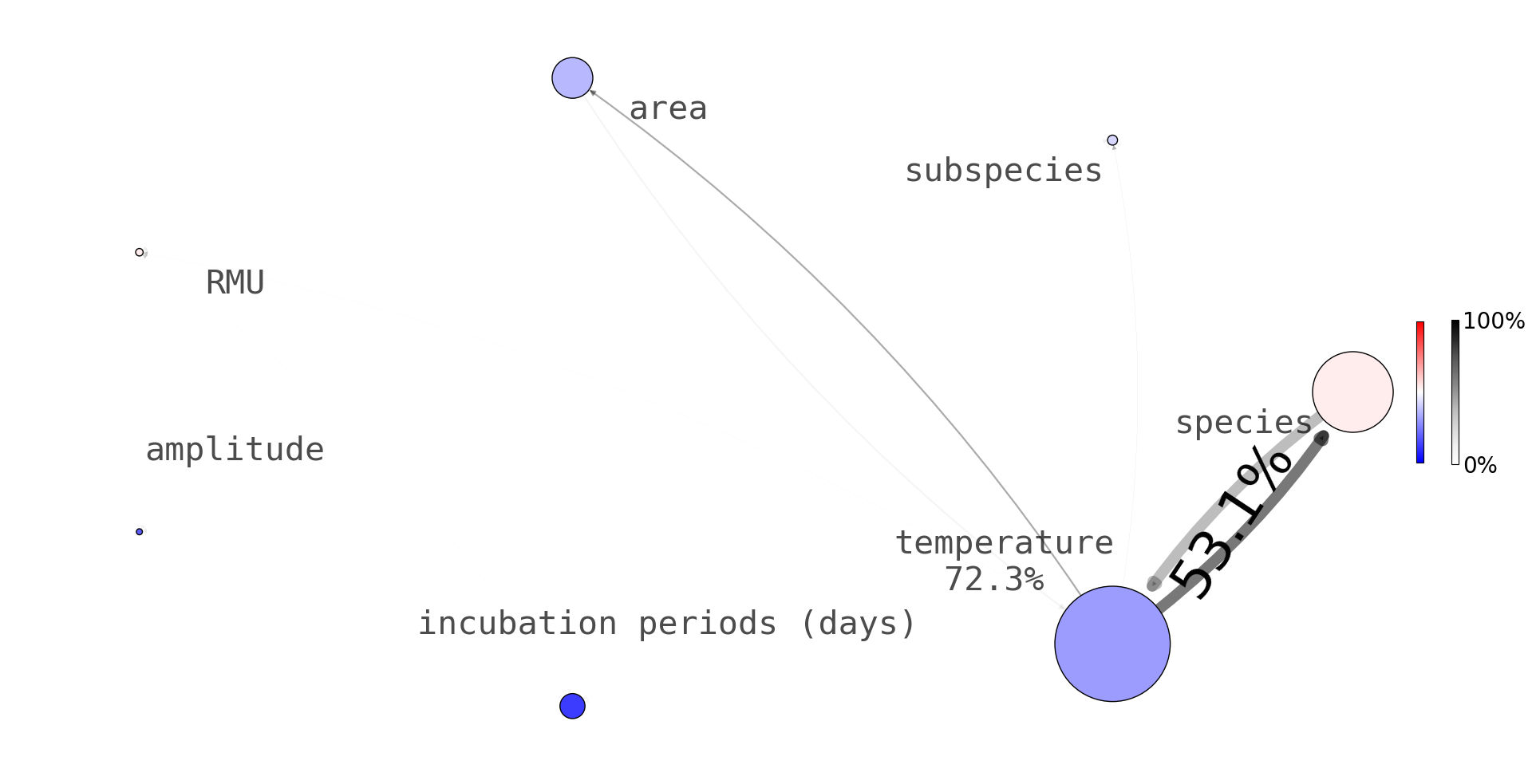}
\caption{Detailed view of Figure~\ref{fig:relations}.}\label{fig:relation2}
\end{figure}

Figure~\ref{fig:relation2} provides an enlarged version of the image displayed in Figure~\ref{fig:relations}. Additionally, Table~\ref{tab:relation} presents detailed numerical values used in Figure~\ref{fig:relations}, rounded to three decimal places. In Table~\ref{tab:relation}, standardized feature importance measures are computed as $\frac{\textnormal{XMDI}_{m}}{0.0167}$, where $0.0167$ represents the sample variance of the response variable (sex ratio of female to total births). Standardized additive effects are calculated as $\frac{\textnormal{XMDI}_{mm}}{\textnormal{XMDI}_{m}}$ for each of the $7$ feature groups. For the interaction matrix, values are obtained as $\frac{\textnormal{XMDI}_{ij}}{\textnormal{XMDI}_{j}}$, where $j$ is the column index of the interaction effect matrix. The names of the feature groups are indicated in the table.

Let us illustrate how to use the table with an example of computing $\textnormal{XMDI}_{73}$, which represents the interaction effect between \texttt{area} and \texttt{temperature}. To compute $\textnormal{XMDI}_{73}$ from Table~\ref{tab:relation}, we follow these steps: 1.) Obtain the value of $\frac{\textnormal{XMDI}_{73}}{\textnormal{XMDI}_{3}}$ from the (7, 3) entry of the interaction matrix. 2.) Get $\frac{\textnormal{XMDI}_{3}}{0.0167}$ from the row of the standardized feature importance.  3.) Multiply the results of steps 1 and 2 by 0.0167 to obtain the final value of $\textnormal{XMDI}_{73}$.

As a result, the interaction effect between \texttt{area} and \texttt{temperature} is estimated to be approximately $0.331 \times 0.009 \times 0.167 \approx 0.005$. Similarly, the interaction effect between \texttt{species} and \texttt{temperature} is approximately 0.032. All other interaction effects, when rounded to three decimal places, are negligible and thus are not reported here. Recall that the thickness of each edge depends on the size of the corresponding interaction effect, while its grayness depends on the corresponding standardized interaction effect. See Section~\ref{Sec3.3b} for details of network diagrams.

\begin{table}[t]
		\begin{center}{
				\begin{tabular}[t]{ |cccccccc|}
    \hline
    & X1 & X2 &  X3 &  X4 & X5 & X6 & X7 \\ \hline          
    Standardized Feature Importance & 0.353 & 0.005 & 0.090 & 0.003 & 0.002 & 0.034 & 0.723 \\ \hline          
    Standardized Additive Effects & 0.462 & 0.580 & 0.637  & 0.461  & 0.815 & 0.882 & 0.695\\ \hline                

  \multicolumn{8}{|c|}{Standardized Interaction Effects}  \\

  From $\Big\backslash$ To& X1 & X2 &  X3 &  X4 & X5 & X6 & X7 \\      
  X1 : \texttt{species} \hspace{3.6cm} & NA & 0.350 & 0.014 & 0.038 & 0.038 & 0.032 & 0.259 \\ 
 X2 : \texttt{subspecies} \hspace{3cm} & 0.001 &  NA & 0.003 & 0.029 &  0.006 &  0.005 &  0.002 \\ 
 X3 : \texttt{area} \hspace{4.3cm} & 0.004 & 0.056 & NA  &  0.056 &  0.048 &  0.032 &  0.041\\ 
 X4 :  \texttt{RMU} \hspace{4.5cm} &  0.000 &  0.016 & 0.002 & NA & 0.026 & 0.017 & 0.001\\ 
 X5 : \texttt{amplitude} \hspace{3.3cm} & 0.000 &  0.002 & 0.001 & 0.017 & NA & 0.001  & 0.000 \\ 
 X6 :  \texttt{incubation periods (days)}& 0.003 & 0.030  & 0.012 & 0.189 & 0.020 &NA & 0.002\\ 
 X7 :  \texttt{temperature} \hspace{2.9cm} &  0.531 & 0.282 & 0.331 &  0.210 & 0.047 &  0.032 &  NA\\ \hline  
			\end{tabular} }
			\caption{ Detailed numerical values used in Figure~\ref{fig:relations}, rounded to three decimal places. Standardized feature importance measures are computed as $\frac{\textnormal{XMDI}_{m}}{0.0167}$, where $0.0167$ represents the sample variance of the response variable (sex ratio). Standardized additive effects are calculated as $\frac{\textnormal{XMDI}_{mm}}{\textnormal{XMDI}_{m}}$ for each of the $7$ feature groups. For the interaction matrix, values are obtained as $\frac{\textnormal{XMDI}_{ij}}{\textnormal{XMDI}_{j}}$, where $j$ is the column index of the interaction effect matrix. The term NA denotes ``not applicable.''} \label{tab:relation}
		\end{center}
	\end{table}

\subsubsection{Additional simulation experiments for Section~\ref{Sec6.1}}\label{Sec6.1.c}

\begin{figure}
 \captionsetup[subfigure]{width=1.0\textwidth}
\centering
 \includegraphics[width=1.0\linewidth]{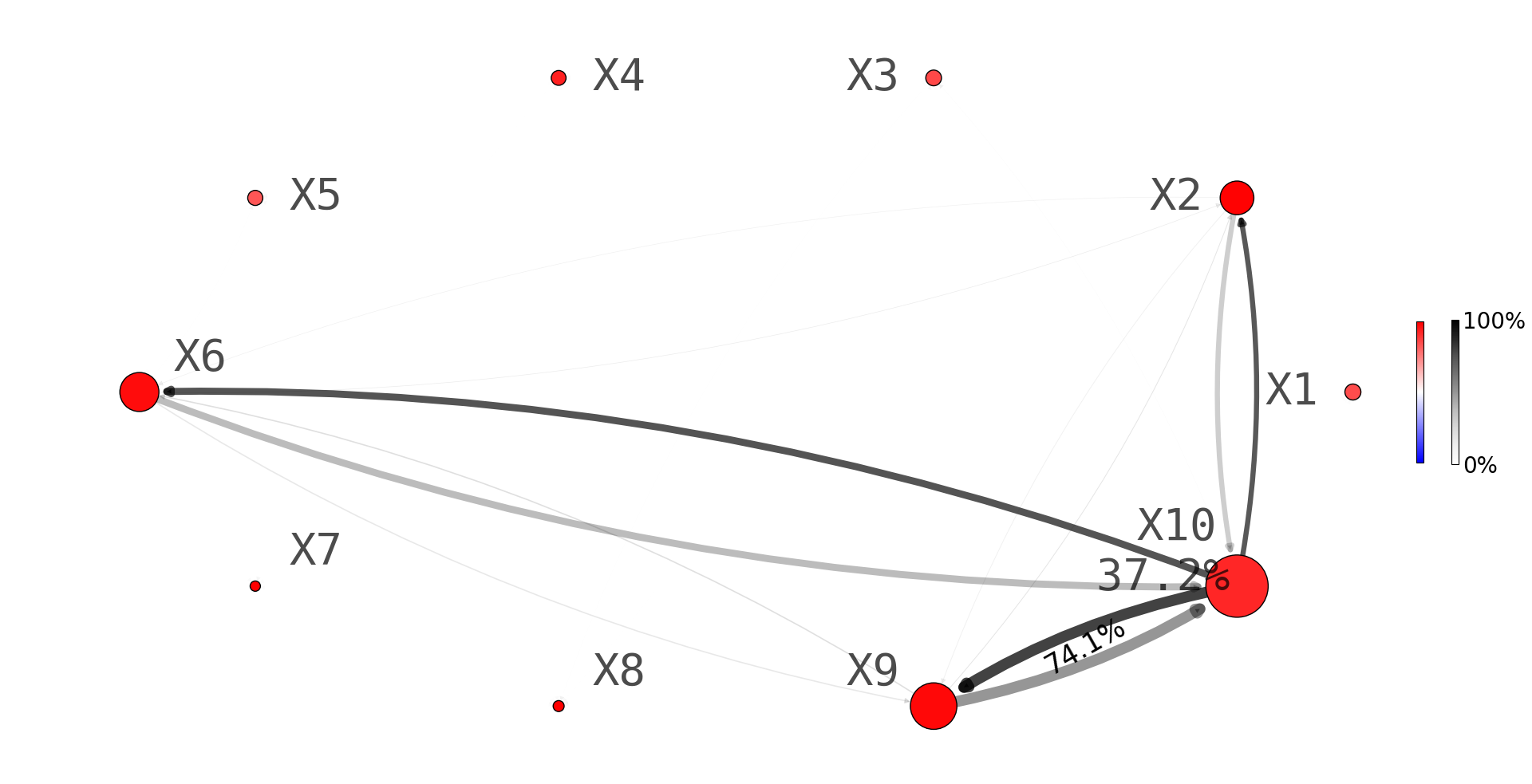}
\caption{Network diagrams for data simulated from Model~\eqref{Y2}  with $p=100$ and $\lambda = 0.1$. The picture here concerns only the first 10 features. }\label{fig:networks2}
\end{figure}

\begin{table}[t]
		\begin{center}
			{
				\begin{tabular}[t]{ | cc| }
                    \hline
                         \multicolumn{2}{|c|}{$\lambda = 0.1$}  \\
                         (I) & (II)   \\
					 \hline   0.95 & 0.97 \\ 
                    \hline  
                    
			\end{tabular} }
			\caption{XMDI of Collaborative Trees Ensemble. (I) The rates of successfully separating significant and insignificant features in Model~\eqref{Y2}, based on the overall feature importance measures. (II)  The rates of $\min_{j\in\{ 2, 6, 9\}} \textnormal{XMDI}_{10, j}$ being the maximum interaction effect among all other pairs.  All rates are derived from $100$ simulation experiments with $p=100$ features.} \label{tab:xmdi_stable_xor}
		\end{center}
	\end{table}

In this section, we present additional experiments similar to those in Section~\ref{Sec5.1} showing that the XMDI can practically distinguish interaction effects stemming from the same feature group. The response model of simulation experiment is given as follows.
\begin{equation}
\begin{split}
        Y & = 2X_{10} + 10\sin{(\pi (X_{2} - 0.5)(X_{10}-0.5))} + 10\sin{(\pi (X_{6} - 0.5)(X_{10}-0.5))} \\
        & \qquad + 10\sin{(\pi (X_{9} - 0.5)(X_{10}-0.5))} + \varepsilon. \label{Y2}
        \end{split}
\end{equation}
The sample size is $1000$ for each experiment, with $\lambda =0.1$ and $p=100$ for Figure~\ref{fig:networks2} and Table~\ref{tab:xmdi_stable_xor}; Figure~\ref{fig:networks2} only displays the results of $X_{1}, \dots, X_{10}$. Here, $p$ denotes the number of features. Other model assumptions are the same as those for Model~\eqref{Y1}.

We use $20$ runs of optimizations for hyperparameter tuning (see Section~\ref{hyper.tuning.1}) and report important hyperparameters: $\textsf{n\_trees} = K = 11$ and $\textsf{n\_bins} = 5$. This challenging learning example involves three XOR components. Hence, we consider a sample of size $1000$, which is larger than the ones of $500$ samples in Section~\ref{Sec5.1}. In Figure~\ref{fig:networks2}, Collaborative Trees Ensemble identifies four relevant features ($X_{2}$, $X_{6}$, $X_{9}$, $X_{10}$) based on their XMDI$_{j}$ importance measure. Furthermore, all interaction effects are visually identified in Figure~\ref{fig:networks2}. On the other hand, the column (II) of Table~\ref{tab:xmdi_stable_xor} shows that Collaborative Trees Ensemble consistently recognizes these three interaction effects as the strongest among all pairs over 100 repeated experiments. Additionally, column (I) of Table~\ref{tab:xmdi_stable_xor} demonstrates that XMDI$_{j}$ effectively differentiates between significant and insignificant features across all experiments. Here, we emphasize again that our goal is to verify empirically whether there exists a threshold for separating significant and insignificant features in each experiment. However, we do not aim at finding such a threshold.

The results presented in Table~\ref{tab:xmdi_stable_xor} complement those of Table~\ref{tab:xmdi_stable}, demonstrating the effectiveness of XMDI in distinguishing interaction effects originating from the same feature. Therefore, the practical performance of XMDI is not contingent on our technical condition requiring tree heredity in Condition~\ref{heredity.1} (but the addition of the weak additive component $2X_{10}$ in Model~\eqref{Y2} does stabilize the numerical results). 
These findings support the application of Collaborative Tree Ensemble in Section~\ref{Sec6.1} for the embryo growth dataset study.

\subsection{Additional details of prediction evaluation for Section~\ref{Sec6.2}}\label{Sec6.2b}

\begin{table}[t]
		\begin{center}
			{
				\begin{tabular}[t]{ |c|ccc|}
                        \hline & \multicolumn{3}{c|}{Win Rates} \\
					 \hline {\tiny CTE} & 0.7528 & 0.8069 & 0.8909 \\ 
                    {\tiny  XGB} & 0.4662 & 0.5986 & 0.7011 \\ 
                    {\tiny  RF} & 0.1612 & 0.2439 & 0.3413  \\ \hline  
                    
			\end{tabular} }

			\caption{For each experiment, an average win rate over $19$ datasets is calculated. From left to right columns, the lowest, the average, and the highest win rates across $10$ experiments are respectively reported for each method.} \label{tab:4}
		\end{center}
	\end{table}

Following the approach outlined in~\citep{grinsztajn2022tree}, we utilize 19 selective datasets from OpenML (\url{https://www.openml.org/}), each containing no more than 10,000 samples, for prediction evaluation based on the goodness-of-fit measure R$^2$ value. The calculation of each R$^2$ value involves splitting the full sample into training, validation, and test sets, constituting 48\%, 32\%, and 20\% of the full dataset, respectively. This division is employed for hyperparameter tuning based on the training and validation samples, as well as prediction loss calculation based on the test sample. A detailed description of the hyperparameter tuning procedure, involving 100 optimization runs for this evaluation task, is provided in Section~\ref{hyper.tuning.1}.

To summarize the comparison, we calculate adjusted win rates for the resulting R$^2$ values. For instance, if three methods (CTE, XGBoost, Random Forests) yield R$^2$ values of $0.13$, $0.135$, and $0.142$, they receive scores of $0$, $\frac{0.135 - 0.13}{0.142 - 0.13}$, and $1$, respectively, rounded to the fourth decimal if necessary. This performance measure, also adopted in~\citep{grinsztajn2022tree} (but their calculation involves more than three models), ensures a fair comparison among the methods across datasets. The term ``adjusted win rate'' is used to account for the interpolation when more than two models are compared. When only two models are compared, win rates (1 if a model wins, 0 otherwise) are calculated based on their resulting R$^2$ scores. For each dataset and method, we conduct 10 independent repeated experiments. We calculate average win rates denoted as $wr_1^{(k)}, \dots, wr_{10}^{(k)}$ for $10$ experiments and the $k$th dataset, with $k\in \{1, \dots, 19\}$. In Table~\ref{tab:4}, the reported win rates for each method represent $\min_{1\le b\le 10} wr_b$, $\frac{1}{10} \times \sum_{1\le b\le 10} wr_{b}$, and $\max_{1\le b\le 10} wr_b$ from left to right, where $wr_{b} = \frac{1}{19} \sum_{k=1}^{19} wr_b^{(k)}$. It is important to note that the entire evaluation procedure, including three sample splits, is repeated independently 10 times for each dataset.

From Table~\ref{tab:4}, Collaborative Trees Ensemble consistently outperforms XBGoost and Random Forests. These benchmark methods, recognized for their strong performance against modern deep learning models in regression and classification prediction tasks~\citep{grinsztajn2022tree}, are surpassed by our tree model. The results underscore our model's superior predictive capabilities for applications with  medium sample sizes. Further experiments targeting regression applications with large sample sizes ($> 10,000$) and classification tasks are reserved for future work.

\begin{figure}[h]
\centering
 \captionsetup[subfigure]{width=0.49\textwidth}
  \includegraphics[width=0.4\linewidth]{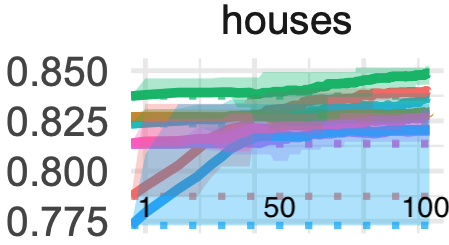}
\caption{A clip from the ``house'' sample in Figure 13 of \citep{grinsztajn2022tree}. The x-axis is the number of optimizations in hyperparameter tuning, ranging from $1$ to $100$ (each column dot is 10 runs), while the y-axis is the R$^2$ scores.  } \label{fig:whydotree}
\end{figure}

In Table~\ref{tab:r_square}, we present the minimum, average, and maximum R$^2$ scores for four models across 10 experiments for each dataset, arranged from left to right. The models included are Collaborative Trees Ensemble, XGBoost, Random Forests, and least squares regression (LS), with the LS serving as a linear benchmark model. It is worth noting that the LS is not included in Table~\ref{tab:4} to prevent potential overestimation of the other models due to interpolation. These results are comparable to Figure 13 of \citep{grinsztajn2022tree}, where a snippet of the results from the sample ``house'' is displayed in Figure~\ref{fig:whydotree}. 

In Figure~\ref{fig:whydotree}, they report all results from the first run to the 100th run of optimization, but they do not give the detailed R$^2$ scores as in our Table~\ref{tab:r_square} since they report many other details. In Figure~\ref{fig:whydotree}, the green line represents the R$^2$ scores of XGBoost, while the brown line represents the results of Random Forests.  We refer to \citep{grinsztajn2022tree} for other details. Our results in Table~\ref{tab:r_square} align closely with theirs up to the 100th column. For instance, the average R$^2$ score results for XGBoost are approximately 0.85 for the "house" sample in both Table~\ref{tab:r_square} and Figure~\ref{fig:whydotree}. However, there is an exception with the sample "delay\_zurich\_transport," where XGBoost exhibits higher variation and a lower average R$^2$ score in Table~\ref{tab:r_square}.

All Python source codes used to replicate these results are available upon request.


\begin{table}[h!]
		\begin{center}
			
				\begin{tabular}[t]{ |c|c| c|c|}
                        \hline & cpu\_act & pol & elevators \\
					 \hline {\tiny CTE} & (0.9848, 0.9872, 0.9894) & (0.9840, 0.9879, 0.9918) & (0.8884, 0.8972, 0.9057) \\ 
                    {\tiny  XGB} & (0.9851, 0.9873, 0.9889) & (0.9861, 0.9897, 0.9943) & (0.8837, 0.8968, 0.9078)  \\ 
                    {\tiny  RF} & (0.9793, 0.9828, 0.9849) & (0.9797, 0.9850, 0.9887) &  (0.7757, 0.8016, 0.8282) \\ 
                    {\tiny  LS} & (0.7029, 0.7170, 0.7523) & (0.4471, 0.4625, 0.4790) & (0.7976, 0.8163, 0.8361) \\\hline  
                    
                    & Ailerons & houses & house\_16H  \\
					 \hline {\tiny CTE} & (0.8369, 0.8592, 0.8719) & (0.8204, 0.8454, 0.8673) & (0.3910, 0.5332, 0.6559) \\ 
                    {\tiny  XGB} & (0.8171, 0.8354, 0.8475) & (0.8040. 0.8352, 0.8566) & (0.3960, 0.5278, 0.6574) \\ 
                    {\tiny  RF} & (0.8147, 0.8395, 0.8503) & (0.7833, 0.8117, 0.8279) & (0.4740, 0.5386, 0.6460) \\ 
                    {\tiny  LS} & (0.7517, 0.8132, 0.8280) & (0.6070, 0.6357, 0.6747) & (0.1624, 0.2422, 0.3259) \\\hline  

                    & Brazilian\_houses & Bike\_Sharing\_Demand & nyc-taxi-green-dec-2016  \\
					 \hline {\tiny CTE} & (0.9776, 0.9952, 0.9998) & (0.6797, 0.6962, 0.7244) & (0.5184, 0.5646, 0.5884) \\ 
                    {\tiny  XGB} &  (0.9791, 0.9954, 0.9996) & (0.6511, 0.6894, 0.7217) & (0.4954, 0.5292, 0.5620) \\ 
                    {\tiny  RF} & (0.9578, 0.9859, 0.9980)  & (0.6625, 0.6863, 0.7043) & (0.4631, 0.5371, 0.5702) \\ 
                    {\tiny  LS} & (0.4390, 0.7615, 0.8399)  & (0.3073, 0.3288, 0.3490) & (0.2636, 0.3067, 0.3522) \\\hline  

                    & sulfur & medical\_charges & MiamiHousing2016 \\
					 \hline {\tiny CTE} & (0.7812, 0.8480, 0.9049) & (0.9755, 0.9792, 0.9846) & (0.9272, 0.9348, 0.9417)\\ 
                    {\tiny  XGB} & (0.8264, 0.8678, 0.8917) & (0.9749, 0.9782, 0.9834) & (0.9206, 0.9305, 0.9350) \\ 
                    {\tiny  RF} & (0.7556, 0.8107, 0.8720) & (0.9748, 0.9784, 0.9843) & (0.8947, 0.9113, 0.9249) \\ 
                    {\tiny  LS} & (0.3379, 0.3874, 0.4364) & (0.7936, 0.8188, 0.8438) & (0.6803, 0.7134, 0.7320) \\ \hline  
                    
                    & yprop\_4\_1 & abalone  & delay\_zurich\_transport \\
					 \hline {\tiny CTE} & (0.0539, 0.0730, 0.0910) & (0.5136, 0.5427, 0.5782) & (0.0113, 0.0236, 0.0393) \\ 
                    {\tiny  XGB} & (0.0182, 0.0599, 0.0955) & (0.4329, 0.5182, 0.5532) & (-0.2871, -0.0324, 0.0178)  \\ 
                    {\tiny  RF} & (0.0537, 0.0827, 0.1020) & (0.4844, 0.5336, 0.5816) & (0.0104, 0.0213, 0.0338) \\ 
                {\tiny  LS} &  (0.0278, 0.0519, 0.0690) & (0.4628, 0.4946, 0.5841) & (0.0013, 0.0059, 0.0115) \\ \hline  
                    & wine\_quality & diamonds & house\_sales \\ \hline 
                    {\tiny CTE} &(0.4989, 0.5328, 0.5607) & (0.9386, 0.9446, 0.9482) & (0.8803, 0.8920, 0.9064) \\
                    {\tiny  XGB} &(0.4954, 0.5389, 0.5759) & (0.9391, 0.9428, 0.9466)  & (0.8788, 0.8880, 0.9017) \\
                    {\tiny  RF} & (0.4014, 0.4885, 0.5390) & (0.9391, 0.9441, 0.9467) & (0.8603, 0.8696, 0.8906) \\
                    {\tiny  LS} & (0.2457, 0.2878, 0.3339) & (0.7999, 0.9122, 0.9384) & (0.7321, 0.7472, 0.7662) \\ 
                    \hline
                    & superconduct & & \\
                    \hline
                    {\tiny CTE} & (0.8963, 0.9065, 0.9152) && \\
                    {\tiny  XGB} & (0.8989, 0.9099, 0.9218) & & \\
                    {\tiny  RF} & (0.8935, 0.9048, 0.9158) & & \\
                    {\tiny  LS} & (0.7155, 0.7342, 0.7523) & &  \\ 
                    \hline
			\end{tabular}

			\caption{ The table displays the minimum, average, and maximum R$^2$ scores across 10 experiments for each of the 19 selective datasets, arranged from left to right. Note that the results of the least square method (LS) are included in this table, but they are excluded from the adjusted win rate calculation.}\label{tab:r_square} 
		\end{center}
	\end{table}	
 \clearpage

\renewcommand{\thesubsection}{B.\arabic{subsection}}
\section{Technical proofs for main paper} \label{technical.proof}

In this section, we present the technical proofs of Theorem~\ref{theorem3} in Section~\ref{Sec4.1qq}, Corollary~\ref{corollary.1} in Section~\ref{Sec4.1qq}, and Theorem~\ref{theorem2} in Section~\ref{Sec4.2qq}. These proofs are given respectively in Sections~\ref{proof.theorem3}--\ref{SecB.4}.

\subsection{Proof of Theorem~\ref{theorem3}}\label{proof.theorem3}

\noindent\textit{Proof of Theorem~\ref{theorem3}: } Let us introduce notation for our proofs. First, recall that the output tree predictor in Algorithm~\ref{algorithm1} is denoted by $\sum_{k=1}^K T_{k}(\vv{x})$. We will introduce another notation for sample tree predictors as used in Section~\ref{Sec3qq} and \eqref{argmax.J.1} below, with subscripts indicating the update round index. This notation is crucial for presenting our technical analysis, for which we need to introduce the following set of notation.

For each $J\subset\{1, \dots, M\}$, let $C_{l, J}$ for $l\in \mathcal{L}(J)$ denote a partition of the feature space (not necessarily a partition of $\mathbb{R}^p$) based on features in $\mathcal{X}(J)$ such that $\sum_{l \in \mathcal{L}(J)} \boldsymbol{1}_{\boldsymbol{X} \in C_{l, J}} = 1$ almost surely. Specifically, it is required that
\begin{equation}
    \label{need.1}
    \{C_{l, J}: l\in \mathcal{L}(J) \} = \{\cap_{m\in J}A_{m}: A_{m} \in \mathcal{C}_{m} \textnormal{ for each } m \in J\},
\end{equation}
 where
 \begin{equation*}
 \mathcal{C}_{m}= 
 \begin{cases}
       \{ \{\vv{x} \in \mathbb{R}^p: x_{j} > 0\} , \{\vv{x} \in \mathbb{R}^p: x_{j} \le 0\}: j\in\mathcal{X}(m) \} \hspace{0.6cm} \textnormal{ if } \texttt{\#}\mathcal{X}(m) = 1,\\
     \{ \{\vv{x} \in \mathbb{R}^p: x_{j} > 0\} : j \in \mathcal{X}(m) \} \hspace{4cm} \textnormal{ if } \texttt{\#}\mathcal{X}(m) > 1.
 \end{cases}
 \end{equation*}

 As an illustrative example, let $J = \{l, k\}\subset\{1, \dots, M\}$, $\mathcal{X}(l) = \{j_{1}\}\subset\{1, \dots, p\}$, and $\mathcal{X}(k) = \{j_{2}, j_{3}\}\subset\{1, \dots, p\}$. Then, we define $\mathcal{L}(J) = \{1, 2, 3, 4\}$, and define $C_{l, J}$'s such that $\{C_{l, J}: l\in \mathcal{L}(J)\} = \big\{\{\vv{x}\in\mathbb{R}^p: x_{j_{1}} > 0, x_{j_{2}} > 0 \}, \{\vv{x}\in\mathbb{R}^p: x_{j_{1}} > 0, x_{j_{3}} > 0 \}, \{\vv{x}\in\mathbb{R}^p: x_{j_{1}} \le 0, x_{j_{2}} > 0 \}, \{\vv{x}\in\mathbb{R}^p: x_{j_{1}} \le 0, x_{j_{3}} > 0 \} \big\}$.
As a result, $\sum_{l \in \mathcal{L}(J)} \boldsymbol{1}_{\boldsymbol{X} \in C_{l, J}} = 1$ almost surely due to Condition~\ref{model.1}. Here, it is worth noting that the precise definition of $\mathcal{L}(J)$ and $C_{l,J}$ is not crucial as long as \eqref{need.1} is satisfied.

Let $\{\widehat{J}_{l}\}_{l=1}^{2K}$ be the sample optimal path of sets given in Section~\ref{Sec4.1qq}. 
For each set $J$ and $0< s\le 2K$, we define functions
$R_{s, J}^{\star}  :\mathbb{R}^p\mapsto\mathbb{R}$, $\widehat{U}_{s, J}:\mathbb{R}^p\mapsto\mathbb{R}$, and $\widehat{U}_{s} :\mathbb{R}^p\mapsto\mathbb{R}$ such that
\begin{equation}
    \label{predictor.trees.b.3}
    \begin{split}
        \widehat{U}_{s, J}(\vv{x}) & = \widehat{U}_{s-1}(\vv{x}) + \sum_{l\in \mathcal{L}(J)} \boldsymbol{1}_{\vv{x} \in C_{l, J}} \frac{\sum_{i\in \{\boldsymbol{X}_{i} \in C_{l, J} \}} \left[Y_{i} - \widehat{U}_{s-1}(\boldsymbol{X}_{i}) \right] }{ \texttt{\#} \{i:\boldsymbol{X}_{i} \in C_{l, J} \} \vee 1 },\\
          \widehat{U}_{s}(\vv{x}) &  = \widehat{U}_{s, \widehat{J}_{s}}(\vv{x}), \\
         R_{s, J}^{\star} (\boldsymbol{X}) & = R_{s-1}^{\star} (\boldsymbol{X})+   \mathbb{E}( f(\boldsymbol{X}) - R_{s-1}^{\star}(\boldsymbol{X}) | \boldsymbol{X}_{\mathcal{X}(J) }),
    \end{split}
\end{equation}
where $R_{l}^{\star}$'s are given as in \eqref{mp.3}, and $\widehat{U}_{s}(\vv{x})$ is recursively defined by \eqref{predictor.trees.b.3} given $\widehat{U}_{0}(\vv{x}) = 0$. In addition, we set $ \widehat{U}_{0, J}(\vv{x}) = R_{0, J}^{\star} (\vv{x}) =0$ for each $\vv{x} \in \mathbb{R}^p$ and every set $J$. It is important to note that $R_{s, \widehat{J}_{s}}^{\star} (\boldsymbol{X}) = R_{s}^{\star} (\boldsymbol{X})$.

To clarify, $\widehat{U}_{s}$ in \eqref{predictor.trees.b.3} represents the sample tree predictor at the (end of the) $s$th round of updates, obtained from Algorithm~\ref{algorithm1} with $s\le 2K$. To see this, let us consider the update at the $s$th round with $s\le K$, and that the feature group to be split has a single binary feature with $\widehat{J}_{s} = \{q\}$. Then, $\{C_{1, \widehat{J}{s}}, C_{2, \widehat{J}{s}} \}  = \{ \{\vv{x}\in \mathbb{R}^p: x_{q}\le 0 \}, \{\vv{x}\in \mathbb{R}^p: x_{q}>0 \} \}$ represent the two child nodes of some root node at \textbf{Step 12} of Algorithm~\ref{algorithm1}. The tree update at \textbf{Step 12} of Algorithm~\ref{algorithm1} is equivalent to the recursive definition in \eqref{predictor.trees.b.3} when considering the residuals $\{Y_{i} - \widehat{U}_{s-1}(\boldsymbol{X}_{i})\}_{i=1}^n$ at the beginning of the $s$th round. The distinction lies in emphasizing the round index, denoted by subscript $s$ in $\widehat{U}_{s}(\vv{x})$,  rather than the tree index, denoted by subscript $k$ in $T_{k}(\vv{x})$. Similar arguments can be applied for general cases with $K<s\le 2K$, but we omit the details for simplicity. Note that this notation for predictors has already been introduced in Section~\ref{Sec3qq}. Furthermore, while the expressions of sample tree predictors presented here may not align with the sample tree predictor in Algorithm~\ref{algorithm1} when $s> 2K$, this discrepancy does not affect our arguments. Our focus is specifically on the first $2K$ updates involving $K$ collaborative decision trees in Theorem~\ref{theorem3}.

Considering the determination of the next feature group to be split in \eqref{gini2}, and the definition of $\widehat{J}_{l}$'s in Section~\ref{Sec4.1qq}, it holds that for $0<s\le 2K$,
\begin{equation}
    \label{argmax.J.1}
    \widehat{J}_{s} = \argmax_{J \in \Theta_{s}} \left\{\left[\sum_{i=1}^n (Y_{i} - \widehat{U}_{s-1}(\boldsymbol{X}_{i}))^2 \right] - \sum_{i=1}^n (Y_{i} - \widehat{U}_{s, J}(\boldsymbol{X}_{i}))^2\right\},
\end{equation}
where $\Theta_{1} = \dots =  \Theta_{K} = \{\{1\}, \dots, \{M\}\}$, and that
\begin{equation*}
    \begin{split}        
        A_{s} & =  \{\{l, k\}: 1\le k\le M, k\not= l\in \widehat{J}_{s}\} \textnormal{ for }  0<  s\le K, \\        
        \Theta_{s}  &= \cup_{l \in Z_{s} } A_{l} \textnormal{ for } K  < s \le 2K, \\
        Z_{s+1} & = Z_{s} \backslash \{\widehat{k}_{s}\}, \textnormal{ for } K  < s < 2K,
    \end{split}
\end{equation*}
where $1\le \widehat{k}_{s}\le K$ is such that $\widehat{J}_{s} \in A_{\widehat{k}_{s}}$ (after $\widehat{J}_{s}$ has been decided) for $K<s\le 2K$, and $Z_{K+1}  = \{1, \dots, K\}$. Ties are broken randomly. The expression within the curly brackets in \eqref{argmax.J.1} represents a split score multiplied by $n$. The specific set restriction for \eqref{argmax.J.1} is imposed by natural tree heredity before the initial $2K$ updates, our update priority outlined in \eqref{gini2}, and our assumption that $\widehat{J}_{l} \neq \widehat{J}_{k}$ for each $\{l, k\}\subset \{1, \dots, 2K\}$. Recall that $\{j, j\} = \{j\}$ according to our definition of sets.  The proof of  \eqref{argmax.J.1} is postponed to Section~\ref{proof.expression.2} after the end of the proof of Theorem~\ref{theorem3}.


 By that  $\widehat{U}_{s, \widehat{J}_{s}}(\boldsymbol{X}) = \widehat{U}_{s}(\boldsymbol{X})$ and $R_{s, \widehat{J}_{s}}^{\star}(\boldsymbol{X}) = R_{s}^{\star}(\boldsymbol{X})$ for each $s\ge 0$, and Lemma~\ref{lemma.3}, it holds that for each $0<s\le 2K$ and every set $J$,
\begin{equation}
\begin{split}\label{equivalence.recur.4.b}
 & n^{-1} \left[ \sum_{i=1}^n (Y_{i} - \widehat{U}_{s-1}(\boldsymbol{X}_{i}))^2 \right] - n^{-1}\sum_{i=1}^n (Y_{i} - \widehat{U}_{s, J}(\boldsymbol{X}_{i}))^2
 \\
 & = \mathbb{E}[(\mathbb{E}(f(\boldsymbol{X}) - R_{s-1}^{\star}(\boldsymbol{X})|\boldsymbol{X}_{\mathcal{X}(J)})  )^2]\\
 &\qquad + D_{2, s-1}(\widehat{J}_{s-1}) - D_{2, s}(J) + D_{1,s-1}(\widehat{J}_{s-1}) - D_{1,s} (J),
 \end{split}
\end{equation}
where
\begin{equation*}
    \begin{split}
        D_{1,s}(J) & = \left(n^{-1}\sum_{i=1}^n (Y_{i} - \widehat{U}_{s, J}(\boldsymbol{X}_{i}))^2\right) - \left(n^{-1}\sum_{i=1}^n (f(\boldsymbol{X}_{i}) - R_{s, J}^{\star}(\boldsymbol{X}_{i}))^2  \right)- n^{-1}\sum_{i=1}^n \varepsilon_{i}^2 ,\\
        D_{2,s}(J) & = \left(n^{-1}\sum_{i=1}^n (f(\boldsymbol{X}_{i}) - R_{s, J}^{\star}(\boldsymbol{X}_{i}))^2 \right)- \mathbb{E}(f(\boldsymbol{X}) - R_{s, J}^{\star}(\boldsymbol{X}) )^2.
    \end{split}
\end{equation*}

We will establish upper bounds for $|D_{1, s}(J)|$'s and $|D_{2, s}(J)|$'s for each $s\ge 0$ and every $J$ with $\texttt{\#}J\le 2$, starting with the expressions of $R_{s, J}^{\star}(\boldsymbol{X})$ and $R_{s}^{\star}(\boldsymbol{X})$ provided in \eqref{dependence.6.b} below. Additionally, we will utilize several other preliminary results, including \eqref{dependence.6.b}--\eqref{event.2} below, to analyze the upper bounds on $|D_{1, s}(J)|$'s and $|D_{2, s}(J)|$'s.

By \eqref{predictor.trees.b.3} and \eqref{mp.3}, it holds that for every set $J$ and each $s>0$,
\begin{equation}
    \begin{split}\label{dependence.6.b}
R_{s, J}^{\star}(\boldsymbol{X}) & = \mathbb{E}(f(\boldsymbol{X}) - R_{s-1}^{\star}(\boldsymbol{X}) |\boldsymbol{X}_{\mathcal{X}(J)}) \\
&\qquad + \sum_{l=1}^{s-1} \mathbb{E}(f(\boldsymbol{X}) - R_{l-1}^{\star}(\boldsymbol{X}) |\boldsymbol{X}_{ \mathcal{X}(\widehat{J}_{l})}) \\
& = \left(\sum_{\vv{c} \in \{0, 1\}^{\texttt{\#}\mathcal{X}(J)}}  a_{s, \vv{c}, J}\times  \boldsymbol{1}_{\boldsymbol{X}_{\mathcal{X}(J)} = \vv{c}}\right) + \sum_{l=1}^{s-1} \sum_{\vv{c} \in \{0, 1\}^{\texttt{\#}\mathcal{X}(\widehat{J}_{l})}}  a_{l, \vv{c}, \widehat{J}_{l}}\times  \boldsymbol{1}_{\boldsymbol{X}_{\mathcal{X}(\widehat{J}_{l})} = \vv{c}} ,\\
R_{s}^{\star}(\boldsymbol{X})
& =  \sum_{l=1}^{s} \sum_{\vv{c} \in \{0, 1\}^{\texttt{\#}\mathcal{X}(\widehat{J}_{l})}}  a_{l, \vv{c}, \widehat{J}_{l}}\times  \boldsymbol{1}_{\boldsymbol{X}_{\mathcal{X}(\widehat{J}_{l})} = \vv{c}},
    \end{split}
\end{equation}
where $a_{l, \vv{c}, J} = \mathbb{E}\{ f(\boldsymbol{X}) - R_{l-1}^{\star}(\boldsymbol{X})  |\boldsymbol{X}_{\mathcal{X}(J)}= \vv{c}\}$ for every set $J$, each $l>0$, and each $\vv{c} \in \{0, 1\}^{\texttt{\#} \mathcal{X}(J)}$. For the completeness of notation, we define $a_{0, \vv{c}, J} = 0$ for all vector $\vv{c}$ and set $J$. Recall that we have defined $\mathbb{E}(h(\boldsymbol{X})|\boldsymbol{X} = \vv{c}) = 0$ when $\mathbb{P}(\boldsymbol{X} = \vv{c}) = 0$ for every $\vv{c} \in \{0, 1\}^{p}$ and every measurable function $h:\mathbb{R}^p\mapsto\mathbb{R}$. Also, a summation over an empty set is defined to be zero. Furthermore, $\sum_{l=i}^{i-k} 1$ is defined to be zero for every integer $i$ and every positive integer $k$.
By Lemma~\ref{lemma.1}, \eqref{dependence.6.b}, and the definition of $\iota_{s}$ in \eqref{iota.1}, for every set $J$ with $\texttt{\#}J \le 2$ and each $s\ge 0$,
\begin{equation}
    \label{dependence.7.b}
    \max_{\vv{c} \in \{0, 1\}^p} |R_{s, J}^{\star}(\vv{c}) | \le 3\sqrt{\mathbb{E}[(f(\boldsymbol{X}))^2]}  \times p_{\min}^{-1} +  \iota_{s-1},
\end{equation}
where we note that $\max_{\vv{c} \in \{0, 1\}^p} |R_{0, J}^{\star}(\vv{c}) |=0$ by the definition in \eqref{predictor.trees.b.3}. In addition, since each feature group contains either a single feature or a group of one-hot indicators, it holds that  for each $l>0$ and each subset $J\subset \{1, \dots, M\}$ with $\texttt{\#} J\le 2$,
\begin{equation}\label{one.hot.1}
\sum_{\vv{c} \in \{0, 1\}^{\texttt{\#}\mathcal{X}(J)}} \boldsymbol{1}\{|a_{l, \vv{c}, J}| >0\} \le M_{\mathcal{X}}^2,
\end{equation}
in which $M_{\mathcal{X}}$ is defined in Condition~\ref{model.1}.

By the definition of $\iota_{l}$'s in \eqref{iota.1},  Lemma~\ref{lemma.1}, and \eqref{one.hot.1}, it holds that for every $J$ with $\texttt{\#}J\le 2$ and each $s\ge 0$,
\begin{equation}
    \label{dependence.8.b}
    \begin{split}
     \sum_{l=1}^{s-1} \sum_{\vv{c} \in \{0, 1\}^{\texttt{\#}\mathcal{X}(\widehat{J}_{l})}}  |a_{l, \vv{c}, \widehat{J}_{l}}|  &\le M_{\mathcal{X}}^2 \iota_{s-1} ,\\
    \left(\sum_{\vv{c} \in \{0, 1\}^{\texttt{\#}\mathcal{X}(J)}}  |a_{s, \vv{c}, J}|  \right) + \sum_{l=1}^{s-1} \sum_{\vv{c} \in \{0, 1\}^{\texttt{\#}\mathcal{X}(\widehat{J}_{l})}}  |a_{l, \vv{c}, \widehat{J}_{l}}|  &\le 3M_{\mathcal{X}}^2  \sqrt{\mathbb{E}[(f(\boldsymbol{X}))^2]}  p_{\min}^{-1} + M_{\mathcal{X}}^2 \iota_{s-1}.
    \end{split}
\end{equation}

Let us define six events, denoted as $E_{41}$ to $E_{46}$, as follows. Here, we recall that $\boldsymbol{X}_{iH}$ represents the feature subset in $H\subset\{1,\dots, p\}$ of the $i$th sample feature vector, which can be represented as $(X_{ij}, j\in H)^{\top}$. Our main results in this proof hold on the event $\cap_{l=1}^6 E_{4l}$, with the probability $\mathbb{P}((\cap_{l=1}^6 E_{4l})^c) = o(1)$ proven in Lemma~\ref{lemma.event.1} in Section~\ref{SecC.5}, where $A^c$ denotes the complementary set of $A$. The intersection operations $\texttt{\#}J \le k$ below are performed over all possible subsets  $J\subset \{1, \dots, M\}$ with $\texttt{\#}J \le k$, where we have $\texttt{\#}\{J: J\subset \{1, \dots, M\}, \texttt{\#}J\le k\} \le M^{k}$. We also note that 
$\sum_{\vv{c} \in \{0, 1\}^{\texttt{\#}\mathcal{X}(J)}} 1 = 2^{\texttt{\#}\mathcal{X}(J)} \le 2^{k \times M_{\mathcal{X}}}$ when $\texttt{\#}J\le k$, and that $M_{\mathcal{X}}$ given in Condition~\ref{model.1} is assumed to be a finite constant.
Similar notation for union operations is employed in the proof of Lemma~\ref{lemma.event.1}.
\begin{equation}
    \begin{split}\label{event.2}
    E_{41}  & =  \left\{\left|\sum_{i=1}^n \varepsilon_{i} f(\boldsymbol{X}_{i})\right| \le t_{1} \right\},\\
    E_{42} & = \cap_{\texttt{\#}J\le 2, \vv{c}\in \{0, 1\}^{\texttt{\#}\mathcal{X}(J)}}\left\{ \left| \sum_{i=1}^n \varepsilon_{i} \boldsymbol{1}_{\boldsymbol{X}_{i\mathcal{X}(J)} = \vv{c}} \right| \le  t_{2 }\right\},\\
    E_{43} & = \cap_{\texttt{\#}J\le 2, \vv{c}\in \{0, 1\}^{\texttt{\#}\mathcal{X}(J)}} \left\{ \left|\sum_{i=1}^n \left(f(\boldsymbol{X}_{i}) \boldsymbol{1}_{\boldsymbol{X}_{i\mathcal{X}(J)} = \vv{c}} -\mathbb{E}(f(\boldsymbol{X}) \boldsymbol{1}_{\boldsymbol{X}_{\mathcal{X}(J)} = \vv{c}} )\right) \right| \le t_{3}\right\},\\
    E_{44} &= \cap_{\texttt{\#}J\le 4, \vv{c}\in \{0, 1\}^{\texttt{\#}\mathcal{X}(J)}} \left\{ \left|\sum_{i=1}^n \left( \boldsymbol{1}_{\boldsymbol{X}_{i\mathcal{X}(J)} = \vv{c}} -\mathbb{P}(\boldsymbol{X}_{\mathcal{X}(J)} = \vv{c} )\right) \right| \le t_{4}\right\},\\
    E_{45}  & =  \left\{\left|\sum_{i=1}^n \varepsilon_{i}^2 \right| \le t_{5} \right\},\\
    E_{46} & = \left\{\left| \sum_{i=1}^n \{(f(\boldsymbol{X}_{i}))^2 - \mathbb{E}[(f(\boldsymbol{X}))^2 ] \} \right| \le t_{6} \right\}.
    \end{split}
\end{equation}

In the following, we deal with establishing an upper bound for $|D_{1, s}(J)|$ in \eqref{equivalence.recur.4.b}. For each $0\le s\le 2K$ and any $J$ with $\texttt{\#}J\le 2$, we deduce that
\begin{equation*}
    \begin{split}
        & \sum_{i=1}^n (Y_{i} - \widehat{U}_{s, J}(\boldsymbol{X}_{i}))^2 \\
        & = \sum_{i=1}^n (f(\boldsymbol{X}_{i}) + \varepsilon_{i} - R_{s, J}^{\star}(\boldsymbol{X}_{i}) + R_{s, J}^{\star}(\boldsymbol{X}_{i}) - \widehat{U}_{s, J}(\boldsymbol{X}_{i}))^2 \\
        & = \left(\sum_{i=1}^n (f(\boldsymbol{X}_{i}) - R_{s, J}^{\star}(\boldsymbol{X}_{i}))^2\right) + \left(\sum_{i=1}^n(R_{s, J}^{\star}(\boldsymbol{X}_{i}) - \widehat{U}_{s, J}(\boldsymbol{X}_{i}))^2\right) + \sum_{i=1}^n \varepsilon_{i}^2 \\ 
        &\qquad + 2\left(\sum_{i=1}^n \varepsilon_{i}(f(\boldsymbol{X}_{i}) -R_{s, J}^{\star}(\boldsymbol{X}_{i})) \right)+ 2\sum_{i=1}^n \varepsilon_{i}(R_{s, J}^{\star}(\boldsymbol{X}_{i}) -\widehat{U}_{s, J}(\boldsymbol{X}_{i})) \\
        &\qquad + 2\sum_{i=1}^n (f(\boldsymbol{X}_{i}) -R_{s, J}^{\star}(\boldsymbol{X}_{i})) (R_{s, J}^{\star}(\boldsymbol{X}_{i}) - \widehat{U}_{s, J}(\boldsymbol{X}_{i})),
    \end{split}
\end{equation*}
which, along with Jensen's inequality, implies that for each $0\le s\le 2K$ and every $\texttt{\#}J\le 2$,
\begin{equation}
    \begin{split}\label{equivalence.recur.3.b}
         |n\times D_{1, s}(J)| & = \left|\left( \sum_{i=1}^n (Y_{i} - \widehat{U}_{s, J}(\boldsymbol{X}_{i}))^2\right) - \left(\sum_{i=1}^n (f(\boldsymbol{X}_{i}) - R_{s, J}^{\star}(\boldsymbol{X}_{i}))^2 \right) - \sum_{i=1}^n \varepsilon_{i}^2 \right| \\
        & \le \left(\sum_{i=1}^n(R_{s, J}^{\star}(\boldsymbol{X}_{i}) - \widehat{U}_{s, J}(\boldsymbol{X}_{i}))^2 \right) + 2\left| \sum_{i=1}^n \varepsilon_{i}(f(\boldsymbol{X}_{i}) - R_{s, J}^{\star}(\boldsymbol{X}_{i})) \right| \\
        &\qquad + 2 \sqrt{\sum_{i=1}^n \varepsilon_{i}^2} \sqrt{\sum_{i=1}^n (R_{s, J}^{\star}(\boldsymbol{X}_{i}) - \widehat{U}_{s, J}(\boldsymbol{X}_{i}))^2}\\
        &\qquad + 2 \sqrt{\sum_{i=1}^n (f(\boldsymbol{X}_{i}) -R_{s, J}^{\star}(\boldsymbol{X}_{i}))^2} \sqrt{\sum_{i=1}^n (R_{s, J}^{\star}(\boldsymbol{X}_{i}) - \widehat{U}_{s, J}(\boldsymbol{X}_{i}))^2}.
    \end{split}
\end{equation}



By \eqref{equivalence.recur.3.b} and the regularity conditions assumed by Theorem~\ref{theorem3}, 
there exists $L_{0}>0$ such that for each $n$ with $\frac{2\log{(n)}}{\sqrt{n}} \le p_{\min}$, each $0\le s\le 2K$, and every set $J$ with $\texttt{\#}J\le 2$, it holds that  on the event $\cap_{l=1}^5 E_{4l} $ with $t_{1}=t_2 = \log{(n)}n^{\frac{1}{2} + \frac{1}{q_{1}}}$, $t_3 = t_4 = \log{(n)}\sqrt{n}$, and $t_{5} = 2n\mathbb{E}\varepsilon^2$,
\begin{equation}
    \begin{split}\label{equivalence.recur.10.b}
        & |n\times D_{1, s}(J)| \\ 
        &\le 
 \left[ 20 p_{\min}^{-2}(\log{n})^2 M_{\mathcal{X}}^4 \left(\sum_{l=1}^s\iota_{l-1}^2\right) + 280 s M_{f}^2 p_{\min}^{-4}(\log{n})^2n^{\frac{2}{q_{1}}} \right] \\
 & \qquad + 2\log{(n)}n^{\frac{1}{2} + \frac{1}{q_{1}}} (3M_{\mathcal{X}}^2p_{\min}^{-1} \sqrt{\mathbb{E}[(f(\boldsymbol{X}))^2]}   + M_{\mathcal{X}}^2 \iota_{s-1} + 1) \\
        & \qquad +  \sqrt{20 p_{\min}^{-2}(\log{n})^2 M_{\mathcal{X}}^4 \left(\sum_{l=1}^s\iota_{l-1}^2\right) + 280 s M_{f}^2 p_{\min}^{-4}(\log{n})^2n^{\frac{2}{q_{1}}} }  \times 2\sqrt{2n\mathbb{E}\varepsilon^2}  \\
        & \qquad + \sqrt{20 p_{\min}^{-2}(\log{n})^2 M_{\mathcal{X}}^4 \left(\sum_{l=1}^s\iota_{l-1}^2\right) + 280 s M_{f}^2 p_{\min}^{-4}(\log{n})^2n^{\frac{2}{q_{1}}} } \\
        & \qquad \qquad \times 2\sqrt{n}\left(\max_{\vv{c} \in \{0, 1\}^{p}}|f(\vv{c})| + 3\sqrt{\mathbb{E}[(f(\boldsymbol{X}))^2]}  \times p_{\min}^{-1} +  \iota_{s-1} \right)\\
        & \le L_{0} \times \bigg(s(\log{(n)})^2 n^{ \frac{2}{q_{1}}} + \left(\sum_{l=1}^s \iota_{l-1}^2\right)( \log{(n)})^2 + n^{\frac{1}{2} + \frac{1}{q_{1}}}\log{(n)}(\iota_{s-1}+1) \sqrt{s\vee 1} \\
        &\qquad + (\iota_{s-1} + 1) \sqrt{\sum_{l=1}^s \iota_{l-1}^2} \sqrt{n}\log{(n)} \bigg),
    \end{split}
\end{equation}
where the detailed derivation for the first inequality is given after the end of the current proof of  Theorem~\ref{theorem3} (see Section~\ref{proof.theorem1.12}), and the the last inequality is due to subadditivity of the square root function.

Let us simplify the upper bound on the RHS of \eqref{equivalence.recur.10.b}.
When $n^{-\frac{1}{2} + \frac{1}{q_{1}}}\log{(n)}\sqrt{s \vee 1}\le 1$ and $\sqrt{\sum_{l=1}^s \iota_{l-1}^2} ( \log{n}) n^{-1/2}  \le 1$, under the same conditions for \eqref{equivalence.recur.10.b},
\begin{equation}
    \begin{split}\label{d_1s}
        |D_{1, s}(J)| & \le L_{0}  n^{-\frac{1}{2} + \frac{1}{q_{1}}}\log{(n)}\sqrt{s\vee 1}(\iota_{s-1} +2) + L_{0} (\iota_{s-1} + 2) \sqrt{\sum_{l=1}^s \iota_{l-1}^2} n^{-\frac{1}{2}}\log{(n)} \\
        & =  L_{0}  n^{-\frac{1}{2}} \log{(n)} \left(n^{ \frac{1}{q_{1}}}\sqrt{s\vee 1} + \sqrt{\sum_{l=1}^s \iota_{l-1}^2} \right) (\iota_{s-1} +2).
    \end{split}
\end{equation}




Next, proceed to deal with the term $D_{2,s}(J)$'s on the RHS of \eqref{equivalence.recur.4.b}. We first deduce that 
\begin{equation}\label{equivalence.recur.8.b}
\begin{split}
    D_{2, s}(J) &= \left( n^{-1}\sum_{i=1}^n (f(\boldsymbol{X}_{i}) - R_{s, J}^{\star}(\boldsymbol{X}_{i}))^2 \right) - \mathbb{E}(f(\boldsymbol{X}) - R_{s, J}^{\star}(\boldsymbol{X}) )^2 \\
    & = n^{-1}\sum_{i=1}^n [(f(\boldsymbol{X}_{i}))^2 - \mathbb{E}(f(\boldsymbol{X}))^2 ]     
     \\
     & \qquad - 2 n^{-1}\sum_{i=1}^n [f(\boldsymbol{X}_{i}) R_{s, J}^{\star}(\boldsymbol{X}_{i}) - \mathbb{E}(f(\boldsymbol{X}) R_{s, J}^{\star}(\boldsymbol{X}) ) ] \\
     & \qquad + n^{-1}\sum_{i=1}^n [ (R_{s, J}^{\star}(\boldsymbol{X}_{i}))^2 - \mathbb{E}( R_{s, J}^{\star}(\boldsymbol{X}) )^2].
\end{split}    
\end{equation}

By \eqref{equivalence.recur.8.b} and the assumed regularity conditions, there exists some sufficiently large $L_{0}>0$ such that  on the event $ E_{43}\cap E_{44}\cap E_{46}$ defined in \eqref{event.2} with $t_{3}= t_{4} = t_{6} = \log{(n)} \sqrt{n}$, it holds that for each $0\le s\le 2K$ and every set $J$ with $\texttt{\#}J\le 2$,
\begin{equation}\label{equivalence.recur.11.b}
\begin{split}
    |D_{2, s}(J)| &\le  \log{(n)} n^{-\frac{1}{2}} + 2\log{(n)}n^{-\frac{1}{2}}(3M_{\mathcal{X}}^2p_{\min}^{-1} \sqrt{\mathbb{E}[(f(\boldsymbol{X}))^2]}  + M_{\mathcal{X}}^2 \iota_{s-1} ) \\
    &\qquad + \log{(n)}n^{-\frac{1}{2}}(3M_{\mathcal{X}}^2p_{\min}^{-1} \sqrt{\mathbb{E}[(f(\boldsymbol{X}))^2]}  + M_{\mathcal{X}}^2 \iota_{s-1} )^2\\
    & \le L_{0}\log{(n)} n^{-\frac{1}{2}} ( \iota_{s-1}^2 + 1).
 \end{split}    
\end{equation}
where the detailed derivation for the first inequality on the RHS of \eqref{equivalence.recur.11.b} is given after the end of the current proof of the Theorem~\ref{theorem3} (see Section~\ref{proof.theorem1.15}), and the second inequality is due to simple calculations and an application of the quadratic mean inequality~\citep{gwanyama2004hm}.



To summarize the results from \eqref{equivalence.recur.4.b}, \eqref{d_1s}, and \eqref{equivalence.recur.11.b}, under the assumed conditions, on the event $\cap_{l=1}^6 E_{4l}$, it holds that for each $0\le s\le 2K$, and every set $J$ with $\texttt{\#}J\le 2$,
\begin{equation*}
    \begin{split}
    |D_{1, s}(J)| & \le L_{0}  n^{-\frac{1}{2}} (\log{n}) \left(n^{ \frac{1}{q_{1}}}\sqrt{s\vee 1} + \sqrt{\sum_{l=1}^s \iota_{l-1}^2} \right) (\iota_{s-1} + 2),\\
    |D_{2, s}(J)| & \le L_{0} (\log{n}) n^{-\frac{1}{2}} ( \iota_{s-1}^2 + 1).
    \end{split}
\end{equation*}
Particularly, for each $0< s \le 2K$ and every set $J$ with $\texttt{\#}J \le 2$,
\begin{equation}\label{particular.1}
    |D_{2, s-1}(\widehat{J}_{s-1})| + |D_{2, s}(J)| + |D_{1,s-1}(\widehat{J}_{s-1})|  + |D_{1,s} (J)| \le d_{n},
\end{equation}
where $d_{n} = L_{1}  n^{-\frac{1}{2}} (\log{n}) (n^{ \frac{1}{q_{1}}}\sqrt{2K} + \sqrt{\sum_{l=1}^{2K} \iota_{l-1}^2} ) (\iota_{2K-1}^2 + \iota_{2K-1} + 3)$ for sufficiently large $L_{1}\ge 1$.
Additionally, $\mathbb{P}((\cap_{l=1}^6 E_{4l})^c) = o(1)$ by Lemma~\ref{lemma.event.1}. We note that the assumption $d_{n} \le 1$ implies the conditions $\sqrt{\sum_{l=1}^{2K} \iota_{l-1}^2} ( \log{n}) n^{-1/2} \le 1$ and $n^{-\frac{1}{2} + \frac{1}{q_{1}}} (\log{n}) \sqrt{2K} \le 1$. These conditions, along with the requirement $\frac{2\log{n}}{\sqrt{n}} \le p_{\min}$, are required for \eqref{d_1s} to hold.

The results \eqref{equivalence.recur.4.b} and \eqref{particular.1} implies that for each $0<s\le 2K$ and every set $J$ with $\texttt{\#}J\le 2$,
\begin{equation}
\begin{split}\label{equivalence.recur.4.c}
 & \Big| \left(n^{-1} \sum_{i=1}^n (Y_{i} - \widehat{U}_{s-1}(\boldsymbol{X}_{i}))^2 \right)- n^{-1}\sum_{i=1}^n (Y_{i} - \widehat{U}_{s, J}(\boldsymbol{X}_{i}))^2\\
& \qquad - \mathbb{E}[(\mathbb{E}(f(\boldsymbol{X}) - R_{s-1}^{\star}(\boldsymbol{X})|\boldsymbol{X}_{\mathcal{X}(J)}))^2  ] \Big|\\
 &\le d_{n}.
 \end{split}
\end{equation}

In light of \eqref{equivalence.recur.4.c}, we now show that  $\widehat{J}_{1}, \dots, \widehat{J}_{K}$  include all significant component in $S_{1}^*$ such that $S_{1}^* \subset \cup_{l=1}^K\widehat{J}_{l}$. By \eqref{equivalence.recur.4.c}, for $0<s\le K$, and those insignificant components $m\in \{1,\dots, M\} \backslash S_{1}^*$,
\begin{equation}
\begin{split}\label{bias.control.1}
    &  \left(n^{-1} \sum_{i=1}^n (Y_{i} - \widehat{U}_{s-1}(\boldsymbol{X}_{i}))^2\right) - n^{-1}\sum_{i=1}^n (Y_{i} - \widehat{U}_{s, \{m\}}(\boldsymbol{X}_{i}))^2  \\
    &\le \mathbb{E}[(\mathbb{E}(f(\boldsymbol{X}) - R_{s-1}^{\star}(\boldsymbol{X})|\boldsymbol{X}_{\mathcal{X}(m)}))^2  ] + d_{n}.
    \end{split}
\end{equation}
Since we have assumed that Algorithm~\ref{algorithm1} makes the initial $2K$ selections  without any repetitions, we do not need to consider the cases with $\widehat{J}_{s}\subset   \cup_{l=1}^{s-1}\widehat{J}_{l}\cap S_{1}^*$ when deciding $\widehat{J}_{s}$ for $0<s\le K$. Meanwhile, by \eqref{equivalence.recur.4.c}, for $0<s\le K$, and those significant components that have not been selected $m \in  S_{1}^*\backslash (\cup_{l=1}^{s-1}\widehat{J}_{l})$,
\begin{equation}
\begin{split}\label{bias.control.2}
    &  \left(n^{-1} \sum_{i=1}^n (Y_{i} - \widehat{U}_{s-1}(\boldsymbol{X}_{i}))^2 \right) - n^{-1}\sum_{i=1}^n (Y_{i} - \widehat{U}_{s, \{m\}}(\boldsymbol{X}_{i}))^2  \\
    &\ge  \mathbb{E}[(\mathbb{E}(f(\boldsymbol{X}) - R_{s-1}^{\star}(\boldsymbol{X})|\boldsymbol{X}_{\mathcal{X}(m)}))^2  ] - d_{n}.
    \end{split}
\end{equation}

From \eqref{bias.control.1} and \eqref{bias.control.2}, it is seen that the analysis of $\mathbb{E}[(\mathbb{E}(f(\boldsymbol{X}) - R_{s-1}^{\star}(\boldsymbol{X})|\boldsymbol{X}_{\mathcal{X}(m)}))^2  ]$ is crucial for our desired results. We can show that for each $m \in  \{1, \dots, M\}\backslash (\cup_{l=1}^{s-1}\widehat{J}_{l})$ and each $0< s\le K$,
\begin{equation}
    \begin{split}\label{bias.control.3.b}
        & \left|\mathbb{E}[(\mathbb{E}(f(\boldsymbol{X}) - R_{s-1}^{\star}(\boldsymbol{X})|\boldsymbol{X}_{\mathcal{X}(m)}))^2  ] - \textnormal{Var}(g_{m}(\boldsymbol{X})) \right|  \le \frac{21\delta_{0}}{(p_{\min})^2}  \mathbb{E}[(f(\boldsymbol{X}))^2],
    \end{split}
\end{equation}
whose proof is deferred to the end of the current proof of Theorem~\ref{theorem3} (see Section~\ref{proof.theorem3.20}).

By \eqref{argmax.J.1}, \eqref{bias.control.1}--\eqref{bias.control.3.b}, Condition~\ref{signal.strength.1}, and other regularity conditions assumed by Theorem~\ref{theorem3}, it holds that on the event $\cap_{l=1}^6 E_{4l}$ with $d_{n} \le 1$ and 
$\frac{2\log{(n)}}{\sqrt{n}} \le p_{\min}$,
\begin{equation}
    \label{consistency.1}
    S_{1}^{\star}\subset\cup_{1\le l\le K}\widehat{J}_{l}.
\end{equation}

We have finished the proof of the results concerning $0<s \le K$, and now we proceed to deal with the case with $K<s \le 2K$. By \eqref{equivalence.recur.4.c}, for $K< s\le 2K$, and those insignificant components $J\subset \{1,\dots, M\}$ with $\texttt{\#}J=2$ and $J \not\in S_{2}^*$,
\begin{equation}
\begin{split}\label{bias.control.1.2k}
    &  \left( n^{-1} \sum_{i=1}^n (Y_{i} - \widehat{U}_{s-1}(\boldsymbol{X}_{i}))^2 \right) - n^{-1}\sum_{i=1}^n (Y_{i} - \widehat{U}_{s, J}(\boldsymbol{X}_{i}))^2  \\
    &\le \mathbb{E}[(\mathbb{E}(f(\boldsymbol{X}) - R_{s-1}^{\star}(\boldsymbol{X})|\boldsymbol{X}_{\mathcal{X}(J)}))^2  ] + d_{n}.
    \end{split}
\end{equation}
Since we have assumed that Algorithm~\ref{algorithm1} makes the initial $2K$ selections  without any repetitions, we do not need to consider those has already been selected when deciding $\widehat{J}_{s}$. Meanwhile, by \eqref{equivalence.recur.4.c}, for $K<s\le 2K$, and those significant components that have not been selected $J \in S_{2}^* \backslash \{ \widehat{J}_{K+1}, \dots, \widehat{J}_{s-1}\}$,
\begin{equation}
\begin{split}\label{bias.control.2.2k}
    &  \left( n^{-1} \sum_{i=1}^n (Y_{i} - \widehat{U}_{s-1}(\boldsymbol{X}_{i}))^2 \right) - n^{-1}\sum_{i=1}^n (Y_{i} - \widehat{U}_{s, J}(\boldsymbol{X}_{i}))^2  \\
    &\ge  \mathbb{E}[ (\mathbb{E}(f(\boldsymbol{X}) - R_{s-1}^{\star}(\boldsymbol{X})|\boldsymbol{X}_{\mathcal{X}(J)}))^2  ] - d_{n}.
    \end{split}
\end{equation}

Let us analyze the term $\mathbb{E}(f(\boldsymbol{X}) - R_{s-1}^{\star}(\boldsymbol{X})|\boldsymbol{X}_{\mathcal{X}(J)})$ on the RHS of \eqref{bias.control.1.2k} and \eqref{bias.control.2.2k} as follows. With the conditions required for \eqref{consistency.1} and Condition~\ref{heredity.1}, it holds that on the event $\cap_{l=1}^6 E_{4l}$, for each $K< s\le K + \texttt{\#}S_{2}^*$, each $J\subset \{1, \dots, M\}$ with $\texttt{\#}J=2$ and $J\not \in \{ \widehat{J}_{K+1}, \dots, \widehat{J}_{s-1}\}$,
\begin{equation}
    \begin{split}\label{bias.control.3.2k.b}
    & \left|\mathbb{E}[(\mathbb{E}(f(\boldsymbol{X}) - R_{s-1}^{\star}(\boldsymbol{X})|\boldsymbol{X}_{\mathcal{X}(J)}))^2  ] - \textnormal{Var}\left[g_{J}(\boldsymbol{X}) - \left(\sum_{j\in J} g_{\{j\}}(\boldsymbol{X}) \right)\right] \right| \\
           & \le 3808 \times \delta_{0} \times {p_{\min}^{-2}}\times  \mathbb{E} [(f(\boldsymbol{X}))^2] \times (\texttt{\#}S_{2}^*)^4,
    \end{split}
\end{equation}
whose proof is postponed to the end of the current proof of Theorem~\ref{theorem3} (see Section~\ref{proof.theorem3.24}).

By \eqref{argmax.J.1}, \eqref{consistency.1}--\eqref{bias.control.3.2k.b}, Condition~\ref{signal.strength.1}, Condition~\ref{heredity.1}, and other regularity conditions assumed by Theorem~\ref{theorem3}, it holds that on the event $\cap_{l=1}^6 E_{4l}$ with $d_{n} \le 1$ and 
$\frac{2\log{(n)}}{\sqrt{n}} \le p_{\min}$,
\begin{equation}\label{consistency.2}
    S_{2}^* \subset \{\widehat{J}_{K+1}, \dots, \widehat{J}_{K+\texttt{\#}S_{2}^*}\}.
\end{equation}

By \eqref{consistency.1}, \eqref{consistency.2}, and Lemma~\ref{lemma.event.1}, we have finished the proof of Theorem~\ref{theorem3}.


\subsubsection{Proof of \eqref{argmax.J.1}}
\label{proof.expression.2}

\noindent\textit{Proof of \eqref{argmax.J.1}: } We commence the proof by noting that $\widehat{J}_{l}$'s are determined based on the selection of feature groups to be split by Collaborative Trees, as is $(\widehat{Q}, \widehat{m})$ in \eqref{gini2}, where $\widehat{Q}$ represents a set of associated nodes. Thus, it is essential to elucidate their relationship as follows.

The definition of $\widehat{J}_{l}$'s has incorporated $(\widehat{Q}, \widehat{m})$ in \eqref{gini2} for the initial $2K$ updates. On the one hand, given $(\widehat{Q}, \widehat{m})$ and the update history preceding the current $s$th round with $s> K$, we derive $\widehat{J}_{s} = \{\widehat{m}, \widehat{m}^{\dagger}\}$, where $\widehat{Q}$ results from the splitting of a root node on the $\widehat{m}^{\dagger}$th feature group. On the other hand, for $s\le K$, $\widehat{J}_{s} = \{\widehat{m}\}$. Therefore, the pair $(\widehat{Q}, \widehat{m})$ that maximizes \eqref{gini2} at the $s$th round defines a pair $(\widehat{J}_{s}, \textnormal{the index of the tree to be split})$. In essence, each pair $(Q, m)$ at the $s$th round is equivalent to a pair of a feature group index set $J$ and the index of the tree to be split in this sense.

In the following, we demonstrate that at the $s$th round, the sum of split scores in the square brackets of \eqref{gini2} for each pair of $(Q, m)$ is equivalent to the term in the curly brackets of \eqref{argmax.J.1} with the corresponding set $J\in \Theta_{s}$, thereby confirming that $\widehat{J}_{s}$ indeed maximizes the RHS of \eqref{argmax.J.1}.

At the onset of the $s$th round of update, where $0 < s \le K$, residuals $\widehat{r}_{i} = Y_{i} - \widehat{U}_{s-1}(\boldsymbol{X}_{i})$ for $i \in \{1, \dots, n\}$ are obtained. Since $s \le K$, one of the root nodes, denoted as $C_{0}$, is split during this round. Let us derive an expression for SplitScore($C_{0}, m$) as in \eqref{gini1} for the $m$th feature group. For simplicity, assume that the $m$th feature group comprises a single binary feature $\mathcal{X}(m) = \{q\}$ for some $1 \le q \le p$. Following the notation in \eqref{gini1}, child nodes $C_{1} = \{\vv{x} \in C_{0}: x_{q} > X_{\widehat{a}q}\}$ and $C_{2} = \{\vv{x} \in C_{0}: x_{q} \le X_{\widehat{a}q}\}$ are established, where $\widehat{a} \in N_{0} = \{i: \boldsymbol{X}_{i} \in C_{0}\} = \{1, \dots, n\}$ is the index maximizing \eqref{gini1}. Consequently, $N_{l}(\widehat{a}) = \{i: \boldsymbol{X}_{i} \in C_{l}\}$ for $l \in \{1, 2\}$.

SplitScore($C_{0}, m$) in \eqref{gini1} can be written as
\begin{equation}
    \begin{split}\label{expression.1}
& \left(\sum_{i\in N_{0}} \widehat{r}_{i}^2\right) -  \sum_{l=1}^2 \sum_{i\in N_{l}(\widehat{a})}  (\widehat{r}_{i} -  \frac{1}{1\vee\texttt{\#} N_{l}(\widehat{a})} \sum_{i\in N_{l}(\widehat{a}) } \widehat{r}_{i}  )^2\\
& = \left(\sum_{i=1}^n \widehat{r}_{i}^2\right) -  \sum_{l=1}^2 \sum_{i\in \{t: \boldsymbol{X}_{t} \in C_{l}\} }  (\widehat{r}_{i} -  \frac{1}{1\vee\texttt{\#} \{i: \boldsymbol{X}_{i} \in C_{l}\}} \sum_{i\in \{t: \boldsymbol{X}_{t} \in C_{l}\} } \widehat{r}_{i}  )^2\\
& = \left(\sum_{i=1}^n \widehat{r}_{i}^2\right) -   \sum_{i=1 }^n \sum_{l=1}^{2} \boldsymbol{1}_{\boldsymbol{X}_{i} \in C_{l}}  (\widehat{r}_{i} -  \frac{1}{1\vee\texttt{\#} \{i: \boldsymbol{X}_{i} \in C_{l}\}} \sum_{i\in \{t: \boldsymbol{X}_{t} \in C_{l}\} } \widehat{r}_{i}  )^2 \\
& = \left(\sum_{i=1}^n \widehat{r}_{i}^2\right) -   \sum_{i=1 }^n    \left[\sum_{l=1}^{2}\boldsymbol{1}_{\boldsymbol{X}_{i} \in C_{l}} \left(\widehat{r}_{i} -  \frac{1}{1\vee\texttt{\#} \{i: \boldsymbol{X}_{i} \in C_{l}\}} \sum_{i\in \{t: \boldsymbol{X}_{t} \in C_{l}\} } \widehat{r}_{i}  \right)\right]^2\\
& = \left(\sum_{i=1}^n \widehat{r}_{i}^2\right) -   \sum_{i=1 }^n     \left(\widehat{r}_{i} -  \sum_{l=1}^{2}\boldsymbol{1}_{\boldsymbol{X}_{i} \in C_{l}} \frac{\sum_{i\in \{t: \boldsymbol{X}_{t} \in C_{l}\} } \widehat{r}_{i}}{1\vee\texttt{\#} \{i: \boldsymbol{X}_{i} \in C_{l}\}}   \right)^2,
    \end{split}
\end{equation}
where the third equality holds since $ \boldsymbol{1}_{\boldsymbol{X}_{i} \in C_{1}} \times \boldsymbol{1}_{\boldsymbol{X}_{i} \in C_{2}} = 0$, and that the square of an indicator function equals the indicator function itself.

It is worth noting that we actually have $X_{\widehat{a}q} = 0$, which is deduced from: 1) if $X_{\widehat{a}q} = 1$, the corresponding split score is zero, and 2) our assumption that all nodes are valid before the first $2K$ updates. Thus, $\{C_{1, J}, C_{2, J}\}= \{C_{1}, C_{2}\}$ with $J = \{m\}$. With this, and $\widehat{r}_{i} = Y_{i} - \widehat{U}_{s-1}(\boldsymbol{X}_{i})$ at the $s$th round of update, the RHS of \eqref{expression.1} has the same expression as in the curly brackets of \eqref{argmax.J.1} with $J=\{m\}$. We have finished the proof for the scenario with $0< s\le K$.

Now, let us consider the cases with $K < s \le 2K$. We aim to calculate the summation of split scores over a set of associated nodes $Q$ in \eqref{gini2}, corresponding to the $m$th feature group. Similar to the previous case, we simplify the analysis by assuming that the nodes in $Q$ are split based on the $m^{\dagger}$th feature group, which has a single feature $\mathcal{X}(m^{\dagger}) = \{q_{1}\}$ for some $1\le q_{1}\le p$. Meanwhile, we assume that the $m$th feature group to be split at the current round also has a single binary feature $\mathcal{X}(m) = \{q_{2}\}$ for some $1\le q_{2}\le p$. Let $J = \{m, m^{\dagger}\}$, and define:
$\{C_{1, J}, C_{2, J}, C_{3, J}, C_{4, J}\} = \{ \{\vv{x} \in \mathbb{R}^p: x_{q_{1}}>0, x_{q_{2}}>0 \}, \{\vv{x} \in \mathbb{R}^p: x_{q_{1}}>0, x_{q_{2}}\le 0 \}, \{\vv{x} \in \mathbb{R}^p: x_{q_{1}}\le 0, x_{q_{2}}>0 \}, \{\vv{x} \in \mathbb{R}^p: x_{q_{1}}\le 0, x_{q_{2}}\le 0 \} \}$. 

Then, the same derivation as in \eqref{expression.1} can be done for showing that the sum of the split scores $\sum_{ (C,k) \in Q}  \textnormal{ SplitScore} (C, m)$  in \eqref{gini2} is equivalent to the expression in the curly bracket of \eqref{argmax.J.1} with $J = \{m, m^{\dagger}\}$. Note that we have $J\in \Theta_{s}$ according to the definition of $\Theta_{s}$ in \eqref{argmax.J.1}. 
For simplicity, we omit the detailed derivation in this scenario.

By the above arguments, the proof of \eqref{argmax.J.1} is complete.

\subsubsection{Proof of \eqref{equivalence.recur.10.b}} \label{proof.theorem1.12}
\noindent\textit{Proof of \eqref{equivalence.recur.10.b}: } Let us begin with establishing an upper bound for the term $\sum_{i=1}^n (f(\boldsymbol{X}_{i}) - R_{s, J}^{\star}(\boldsymbol{X}_{i}))^2$ on the RHS of \eqref{equivalence.recur.3.b} as follows. By \eqref{dependence.7.b}, we have that for each $s \ge 0$ and every set $J$ with $\texttt{\#} J \le 2$,
\begin{equation}
    \begin{split}
        \label{equivalence.recur.5.b}
        \sum_{i=1}^n (f(\boldsymbol{X}_{i}) - R_{s, J}^{\star}(\boldsymbol{X}_{i}))^2 \le n(\max_{\vv{c} \in \{0, 1\}^{p}}|f(\vv{c})| + 3\sqrt{\mathbb{E}[(f(\boldsymbol{X}))^2]}  \times p_{\min}^{-1} +  \iota_{s-1} )^2.
    \end{split}
\end{equation}


Next, we establish upper bounds for $\sum_{i=1}^n \varepsilon_{i} (f(\boldsymbol{X}_{i}) - R_{s, J}^{\star}(\boldsymbol{X}_{i}))$ as follows. By \eqref{dependence.6.b},
{\small \begin{equation}
    \begin{split}
        \label{equivalence.recur.6.b}
        & \left |\sum_{i=1}^n \varepsilon_{i}R_{s, J}^{\star}(\boldsymbol{X}_{i})\right| \\
        & = 
        \left| \sum_{i=1}^n \varepsilon_{i} \left[\left(\sum_{\vv{c} \in \{0, 1\}^{\texttt{\#}\mathcal{X}(J)}}  a_{s, \vv{c}, J}\times  \boldsymbol{1}_{\boldsymbol{X}_{i\mathcal{X}(J)} = \vv{c}} \right) + \sum_{l=1}^{s-1} \sum_{\vv{c} \in \{0, 1\}^{\texttt{\#}\mathcal{X}(\widehat{J}_{l})}}  a_{l, \vv{c}, \widehat{J}_{l}}\times  \boldsymbol{1}_{\boldsymbol{X}_{i\mathcal{X}(\widehat{J}_{l})} = \vv{c}} \right] \right| \\
        & =  \left| \left(\sum_{\vv{c} \in \{0, 1\}^{\texttt{\#}\mathcal{X}(J)}}  a_{s, \vv{c}, J} \sum_{i=1}^n \varepsilon_{i} \boldsymbol{1}_{\boldsymbol{X}_{i\mathcal{X}(J)} = \vv{c}}\right) + \left(\sum_{l=1}^{s-1} \sum_{\vv{c} \in \{0, 1\}^{\texttt{\#}\mathcal{X}(\widehat{J}_{l})}}  a_{l, \vv{c}, \widehat{J}_{l}} \sum_{i=1}^n \varepsilon_{i} \boldsymbol{1}_{\boldsymbol{X}_{i\mathcal{X}(\widehat{J}_{l})} = \vv{c}} \right) \right| \\
        & \le \left(\sum_{\vv{c} \in \{0, 1\}^{\texttt{\#}\mathcal{X}(J)}}  |a_{s, \vv{c}, J} |\left| \sum_{i=1}^n \varepsilon_{i} \boldsymbol{1}_{\boldsymbol{X}_{i\mathcal{X}(J)} = \vv{c}} \right| \right) \\
        &\qquad + \left(\sum_{l=1}^{s-1} \sum_{\vv{c} \in \{0, 1\}^{\texttt{\#}\mathcal{X}(\widehat{J}_{l})}}  | a_{l, \vv{c}, \widehat{J}_{l}}| \left| \sum_{i=1}^n \varepsilon_{i} \boldsymbol{1}_{\boldsymbol{X}_{i\mathcal{X}(\widehat{J}_{l})} = \vv{c}} \right| \right),
    \end{split}
\end{equation}}%
where $a_{l, \vv{c}, J} = \mathbb{E}\{ f(\boldsymbol{X}) - R_{l-1}^{\star}(\boldsymbol{X})  |\boldsymbol{X}_{\mathcal{X}(J)}= \vv{c}\}$ for every set $J$, each $l>0$, and each $\vv{c} \in \{0, 1\}^{\texttt{\#}\mathcal{X}(J)}$. For the completeness of notation, we define $a_{0, \vv{c}, J} = 0$ for all vector $\vv{c}$ and set $J$. Also, recall that summation over an empty set is defined to be zero.

By \eqref{equivalence.recur.6.b} and \eqref{dependence.8.b}, on the event $E_{41}\cap E_{42}$ defined in \eqref{event.2} with $t_{1}=t_2 = \log{(n)}n^{\frac{1}{2} + \frac{1}{q_{1}}}$, it holds that for each $s \ge 0$ and every set $J$ with $\texttt{\#} J \le 2$,
\begin{equation}
    \label{event.1}
    \begin{split}        
    & |\sum_{i=1}^n \varepsilon_{i} (f(\boldsymbol{X}_{i}) - R_{s, J}^{\star}(\boldsymbol{X}_{i}))| \\
    & \le t_{1} + t_{2} (M_{\mathcal{X}}^2 \times 3\sqrt{\mathbb{E}[(f(\boldsymbol{X}))^2]}  \times p_{\min}^{-1} + M_{\mathcal{X}}^2 \iota_{s-1})\\
    & = \log{(n)}n^{\frac{1}{2} + \frac{1}{q_{1}}} (3M_{\mathcal{X}}^2p_{\min}^{-1} \sqrt{\mathbb{E}[(f(\boldsymbol{X}))^2]}   + M_{\mathcal{X}}^2 \iota_{s-1} + 1).
    \end{split}
\end{equation}


Next, we proceed to establish an upper bound for the term $\sum_{i=1}^n (R_{s, J}^{\star}(\boldsymbol{X}_{i}) - \widehat{U}_{s, J}(\boldsymbol{X}_{i}))^2$. First, when $s=0$, by the definitions of $R_{s, J}^{\star}(\boldsymbol{X}_{i})$ and $\widehat{U}_{s, J}(\boldsymbol{X}_{i})$ in \eqref{predictor.trees.b.3},
\begin{equation}\label{equivalence.recur.12.b}
    \sum_{i=1}^n (R_{s, J}^{\star}(\boldsymbol{X}_{i}) - \widehat{U}_{s, J}(\boldsymbol{X}_{i}))^2 = 0.
\end{equation}
Meanwhile, when $s>0$, by \eqref{predictor.trees.b.3} and the model assumption of $Y= f(\boldsymbol{X}) + \varepsilon$, we deduce that
\begin{equation}
    \begin{split}\label{equivalence.recur.2.b}
        & \sum_{i=1}^n (R_{s, J}^{\star}(\boldsymbol{X}_{i}) - \widehat{U}_{s, J}(\boldsymbol{X}_{i}))^2 \\
        &=  \sum_{i=1}^n \bigg(R_{s-1}^{\star} (\boldsymbol{X}_{i})+  \mathbb{E}( f(\boldsymbol{X}_{i}) - R_{s-1}^{\star}(\boldsymbol{X}_{i}) | \boldsymbol{X}_{i\mathcal{X}(J)}) \\
        &\qquad - \widehat{U}_{s-1}(\boldsymbol{X}_{i}) - \sum_{l\in \mathcal{L}(J)} \boldsymbol{1}_{\boldsymbol{X}_{i} \in C_{l, J}} \frac{\sum_{i:\boldsymbol{X}_{i} \in C_{l, J} } Y_{i} - \widehat{U}_{s-1}(\boldsymbol{X}_{i}) }{ \texttt{\#} \{i:\boldsymbol{X}_{i} \in C_{l, J} \} \vee 1 }  \bigg)^2 \\
        &=  \sum_{i=1}^n \bigg(R_{s-1}^{\star} (\boldsymbol{X}_{i}) - \left( \sum_{l\in \mathcal{L}(J)} \boldsymbol{1}_{\boldsymbol{X}_{i} \in C_{l, J}} \frac{\sum_{i:\boldsymbol{X}_{i} \in C_{l, J} }  R_{s-1}^{\star}(\boldsymbol{X}_{i}) }{ \texttt{\#} \{i:\boldsymbol{X}_{i} \in C_{l, J} \} \vee 1 } \right) \\
        &\qquad + \left(\sum_{l\in \mathcal{L}(J)} \boldsymbol{1}_{\boldsymbol{X}_{i} \in C_{l, J}} \frac{\sum_{i:\boldsymbol{X}_{i} \in C_{l, J} }  R_{s-1}^{\star}(\boldsymbol{X}_{i}) }{ \texttt{\#} \{i:\boldsymbol{X}_{i} \in C_{l, J} \} \vee 1 }\right)  -  \mathbb{E}( R_{s-1}^{\star}(\boldsymbol{X}_{i}) | \boldsymbol{X}_{i\mathcal{X}(J)}) \\
        &\qquad +  \mathbb{E}( f(\boldsymbol{X}_{i})  | \boldsymbol{X}_{i\mathcal{X}(J)}) - \sum_{l\in \mathcal{L}(J)} \boldsymbol{1}_{\boldsymbol{X}_{i} \in C_{l, J}} \frac{\sum_{i:\boldsymbol{X}_{i} \in C_{l, J} } f(\boldsymbol{X}_{i})  }{ \texttt{\#} \{i:\boldsymbol{X}_{i} \in C_{l, J} \} \vee 1 } \\
        &\qquad - \sum_{l\in \mathcal{L}(J)} \boldsymbol{1}_{\boldsymbol{X}_{i} \in C_{l, J}} \frac{\sum_{i:\boldsymbol{X}_{i} \in C_{l, J} } \varepsilon_{i}  }{ \texttt{\#} \{i:\boldsymbol{X}_{i} \in C_{l, J} \} \vee 1 }\\
        &\qquad - \widehat{U}_{s-1}(\boldsymbol{X}_{i}) + \sum_{l\in \mathcal{L}(J)} \boldsymbol{1}_{\boldsymbol{X}_{i} \in C_{l, J}} \frac{\sum_{i:\boldsymbol{X}_{i} \in C_{l, J} }  \widehat{U}_{s-1}(\boldsymbol{X}_{i}) }{ \texttt{\#} \{i:\boldsymbol{X}_{i} \in C_{l, J} \} \vee 1 }  \bigg)^2.
    \end{split}
\end{equation}
In addition, we have that for every measurable $q:\{0, 1\}^{\texttt{\#}\mathcal{X}(J)} \mapsto\mathbb{R}$,
\begin{equation}
    \begin{split}
&  \sum_{i=1}^n q(\boldsymbol{X}_{i\mathcal{X}(J)}) \bigg( R_{s-1}^{\star} (\boldsymbol{X}_{i}) -  \sum_{l\in \mathcal{L}(J)} \boldsymbol{1}_{\boldsymbol{X}_{i} \in C_{l, J}} \frac{\sum_{i:\boldsymbol{X}_{i} \in C_{l, J} }  R_{s-1}^{\star}(\boldsymbol{X}_{i}) }{ \texttt{\#} \{i:\boldsymbol{X}_{i} \in C_{l, J} \} \vee 1 } \\
& \qquad - \widehat{U}_{s-1}(\boldsymbol{X}_{i}) + \sum_{l\in \mathcal{L}(J)} \boldsymbol{1}_{\boldsymbol{X}_{i} \in C_{l, J}} \frac{\sum_{i:\boldsymbol{X}_{i} \in C_{l, J} }  \widehat{U}_{s-1}(\boldsymbol{X}_{i}) }{ \texttt{\#} \{i:\boldsymbol{X}_{i} \in C_{l, J} \} \vee 1 } \bigg)  \\
  &  = \sum_{i=1}^n \sum_{l \in \mathcal{L}(J)} \boldsymbol{1}\{\boldsymbol{X}_{i} \in C_{l, J}\} \times q(\boldsymbol{X}_{i\mathcal{X}(J)}) \bigg( R_{s-1}^{\star} (\boldsymbol{X}_{i}) -   \frac{\sum_{i:\boldsymbol{X}_{i} \in C_{l, J} }  R_{s-1}^{\star}(\boldsymbol{X}_{i}) }{ \texttt{\#} \{i:\boldsymbol{X}_{i} \in C_{l, J} \} \vee 1 } \\
  & \qquad - \widehat{U}_{s-1}(\boldsymbol{X}_{i}) +  \frac{\sum_{i:\boldsymbol{X}_{i} \in C_{l, J} }  \widehat{U}_{s-1}(\boldsymbol{X}_{i}) }{ \texttt{\#} \{i:\boldsymbol{X}_{i} \in C_{l, J} \} \vee 1 } \bigg)  \\
  &  =   \sum_{l \in \mathcal{L}(J)} q_{l, J} \sum_{i=1}^n\boldsymbol{1}\{\boldsymbol{X}_{i} \in C_{l, J}\} \bigg( R_{s-1}^{\star} (\boldsymbol{X}_{i}) -   \frac{\sum_{i:\boldsymbol{X}_{i} \in C_{l, J} }  R_{s-1}^{\star}(\boldsymbol{X}_{i}) }{ \texttt{\#} \{i:\boldsymbol{X}_{i} \in C_{l, J} \} \vee 1 } \\
  & \qquad - \widehat{U}_{s-1}(\boldsymbol{X}_{i}) +  \frac{\sum_{i:\boldsymbol{X}_{i} \in C_{l, J} }  \widehat{U}_{s-1}(\boldsymbol{X}_{i}) }{ \texttt{\#} \{i:\boldsymbol{X}_{i} \in C_{l, J} \} \vee 1 } \bigg) \\
  & = 0,
    \end{split}
\end{equation}
where the second equality holds due to an interchange of the two outer summations and that $q_{l, J} = q(\boldsymbol{X}_{i\mathcal{X}(J)})$  when $\boldsymbol{X}_{i} \in C_{l, J}$ for each $l\in\mathcal{L}(J)$. Moreover, it holds that 
\begin{equation}
    \begin{split}\label{equivalence.recur.1.b}
        & \sum_{i=1}^n \bigg(R_{s-1}^{\star} (\boldsymbol{X}_{i}) - \sum_{l\in \mathcal{L}(J)} \boldsymbol{1}_{\boldsymbol{X}_{i} \in C_{l, J}} \frac{\sum_{i:\boldsymbol{X}_{i} \in C_{l, J} }  R_{s-1}^{\star}(\boldsymbol{X}_{i}) }{ \texttt{\#} \{i:\boldsymbol{X}_{i} \in C_{l, J} \} \vee 1 } \\
        &\qquad - \widehat{U}_{s-1}(\boldsymbol{X}_{i}) + \sum_{l\in \mathcal{L}(J)} \boldsymbol{1}_{\boldsymbol{X}_{i} \in C_{l, J}} \frac{\sum_{i:\boldsymbol{X}_{i} \in C_{l, J} }  \widehat{U}_{s-1}(\boldsymbol{X}_{i}) }{ \texttt{\#} \{i:\boldsymbol{X}_{i} \in C_{l, J} \} \vee 1 }  \bigg)^2\\
        & = \sum_{l\in \mathcal{L}(J)} \sum_{i: \boldsymbol{X}_{i} \in C_{l, J}} \bigg(R_{s-1}^{\star} (\boldsymbol{X}_{i}) -  \frac{\sum_{i:\boldsymbol{X}_{i} \in C_{l, J} }  R_{s-1}^{\star}(\boldsymbol{X}_{i}) }{ \texttt{\#} \{i:\boldsymbol{X}_{i} \in C_{l, J} \} \vee 1 } \\
        &\qquad - \widehat{U}_{s-1}(\boldsymbol{X}_{i}) +  \frac{\sum_{i:\boldsymbol{X}_{i} \in C_{l, J} }  \widehat{U}_{s-1}(\boldsymbol{X}_{i}) }{ \texttt{\#} \{i:\boldsymbol{X}_{i} \in C_{l, J} \} \vee 1 }  \bigg)^2 \\        
        & \le \sum_{l\in \mathcal{L}(J)} \sum_{i: \boldsymbol{X}_{i} \in C_{l, J}} \left( R_{s-1}^{\star} (\boldsymbol{X}_{i}) - \widehat{U}_{s-1}(\boldsymbol{X}_{i})   \right)^2  \\
        & = \sum_{i=1}^n  \left( R_{s-1}^{\star} (\boldsymbol{X}_{i}) - \widehat{U}_{s-1}(\boldsymbol{X}_{i})   \right)^2,
    \end{split}    
\end{equation}
where the first inequality follows from that $\sum_{i=1}^n (b_{i} - n^{-1}\sum_{i=1}^n b_{i })^2 \le \sum_{i=1}^n b_{i}^2$ for all real $b_{i}$'s.

By \eqref{equivalence.recur.2.b}--\eqref{equivalence.recur.1.b} and Minkowski's inequality, it holds that for each $s> 0$ and every set $J$ with $\texttt{\#}J\le 2$, 
\begin{equation}
    \begin{split}\label{recursive.1}
        \sum_{i=1}^n (R_{s, J}^{\star}(\boldsymbol{X}_{i}) - \widehat{U}_{s, J}(\boldsymbol{X}_{i}))^2 \le \sum_{i=1}^n (R_{s-1}^{\star}(\boldsymbol{X}_{i}) - \widehat{U}_{s-1}(\boldsymbol{X}_{i}))^2 + \left(\sum_{l=1}^3 E_{l}\right)^2,
    \end{split}
\end{equation}
where
\begin{equation*}
    \begin{split}
        E_{1} & = \sqrt{\sum_{i=1}^n \left( \left(\sum_{l\in \mathcal{L}(J)} \boldsymbol{1}_{\boldsymbol{X}_{i} \in C_{l, J}} \frac{\sum_{i:\boldsymbol{X}_{i} \in C_{l, J} }  R_{s-1}^{\star}(\boldsymbol{X}_{i}) }{ \texttt{\#} \{i:\boldsymbol{X}_{i} \in C_{l, J}  \} \vee 1 } \right) -  \mathbb{E}( R_{s-1}^{\star}(\boldsymbol{X}_{i}) | \boldsymbol{X}_{i\mathcal{X}(J)}) \right)^2 }, \\
        E_{2} & = \sqrt{ \sum_{i=1}^n \left(  \mathbb{E}( f(\boldsymbol{X}_{i})  | \boldsymbol{X}_{i\mathcal{X}(J)}) - \sum_{l\in \mathcal{L}(J)} \boldsymbol{1}_{\boldsymbol{X}_{i} \in C_{l, J}} \frac{\sum_{i:\boldsymbol{X}_{i} \in C_{l, J} } f(\boldsymbol{X}_{i})  }{ \texttt{\#} \{i:\boldsymbol{X}_{i} \in C_{l, J} \} \vee 1 } \right)^2} , \\
        E_{3} & = \sqrt{ \sum_{i=1}^n\left( \sum_{l\in \mathcal{L}(J)} \boldsymbol{1}_{\boldsymbol{X}_{i} \in C_{l, J}} \frac{\sum_{i:\boldsymbol{X}_{i} \in C_{l, J} } \varepsilon_{i}  }{ \texttt{\#} \{i:\boldsymbol{X}_{i} \in C_{l, J} \} \vee 1 } \right)^2}.
    \end{split}
\end{equation*}

We now establish bounds for $E_{1}, E_{2}, E_{3}$ as follows. For $E_{1}$, note that
\begin{equation}
    \begin{split}\label{dependence.4.b}
        & \left(\sum_{l\in \mathcal{L}(J)} \boldsymbol{1}_{\boldsymbol{X}_{i} \in C_{l, J}} \frac{\sum_{i:\boldsymbol{X}_{i} \in C_{l, J} }  R_{s-1}^{\star}(\boldsymbol{X}_{i}) }{ \texttt{\#} \{i:\boldsymbol{X}_{i} \in C_{l, J}  \} \vee 1 } \right) -  \mathbb{E}( R_{s-1}^{\star}(\boldsymbol{X}_{i}) | \boldsymbol{X}_{i\mathcal{X}(J)}) \\
        & = \sum_{l\in \mathcal{L}(J)}  \boldsymbol{1}_{\boldsymbol{X}_{i} \in C_{l, J}} \left(\frac{\sum_{i=1 }^{n}  \boldsymbol{1}_{\boldsymbol{X}_{i} \in C_{l, J}} R_{s-1}^{\star}(\boldsymbol{X}_{i}) }{ \texttt{\#} \{i:\boldsymbol{X}_{i} \in C_{l, J}  \} \vee 1 }  -  \mathbb{E}( R_{s-1}^{\star}(\boldsymbol{X}_{i}) | \boldsymbol{X}_{i} \in C_{l, J}) \right),
    \end{split}
\end{equation}
where we rely on that $\mathbb{E}( R_{s-1}^{\star}(\boldsymbol{X}_{i}) | \boldsymbol{X}_{i\mathcal{X}(J)})= \sum_{l\in \mathcal{L}(J)} \boldsymbol{1}_{\boldsymbol{X}_{i} \in C_{l, J}} \mathbb{E}( R_{s-1}^{\star}(\boldsymbol{X}_{i}) | \boldsymbol{X}_{i} \in C_{l, J}) $ almost surely for every set $J$ with $\texttt{\#}J\le 2$.
Below, we deal with terms inside the summation on the RHS of \eqref{dependence.4.b}. By \eqref{dependence.6.b}, for every set $J$ with $\texttt{\#}J \le 2$, each $s>0$, and each $l\in \mathcal{L}(J)$,
{\small \begin{equation}
    \begin{split}\label{E1.1}
        & \bigg| \frac{\sum_{i=1 }^n \boldsymbol{1}_{\boldsymbol{X}_{i} \in C_{l, J}} R_{s-1}^{\star}(\boldsymbol{X}_{i}) }{ \texttt{\#} \{i:\boldsymbol{X}_{i} \in C_{l, J}  \} \vee 1 }  -  \frac{\mathbb{E}( R_{s-1}^{\star}(\boldsymbol{X}) \boldsymbol{1}_{\boldsymbol{X} \in C_{l, J} })}{\mathbb{P}(\boldsymbol{X} \in C_{l, J})} \bigg|\\
        & = \bigg|\left( n^{-1}\sum_{i=1 }^n \left( \boldsymbol{1}_{\boldsymbol{X}_{i} \in C_{l, J}} R_{s-1}^{\star}(\boldsymbol{X}_{i}) -  \mathbb{E}( R_{s-1}^{\star}(\boldsymbol{X}) \boldsymbol{1}_{\boldsymbol{X} \in C_{l, J} }) \right) \right)\frac{n }{ \texttt{\#} \{i:\boldsymbol{X}_{i} \in C_{l, J}  \} \vee 1 }  \\
        & \qquad -  \mathbb{E}( R_{s-1}^{\star}(\boldsymbol{X}) \boldsymbol{1}_{\boldsymbol{X} \in C_{l, J} }) \left(\frac{1 }{\mathbb{P}(\boldsymbol{X} \in C_{l, J})} - \frac{n }{ \texttt{\#} \{i:\boldsymbol{X}_{i} \in C_{l, J}  \} \vee 1 }\right) \bigg|\\
        & = \bigg| \frac{n }{ \texttt{\#} \{i:\boldsymbol{X}_{i} \in C_{l, J}  \} \vee 1 }\\
        & \ \  \times \left( n^{-1}\sum_{i=1 }^n \left( \sum_{l=1}^{s-1} \sum_{\vv{c} \in \{0, 1\}^{\texttt{\#}\mathcal{X}(\widehat{J}_{l})}}  a_{l, \vv{c}, \widehat{J}_{l}}\times  \Big[ \boldsymbol{1}_{\boldsymbol{X}_{i\mathcal{X}(\widehat{J}_{l})} = \vv{c}} \boldsymbol{1}_{\boldsymbol{X}_{i} \in C_{l, J}}  -  \mathbb{E}( \boldsymbol{1}_{\boldsymbol{X}_{\mathcal{X}(\widehat{J}_{l})} = \vv{c}} \boldsymbol{1}_{\boldsymbol{X} \in C_{l, J} })  \Big]  \right) \right)  \\
        & \qquad -  \mathbb{E}( R_{s-1}^{\star}(\boldsymbol{X}) \boldsymbol{1}_{\boldsymbol{X} \in C_{l, J} }) \left(\frac{1 }{\mathbb{P}(\boldsymbol{X} \in C_{l, J})} - \frac{n }{ \texttt{\#} \{i:\boldsymbol{X}_{i} \in C_{l, J}  \} \vee 1 }\right) \bigg| \\
        & \le \frac{n }{ \texttt{\#} \{i:\boldsymbol{X}_{i} \in C_{l, J}  \} \vee 1 }  \\
        & \ \  \times  \left( \sum_{l=1}^{s-1} \sum_{\vv{c} \in \{0, 1\}^{\texttt{\#}\mathcal{X}(\widehat{J}_{l})}}  |a_{l, \vv{c}, \widehat{J}_{l}}| \times \left| n^{-1}\sum_{i=1 }^n  \Big[ \boldsymbol{1}_{\boldsymbol{X}_{i\mathcal{X}(\widehat{J}_{l})} = \vv{c}} \boldsymbol{1}_{\boldsymbol{X}_{i} \in C_{l, J}}  -  \mathbb{E}( \boldsymbol{1}_{\boldsymbol{X}_{\mathcal{X}(\widehat{J}_{l})} = \vv{c}} \boldsymbol{1}_{\boldsymbol{X} \in C_{l, J} }) \Big] \right|  \right)   \\
        & \qquad +  \mathbb{E}( |R_{s-1}^{\star}(\boldsymbol{X}) |) \left|\frac{1 }{\mathbb{P}(\boldsymbol{X} \in C_{l, J})} - \frac{n }{ \texttt{\#} \{i:\boldsymbol{X}_{i} \in C_{l, J}  \} \vee 1 }\right|.
    \end{split}
\end{equation}}%

To deal with the RHS of \eqref{E1.1}, we first derive that for each $J\subset\{1, \dots, M\}$ with $\texttt{\#}J\le 2$ and every $l\in\mathcal{L}(J)$,
\begin{equation}
    \label{lower.bound.pmin.1}
    \mathbb{P}(\boldsymbol{X} \in C_{l, J})\ge p_{\min},
\end{equation} 
due to the definitions of $C_{l, J}$'s and $p_{\min}$. By \eqref{lower.bound.pmin.1}, it holds that on the event $E_{44}$, for every set $J\subset \{1, \dots, M\}$ with $\texttt{\#}J\le 2$ and each $l\in\mathcal{L}(J)$,
\begin{equation}\label{lower.bound.1}
    \texttt{\#} \{i:\boldsymbol{X}_{i} \in C_{l, J} \} \times n^{-1} \ge p_{\min} - \frac{t_{4}}{n},
\end{equation}
where $E_{44}$  is defined in \eqref{event.2}, and $p_{\min}$ is defined in Condition~\ref{signal.strength.1}.

By \eqref{dependence.8.b}, \eqref{lower.bound.1}, Lemma~\ref{lemma.2}, and \eqref{lower.bound.pmin.1}, it holds that on the event $E_{44}$, for each $s> 0$ and every set $J$ with $\texttt{\#}J\le 2$,
\begin{equation}
    \begin{split}\label{upper.bound.1}
        \textnormal{RHS of \eqref{E1.1}} & \le \frac{t_{4}n^{-1}\times M_{\mathcal{X}}^2\iota_{s-1}   }{p_{\min} - \frac{t_{4}}{n} } +  \frac{t_{4}n^{-1} \times 2\sqrt{\mathbb{E}[(f(\boldsymbol{X}))^2]} }{p_{\min}(p_{\min} - \frac{t_{4}}{n})}.
    \end{split}
\end{equation}

Next, we proceed to deal with $E_{2}$. Since  
$$\mathbb{E}( f(\boldsymbol{X}_{i}) | \boldsymbol{X}_{i\mathcal{X}(J)})= \sum_{l\in \mathcal{L}(J)} \boldsymbol{1}_{\boldsymbol{X}_{i} \in C_{l, J}} \mathbb{E}( f(\boldsymbol{X}_{i}) | \boldsymbol{X}_{i} \in C_{l, J}) $$ 
for every set $J$ with $\texttt{\#}J\le 2$, we deduce that for each $s>0$,
\begin{equation}
    \begin{split}\label{upper.bound.2.b}
        & \left|  \mathbb{E}( f(\boldsymbol{X}_{i})  | \boldsymbol{X}_{i\mathcal{X}(J)}) - \sum_{l\in \mathcal{L}(J)} \boldsymbol{1}_{\boldsymbol{X}_{i} \in C_{l, J}} \frac{\sum_{i:\boldsymbol{X}_{i} \in C_{l, J} } f(\boldsymbol{X}_{i})  }{ \texttt{\#} \{i:\boldsymbol{X}_{i} \in C_{l, J} \} \vee 1 } \right| \\
        & = \bigg| \sum_{l\in \mathcal{L}(J)} \boldsymbol{1}_{\boldsymbol{X}_{i} \in C_{l, J}} \left( \mathbb{E}( f(\boldsymbol{X}_{i})  | \boldsymbol{X}_{i} \in C_{l, J}) -  \frac{\sum_{i:\boldsymbol{X}_{i} \in C_{l, J} } f(\boldsymbol{X}_{i})  }{ \texttt{\#} \{i:\boldsymbol{X}_{i} \in C_{l, J} \} \vee 1 } \right) \bigg|\\
        & = \bigg| \sum_{l\in \mathcal{L}(J)} \boldsymbol{1}_{\boldsymbol{X}_{i} \in C_{l, J}} \bigg[  \left( \mathbb{E}( f(\boldsymbol{X}_{i})  \boldsymbol{1}_{\boldsymbol{X}_{i} \in C_{l, J}}) \right)\times \left(\frac{  1  }{ \mathbb{P}(\boldsymbol{X} \in C_{l, J})  }  - \frac{  n  }{ \texttt{\#} \{i:\boldsymbol{X}_{i} \in C_{l, J} \} \vee 1 } \right)  \\
        & \qquad -  \frac{ \sum_{i=1}^n [ \boldsymbol{1}_{\boldsymbol{X}_{i} \in C_{l, J} }f(\boldsymbol{X}_{i})  - \mathbb{E}( f(\boldsymbol{X}_{i})  \boldsymbol{1}_{\boldsymbol{X}_{i} \in C_{l, J}}) ] }{ \texttt{\#} \{i:\boldsymbol{X}_{i} \in C_{l, J} \} \vee 1 } \bigg] \bigg|\\
        &\le \sum_{l\in \mathcal{L}(J)} \boldsymbol{1}_{\boldsymbol{X}_{i} \in C_{l, J}} \bigg(  \left( \mathbb{E}( |f(\boldsymbol{X}_{i})|  \boldsymbol{1}_{\boldsymbol{X}_{i} \in C_{l, J}}) \right)\times \left|\frac{  1  }{ \mathbb{P}(\boldsymbol{X} \in C_{l, J})  }  - \frac{  n  }{ \texttt{\#} \{i:\boldsymbol{X}_{i} \in C_{l, J} \} \vee 1 } \right|  \\
        & \qquad +  \frac{ \left| \sum_{i=1}^n [ \boldsymbol{1}_{\boldsymbol{X}_{i} \in C_{l, J} }f(\boldsymbol{X}_{i})  - \mathbb{E}( f(\boldsymbol{X}_{i})  \boldsymbol{1}_{\boldsymbol{X}_{i} \in C_{l, J}}) ]\right| }{ \texttt{\#} \{i:\boldsymbol{X}_{i} \in C_{l, J} \} \vee 1 } \bigg).
    \end{split}
\end{equation}

By \eqref{lower.bound.1}, and that $\sum_{l\in \mathcal{L}(J)} \boldsymbol{1}_{\boldsymbol{X}_{i} \in C_{l, J}} = 1$, on the event $E_{43}\cap E_{44}$ defined in \eqref{event.2}, for every set $J$ with $\texttt{\#}J\le 2$ and each $s>0$,
\begin{equation}
    \begin{split}
        \label{upper.bound.2}
        \textnormal{RHS of \eqref{upper.bound.2.b}} \le  \frac{\mathbb{E}( |f(\boldsymbol{X})| ) t_{4}n^{-1} }{p_{\min}(p_{\min} - \frac{t_{4}}{n})} + \frac{t_{3}n^{-1} }{p_{\min} - \frac{t_{4}}{n}}.
    \end{split}
\end{equation}

For $E_{3}$, it holds that on the event $E_{42}\cap E_{44}$, 
\begin{equation}
    \begin{split}\label{upper.bound.3}
        \left| \frac{\sum_{i:\boldsymbol{X}_{i} \in C_{l, J} } \varepsilon_{i}  }{ \texttt{\#} \{i:\boldsymbol{X}_{i} \in C_{l, J} \} \vee 1 }\right| & = \left| \frac{\sum_{i=1 }^n \boldsymbol{1}_{\boldsymbol{X}_{i} \in C_{l, J}} \varepsilon_{i}  }{ \texttt{\#} \{i:\boldsymbol{X}_{i} \in C_{l, J} \} \vee 1 } \right|\\
        & \le \frac{t_{2}n^{-1}}{p_{\min} - \frac{t_{4}}{n}},
    \end{split}
\end{equation}
where the inequality holds due to \eqref{lower.bound.1} and \eqref{lower.bound.pmin.1}.

By \eqref{equivalence.recur.12.b}, that $1 = \sum_{l\in \mathcal{L}(J)}  \boldsymbol{1}_{\boldsymbol{X}_{i} \in C_{l, J}}$, 
\eqref{dependence.4.b}, \eqref{upper.bound.1}, \eqref{upper.bound.2}, \eqref{upper.bound.3}, a recursive application of \eqref{recursive.1}, Jensen's inequality, and the choice with $t_2 = \log{(n)}n^{\frac{1}{2} + \frac{1}{q_{1}}}$ and $t_3 = t_4 = \log{(n)}\sqrt{n}$, it holds that on $\cap_{l=2}^4 E_{4l}$, for each $n$ with $\frac{2\log{(n)}}{\sqrt{n}} \le p_{\min}$, each $s\ge 0$, and every set $J$ with $\texttt{\#}J\le 2$,
\begin{equation}
    \begin{split}\label{recursive.2}
        &\sum_{i=1}^n (R_{s, J}^{\star}(\boldsymbol{X}_{i}) - \widehat{U}_{s, J}(\boldsymbol{X}_{i}))^2 \\
        & \le \sum_{l=1}^s \bigg( \frac{t_{4}n^{-1/2}\times M_{\mathcal{X}}^2\iota_{l-1}  }{p_{\min} - \frac{t_{4}}{n} } +  \frac{t_{4}n^{-1/2} \times2\sqrt{\mathbb{E}[(f(\boldsymbol{X}))^2]} }{p_{\min}(p_{\min} - \frac{t_{4}}{n})} \\
        & \qquad + \frac{\mathbb{E}( |f(\boldsymbol{X})| ) t_{4}n^{-1/2} }{p_{\min}(p_{\min} - \frac{t_{4}}{n})} + \frac{t_{3}n^{-1/2} }{p_{\min} - \frac{t_{4}}{n}} + \frac{t_{2}n^{-1/2}}{p_{\min} - \frac{t_{4}}{n}}\bigg)^2 \\
        & \le 5\sum_{l=1}^s \bigg( 4 p_{\min}^{-2}t_{4}^2n^{-1}\times M_{\mathcal{X}}^4\iota_{l-1}^2 +  16 t_{4}^2n^{-1} p_{\min}^{-4}\mathbb{E}[(f(\boldsymbol{X}))^2]  \\
        & \qquad + 4 p_{\min}^{-4} [\mathbb{E}( |f(\boldsymbol{X})| )]^2 t_{4}^2 n^{-1} + 4 p_{\min}^{-2}t_{3}^2n^{-1} + 4p_{\min}^{-2}t_{2}^2n^{-1}\bigg) \\
        & \le 20 p_{\min}^{-2}(\log{n})^2 M_{\mathcal{X}}^4 \left(\sum_{l=1}^s\iota_{l-1}^2\right) + 280 s M_{f}^2 p_{\min}^{-4}(\log{n})^2n^{\frac{2}{q_{1}}},
    \end{split}
\end{equation}
where the second inequality is due to the quadratic mean inequality~\citep{gwanyama2004hm} (i.e., $(\sum_{i=1}^k a_{i})^2 \le k\sum_{i=1}^k a_{i}^2$). The condition $\frac{2\log{(n)}}{\sqrt{n}} \le p_{\min}$ is used for simplifying $p_{\min} - \frac{t_{4}}{n}\ge \frac{p_{\min}}{2}$ with $t_4 = \log{(n)}\sqrt{n}$.

By \eqref{equivalence.recur.5.b}, \eqref{event.1}, \eqref{recursive.2}, and the definition of $E_{45}$, we conclude the first inequality of \eqref{equivalence.recur.10.b}, and hence finish the proof of \eqref{equivalence.recur.10.b}.

\subsubsection{Proof of \eqref{equivalence.recur.11.b}}\label{proof.theorem1.15}

    \noindent\textit{Proof of \eqref{equivalence.recur.11.b}: } In what follows, we derive upper bounds for \eqref{equivalence.recur.11.b}, and begin with the term $n^{-1}\sum_{i=1}^n [f(\boldsymbol{X}_{i}) R_{s, J}^{\star}(\boldsymbol{X}_{i}) - \mathbb{E}(f(\boldsymbol{X}) R_{s, J}^{\star}(\boldsymbol{X}) ) ]$ on the RHS of \eqref{equivalence.recur.8.b}. By \eqref{dependence.6.b} and the definition that $R_{0, J}^{\star}(\boldsymbol{X}_{i}) = 0$ in \eqref{predictor.trees.b.3}, for each $s\ge 0$ and every set $J\subset\{1, \dots, M\}$ with $\texttt{\#}J\le 2$,
\begin{equation*}
    \begin{split}
         & \sum_{i=1}^n f(\boldsymbol{X}_{i})(R_{s, J}^{\star}(\boldsymbol{X}_{i}))  \\
         & = 
         \sum_{i=1}^n  f(\boldsymbol{X}_{i})\left[\left(\sum_{\vv{c} \in \{0, 1\}^{\texttt{\#}\mathcal{X}(J)}}  a_{s, \vv{c}, J}\times  \boldsymbol{1}_{\boldsymbol{X}_{i\mathcal{X}(J)} = \vv{c}} \right) + \sum_{l=1}^{s-1} \sum_{\vv{c} \in \{0, 1\}^{\texttt{\#}\mathcal{X}(\widehat{J}_{l})}}  a_{l, \vv{c}, \widehat{J}_{l}}\times  \boldsymbol{1}_{\boldsymbol{X}_{i\mathcal{X}(\widehat{J}_{l})} = \vv{c}} \right] \\
         & = \sum_{i=1}^n  \left[\left(\sum_{\vv{c} \in \{0, 1\}^{\texttt{\#}\mathcal{X}(J)}}  a_{s, \vv{c}, J}\times  f(\boldsymbol{X}_{i}) \boldsymbol{1}_{\boldsymbol{X}_{i\mathcal{X}(J)} = \vv{c}} \right) + \sum_{l=1}^{s-1} \sum_{\vv{c} \in \{0, 1\}^{\texttt{\#}\mathcal{X}(\widehat{J}_{l})}}  a_{l, \vv{c}, \widehat{J}_{l}}\times  f(\boldsymbol{X}_{i}) \boldsymbol{1}_{\boldsymbol{X}_{i\mathcal{X}(\widehat{J}_{l})} = \vv{c}} \right],
    \end{split}
\end{equation*}
which implies that 
\begin{equation}\label{equivalence.recur.9.b}
    \begin{split}
        & \sum_{i=1}^n [f(\boldsymbol{X}_{i}) R_{s, J}^{\star}(\boldsymbol{X}_{i}) - \mathbb{E}(f(\boldsymbol{X}) R_{s, J}^{\star}(\boldsymbol{X}) ) ] \\
        & =  \sum_{i=1}^n  \bigg[\left(\sum_{\vv{c} \in \{0, 1\}^{\texttt{\#}\mathcal{X}(J)}}  a_{s, \vv{c}, J}\times  \left(f(\boldsymbol{X}_{i}) \boldsymbol{1}_{\boldsymbol{X}_{i\mathcal{X}(J)} = \vv{c}} - \mathbb{E} (f(\boldsymbol{X}) \boldsymbol{1}_{\boldsymbol{X}_{\mathcal{X}(J)} = \vv{c}})\right) \right) \\
        & \qquad + \sum_{l=1}^{s-1} \sum_{\vv{c} \in \{0, 1\}^{\texttt{\#}\mathcal{X}(\widehat{J}_{l})}}  a_{l, \vv{c}, \widehat{J}_{l}}\times  \left( f(\boldsymbol{X}_{i}) \boldsymbol{1}_{\boldsymbol{X}_{i\mathcal{X}(\widehat{J}_{l})} = \vv{c}} - \mathbb{E} (f(\boldsymbol{X}) \boldsymbol{1}_{\boldsymbol{X}_{\mathcal{X}(\widehat{J}_{l})} = \vv{c}})\right) \bigg] \\
        & = \left(\sum_{\vv{c} \in \{0, 1\}^{\texttt{\#}\mathcal{X}(J)}}  a_{s, \vv{c}, J}\times \sum_{i=1}^n \left(f(\boldsymbol{X}_{i}) \boldsymbol{1}_{\boldsymbol{X}_{i\mathcal{X}(J)} = \vv{c}} - \mathbb{E} (f(\boldsymbol{X}) \boldsymbol{1}_{\boldsymbol{X}_{\mathcal{X}(J)} = \vv{c}})\right) \right) \\
        & \qquad + \sum_{l=1}^{s-1} \sum_{\vv{c} \in \{0, 1\}^{\texttt{\#}\mathcal{X}(\widehat{J}_{l})}}  a_{l, \vv{c}, \widehat{J}_{l}}\times  \sum_{i=1}^n\left( f(\boldsymbol{X}_{i}) \boldsymbol{1}_{\boldsymbol{X}_{i\mathcal{X}(\widehat{J}_{l})} = \vv{c}} - \mathbb{E} (f(\boldsymbol{X}) \boldsymbol{1}_{\boldsymbol{X}_{\mathcal{X}(\widehat{J}_{l})} = \vv{c}})\right) \\
        & \le \left(\sum_{\vv{c} \in \{0, 1\}^{\texttt{\#}\mathcal{X}(J)}}  |a_{s, \vv{c}, J}| \times \left|\sum_{i=1}^n \left(f(\boldsymbol{X}_{i}) \boldsymbol{1}_{\boldsymbol{X}_{i\mathcal{X}(J)} = \vv{c}} - \mathbb{E} (f(\boldsymbol{X}) \boldsymbol{1}_{\boldsymbol{X}_{\mathcal{X}(J)} = \vv{c}})\right) \right| \right) \\
        & \qquad + \sum_{l=1}^{s-1} \sum_{\vv{c} \in \{0, 1\}^{\texttt{\#}\mathcal{X}(\widehat{J}_{l})}}  |a_{l, \vv{c}, \widehat{J}_{l}}| \times  \left|\sum_{i=1}^n\left( f(\boldsymbol{X}_{i}) \boldsymbol{1}_{\boldsymbol{X}_{i\mathcal{X}(\widehat{J}_{l})} = \vv{c}} - \mathbb{E} (f(\boldsymbol{X}) \boldsymbol{1}_{\boldsymbol{X}_{\mathcal{X}(\widehat{J}_{l})} = \vv{c}})\right) \right|,
    \end{split}
\end{equation}
where the second equality is due to an interchange of summations. By \eqref{equivalence.recur.9.b}, \eqref{dependence.8.b}, and that $R_{0, J}^{\star}(\boldsymbol{X}) = 0$ as defined in \eqref{predictor.trees.b.3}, it holds that for each $s\ge 0$ and every set $J$ with $\texttt{\#}J\le 2$, on the event $E_{43}$ defined in \eqref{event.2}, 
\begin{equation}\label{equivalence.recur.9.2.b}
\left| \sum_{i=1}^n [f(\boldsymbol{X}_{i}) R_{s, J}^{\star}(\boldsymbol{X}_{i}) - \mathbb{E}(f(\boldsymbol{X}) R_{s, J}^{\star}(\boldsymbol{X}) ) ] \right| \le t_{3}(3M_{\mathcal{X}}^2p_{\min}^{-1} \sqrt{\mathbb{E}[(f(\boldsymbol{X}))^2]}  + M_{\mathcal{X}}^2 \iota_{s-1} ).
\end{equation}

Next, we establish an upper bound for the term $n^{-1}\sum_{i=1}^n [ (R_{s, J}^{\star}(\boldsymbol{X}_{i}))^2 - \mathbb{E}( R_{s, J}^{\star}(\boldsymbol{X}) )^2]$ on the RHS of \eqref{equivalence.recur.8.b}. To simplify the derivation below, let us denote $\widehat{J}_{1}, \dots, \widehat{J}_{s-1}, J$ respectively by $H_{1}, \dots, H_{s-1}, H_{s}$. Then, by \eqref{dependence.6.b} and the definition that $R_{0, J}^{\star}(\boldsymbol{X}_{i}) = 0$ in \eqref{predictor.trees.b.3}, for each $s\ge 0$ and every set $J$ with $\texttt{\#}J\le 2$,
\begin{equation*}
    \begin{split}
         & \sum_{i=1}^n (R_{s, J}^{\star}(\boldsymbol{X}_{i}))^2  \\
         & = 
         \sum_{i=1}^n  \left[\left(\sum_{\vv{c} \in \{0, 1\}^{\texttt{\#}\mathcal{X}(J)}}  a_{s, \vv{c}, J}\times  \boldsymbol{1}_{\boldsymbol{X}_{i\mathcal{X}(J)} = \vv{c}} \right) + \sum_{l=1}^{s-1} \sum_{\vv{c} \in \{0, 1\}^{\texttt{\#}\mathcal{X}(\widehat{J}_{l})}}  a_{l, \vv{c}, \widehat{J}_{l}}\times  \boldsymbol{1}_{\boldsymbol{X}_{i\mathcal{X}(\widehat{J}_{l})} = \vv{c}} \right]^2\\
        & =  \sum_{i=1}^n  \left( \sum_{l=1}^{s} \sum_{\vv{c} \in \{0, 1\}^{\texttt{\#} \mathcal{X}(H_{l}) }}  a_{l, \vv{c}, H_{l}}\times  \boldsymbol{1}_{\boldsymbol{X}_{i \mathcal{X}(H_{l})} = \vv{c}} \right)^2 ,
    \end{split}
\end{equation*}
which implies that 
\begin{equation}
    \begin{split}
        \label{equivalence.recur.7.bb}
        & \left|\sum_{i=1}^n \{(R_{s, J}^{\star}(\boldsymbol{X}_{i}))^2 - \mathbb{E} [(R_{s, J}^{\star}(\boldsymbol{X}))^2] \}\right| \\
        & = \bigg| \sum_{i=1}^n   \sum_{l_{2}=1}^{s} \sum_{l_{1}=1}^{s} \sum_{\vv{c}_{2} \in \{0, 1\}^{\texttt{\#} \mathcal{X}(H_{l_{2}})}} \sum_{\vv{c}_{1} \in \{0, 1\}^{\texttt{\#} \mathcal{X}(H_{l_{1}})}}  \bigg[a_{l_{1}, \vv{c}_{1}, H_{l_{1}}} \times a_{l_{2}, \vv{c}_{2}, H_{l_{2}}}  \\
        &\qquad \times \left(\boldsymbol{1}_{\boldsymbol{X}_{i \mathcal{X}(H_{l_{1}})} = \vv{c}_{1}} \boldsymbol{1}_{\boldsymbol{X}_{i \mathcal{X}(H_{l_{2}})} = \vv{c}_{2}}  - \mathbb{P} (\boldsymbol{X}_{\mathcal{X}(H_{l_{1}})} = \vv{c}_{1}, \boldsymbol{X}_{\mathcal{X}(H_{l_{2}})} = \vv{c}_{2} )  \right) \bigg]  \bigg|\\
        & = \bigg|  \sum_{l_{2}=1}^{s} \sum_{l_{1}=1}^{s} \sum_{\vv{c}_{2} \in \{0, 1\}^{\texttt{\#} \mathcal{X}(H_{l_{2}})}} \sum_{\vv{c}_{1} \in \{0, 1\}^{\texttt{\#} \mathcal{X}(H_{l_{1}})}}  \bigg[a_{l_{1}, \vv{c}_{1}, H_{l_{1}}} \times a_{l_{2}, \vv{c}_{2}, H_{l_{2}}}  \\
        &\qquad \times \sum_{i=1}^n  
 \left(\boldsymbol{1}_{\boldsymbol{X}_{i \mathcal{X}(H_{l_{1}})} = \vv{c}_{1}} \boldsymbol{1}_{\boldsymbol{X}_{i \mathcal{X}(H_{l_{2}})} = \vv{c}_{2}}  - \mathbb{P} (\boldsymbol{X}_{\mathcal{X}(H_{l_{1}})} = \vv{c}_{1}, \boldsymbol{X}_{\mathcal{X}(H_{l_{2}})} = \vv{c}_{2} )  \right) \bigg]  \bigg| \\
 & \le   \sum_{l_{2}=1}^{s} \sum_{l_{1}=1}^{s} \sum_{\vv{c}_{2} \in \{0, 1\}^{\texttt{\#} \mathcal{X}(H_{l_{2}})}} \sum_{\vv{c}_{1} \in \{0, 1\}^{\texttt{\#} \mathcal{X}(H_{l_{1}})}}  \bigg[ |a_{l_{1}, \vv{c}_{1}, H_{l_{1}}} \times  a_{l_{2}, \vv{c}_{2}, H_{l_{2}}}| \\
        &\qquad \times  \left|\sum_{i=1}^n  
 \left(\boldsymbol{1}_{\boldsymbol{X}_{i \mathcal{X}(H_{l_{1}})} = \vv{c}_{1}} \boldsymbol{1}_{\boldsymbol{X}_{i \mathcal{X}(H_{l_{2}})} = \vv{c}_{2}}  - \mathbb{P} (\boldsymbol{X}_{\mathcal{X}(H_{l_{1}})} = \vv{c}_{1}, \boldsymbol{X}_{\mathcal{X}(H_{l_{2}})} = \vv{c}_{2} )  \right) \right| \bigg].
    \end{split}
\end{equation}

By the observation that $\mathbb{P} (\boldsymbol{X}_{\mathcal{X}(H_{l_{1}})} = \vv{c}_{1}, \boldsymbol{X}_{\mathcal{X}(H_{l_{2}})} = \vv{c}_{2} )=0$ implies that $\boldsymbol{1}\{\boldsymbol{X}_{i \mathcal{X}(H_{l_{1}})} = \vv{c}_{1}\} \times \boldsymbol{1}\{\boldsymbol{X}_{i \mathcal{X}(H_{l_{2}})} = \vv{c}_{2}\} = 0$ almost surely, we deduce that
\begin{equation}
    \begin{split}
        \label{equivalence.recur.7.b}
        & \text{RHS of \eqref{equivalence.recur.7.bb}}\\
 & =   \sum_{l_{2}=1}^{s} \sum_{l_{1}=1}^{s} \sum_{\vv{c}_{2} \in \{0, 1\}^{\texttt{\#} \mathcal{X}(H_{l_{2}})}} \sum_{\vv{c}_{1} \in \{0, 1\}^{\texttt{\#} \mathcal{X}(H_{l_{1}})}}  \bigg[ |a_{l_{1}, \vv{c}_{1}, H_{l_{1}}} \times  a_{l_{2}, \vv{c}_{2}, H_{l_{2}}}| \\
 &\qquad \times \boldsymbol{1}\{ \mathbb{P} (\boldsymbol{X}_{\mathcal{X}(H_{l_{1}})} = \vv{c}_{1}, \boldsymbol{X}_{\mathcal{X}(H_{l_{2}})} = \vv{c}_{2} )>0 \}\\
        &\qquad \times  \left|\sum_{i=1}^n  
 \left(\boldsymbol{1}_{\boldsymbol{X}_{i \mathcal{X}(H_{l_{1}})} = \vv{c}_{1}} \boldsymbol{1}_{\boldsymbol{X}_{i \mathcal{X}(H_{l_{2}})} = \vv{c}_{2}}  - \mathbb{P} (\boldsymbol{X}_{\mathcal{X}(H_{l_{1}})} = \vv{c}_{1}, \boldsymbol{X}_{\mathcal{X}(H_{l_{2}})} = \vv{c}_{2} )  \right) \right| \bigg].
    \end{split}
\end{equation}
The consideration of $\{\boldsymbol{X}_{\mathcal{X}(H_{l_{1}})} = \vv{c}_{1}, \boldsymbol{X}_{\mathcal{X}(H_{l_{2}})} = \vv{c}_{2}\}$ with positive probability measure leads us to control the difference term on the RHS of \eqref{equivalence.recur.7.b} based on the event $E_{44}$ defined in \eqref{event.2}. To see this, note that for every $\vv{c}_{1} \in \{0, 1\}^{\texttt{\#} \mathcal{X}(H_{l_{1}})}$ and $\vv{c}_{2} \in \{0, 1\}^{\texttt{\#} \mathcal{X}(H_{l_{2}})}$ with $\mathbb{P} (\boldsymbol{X}_{\mathcal{X}(H_{l_{1}})} = \vv{c}_{1}, \boldsymbol{X}_{\mathcal{X}(H_{l_{2}})} = \vv{c}_{2} )>0$, the case with $J = H_{l_{1}}\cup H_{l_{2}}$ and $\vv{c} = (\vv{c}_{1}^{\top}, \vv{c}_{2}^{\top})^{\top}$ is considered by the event $E_{44}$.

In addition to the above observation, to deal with the RHS of \eqref{equivalence.recur.7.b}, we need the following result \eqref{equivalence.recur.7.2.b}. By the definition of $H_{l}$'s and \eqref{dependence.8.b}, it holds that for each $s\ge0$ and every set $J$ with $\texttt{\#}J\le 2$,
\begin{equation}
    \begin{split}\label{equivalence.recur.7.2.b}
        & \sum_{l_{2}=1}^{s} \sum_{l_{1}=1}^{s} \sum_{\vv{c}_{2} \in \{0, 1\}^{\texttt{\#} \mathcal{X}(H_{l_{2}})}} \sum_{\vv{c}_{1} \in \{0, 1\}^{\texttt{\#} \mathcal{X}(H_{l_{1}})}}   \bigg( |a_{l_{1}, \vv{c}_{1}, H_{l_{1}}} \times a_{l_{2}, \vv{c}_{2}, H_{l_{2}}}| \\
        & \qquad\qquad \times \boldsymbol{1}\{ \mathbb{P} (\boldsymbol{X}_{\mathcal{X}(H_{l_{1}})} = \vv{c}_{1}, \boldsymbol{X}_{\mathcal{X}(H_{l_{2}})} = \vv{c}_{2} )>0 \} \bigg) \\ 
        & = \left(    \sum_{\vv{c} \in \{0, 1\}^{\texttt{\#}\mathcal{X}(J)}}  |a_{s, \vv{c}, J}|  + \sum_{l=1}^{s-1} \sum_{\vv{c} \in \{0, 1\}^{\texttt{\#}\mathcal{X}(\widehat{J}_{l})}}  |a_{l, \vv{c}, \widehat{J}_{l}}| \right)^2\\
        &\le (M_{\mathcal{X}}^2 \times 3\sqrt{\mathbb{E}[(f(\boldsymbol{X}))^2]}  \times p_{\min}^{-1} + M_{\mathcal{X}}^2 \iota_{s-1} )^2.
    \end{split}
\end{equation}

By \eqref{equivalence.recur.7.b}--\eqref{equivalence.recur.7.2.b} and that $R_{0, J}^{\star}(\boldsymbol{X}) = 0$ as defined in \eqref{predictor.trees.b.3}, it holds that on $E_{44}$, for each $s\ge 0$ and every set $J$ with $\texttt{\#}J\le 2$,
\begin{equation}\label{equivalence.recur.7.3.b}
    \left|\sum_{i=1}^n \{(R_{s, J}^{\star}(\boldsymbol{X}_{i}))^2 - \mathbb{E} [(R_{s, J}^{\star}(\boldsymbol{X}))^2] \}\right| \le t_{4}(3M_{\mathcal{X}}^2 p_{\min}^{-1} \sqrt{\mathbb{E}[(f(\boldsymbol{X}))^2]}  + M_{\mathcal{X}}^2 \iota_{s-1} )^2.
\end{equation}

By \eqref{equivalence.recur.9.2.b}, \eqref{equivalence.recur.7.3.b}, the definition of $E_{46}$, and the choice of $t_{3}= t_{4} = t_{6} = \log{(n)} \sqrt{n}$, we conclude the desired result of \eqref{equivalence.recur.11.b}.

\subsubsection{Proof of \eqref{bias.control.3.b}}\label{proof.theorem3.20}

    \noindent\textit{Proof of \eqref{bias.control.3.b}: } At the $s$th round with $0<s\le K$, let us consider $J = \{m\}$ for any $m\in  \{1, \dots ,M\}\backslash (\cup_{l=1}^{s-1}\widehat{J}_{l})$.   We deduce that 
\begin{equation}
    \begin{split}\label{dependence.13}
& \left( \mathbb{E}[ f(\boldsymbol{X}) - R_{s-1}^{\star}(\boldsymbol{X})  |\boldsymbol{X}_{\mathcal{X}(m)} ] \right)^2 \\
& = \left\{ \mathbb{E}\{ f(\boldsymbol{X}) - \mathbb{E}[f(\boldsymbol{X}) | \boldsymbol{X}_{-\mathcal{X}(m)} ] + \mathbb{E}[f(\boldsymbol{X}) | \boldsymbol{X}_{-\mathcal{X}(m)} ] - R_{s-1}^{\star}(\boldsymbol{X})  |\boldsymbol{X}_{\mathcal{X}(m)} \}\right\}^2   \\
& =  (g_{m}(\boldsymbol{X}) )^2 - 2g_{m}(\boldsymbol{X}) \times \mathbb{E}\left\{ \mathbb{E}[f(\boldsymbol{X}) | \boldsymbol{X}_{-\mathcal{X}(m)} ]  - R_{s-1}^{\star}(\boldsymbol{X}) |\boldsymbol{X}_{\mathcal{X}(m)}\right\} \\
& \qquad+ \left(\mathbb{E}\{ \mathbb{E}[f(\boldsymbol{X}) | \boldsymbol{X}_{-\mathcal{X}(m)} ]  - R_{s-1}^{\star}(\boldsymbol{X}) |\boldsymbol{X}_{\mathcal{X}(m)}\}\right)^2,
    \end{split}
\end{equation}
where we recall that $g_{m}(\boldsymbol{X}) = \mathbb{E}\{ f(\boldsymbol{X}) - \mathbb{E}[f(\boldsymbol{X}) | \boldsymbol{X}_{-\mathcal{X}(m)} ] | \boldsymbol{X}_{\mathcal{X}(m)}\}$, which is defined in Section~\ref{Sec3qq}.

Next, we proceed to deal with terms on the RHS of \eqref{dependence.13}. By the assumption that $\mathbb{E}(f(\boldsymbol{X})) = 0$, Lemma~\ref{lemma.2}, the law of total expectation, Jensen's inequality, and Lemma~\ref{lemma.5}, it holds that almost surely,
\begin{equation}
    \begin{split}\label{dependence.12}
    &\bigg|\mathbb{E}\left\{ \mathbb{E}[f(\boldsymbol{X}) | \boldsymbol{X}_{-\mathcal{X}(m)} ]  - R_{s-1}^{\star}(\boldsymbol{X}) \Big|\boldsymbol{X}_{\mathcal{X}(m)} \right\}  \bigg|\\
       & = \bigg|\mathbb{E}\left\{ \mathbb{E}[f(\boldsymbol{X}) | \boldsymbol{X}_{-\mathcal{X}(m)} ]  - R_{s-1}^{\star}(\boldsymbol{X}) \Big|\boldsymbol{X}_{\mathcal{X}(m)} \right\}  \\
       & \qquad - \mathbb{E}\left\{ \mathbb{E}[f(\boldsymbol{X}) | \boldsymbol{X}_{-\mathcal{X}(m)} ]  - R_{s-1}^{\star}(\boldsymbol{X}) \right\} \bigg|\\
        & \le \frac{\delta_{0}}{p_{\min}} \mathbb{E}\left| \mathbb{E}[f(\boldsymbol{X}) | \boldsymbol{X}_{-\mathcal{X}(m)} ]  - R_{s-1}^{\star}(\boldsymbol{X})  \right| \\
        & \le \frac{\delta_{0}}{p_{\min}} (\mathbb{E}|f(\boldsymbol{X}) |  + \mathbb{E}|R_{s-1}^{\star}(\boldsymbol{X})|  )\\
        & \le \frac{3\delta_{0}}{p_{\min}}  \sqrt{\mathbb{E}[(f(\boldsymbol{X}))^2]},
    \end{split}
\end{equation}
where we rely on that $\mathbb{E}[f(\boldsymbol{X}) | \boldsymbol{X}_{-\mathcal{X}(m)} ] - R_{s-1}^{\star}(\boldsymbol{X})$ is a measurable function of $\boldsymbol{X}_{-\mathcal{X}(m)}$ for $0<s\le K$ due to our assumption that $\widehat{J}_{l} \not = \widehat{J}_{k}$ for each $1\le l< k\le 2K$.

By \eqref{dependence.13}, \eqref{dependence.12}, and that $\mathbb{E} [(g_{m}(\boldsymbol{X}) )^2] = \textnormal{Var}(g_{m}(\boldsymbol{X}))$ by the definition of $g_{m}(\boldsymbol{X})$, it holds that for each $0< s\le K$ and  each $m \in  \{1, \dots, M\}\backslash (\cup_{l=1}^{s-1}\widehat{J}_{l})$,
\begin{equation}
    \begin{split}\label{bias.control.3}
        & \left|\mathbb{E}[(\mathbb{E}(f(\boldsymbol{X}) - R_{s-1}^{\star}(\boldsymbol{X})|\boldsymbol{X}_{\mathcal{X}(m)}))^2  ] - \textnormal{Var}(g_{m}(\boldsymbol{X})) \right| \\
        & \le \frac{6\delta_{0}}{p_{\min}}  \sqrt{\mathbb{E}[(f(\boldsymbol{X}))^2]} (\mathbb{E}|g_{m}(\boldsymbol{X})|) + \frac{9\delta_{0}^2}{(p_{\min})^2}  \mathbb{E}[(f(\boldsymbol{X}))^2]\\
        & \le \frac{21\delta_{0}}{(p_{\min})^2}  \mathbb{E}[(f(\boldsymbol{X}))^2],
    \end{split}
\end{equation}
where we use the fact that $\delta_{0}\le 1$ and $p_{\min}\le 1$, and the result due to Jensen's inequality and the definition of $g_{m}(\boldsymbol{X})$ that
\begin{equation}
    \begin{split}\label{bias.control.6} 
        \mathbb{E}|g_{m}(\boldsymbol{X})| & = \mathbb{E}\left|\mathbb{E}(f(\boldsymbol{X})  - \mathbb{E} (f(\boldsymbol{X})| \boldsymbol{X}_{-\mathcal{X}(m)}) |\boldsymbol{X}_{\mathcal{X}(m)})\right|\\
        & \le \mathbb{E}\left[\mathbb{E}(\left|f(\boldsymbol{X}) \right| +\left| \mathbb{E} (f(\boldsymbol{X})| \boldsymbol{X}_{-\mathcal{X}(m)}) \right||\boldsymbol{X}_{\mathcal{X}(m)})\right]\\
        & \le \mathbb{E}\left|f(\boldsymbol{X}) \right| 
 + \mathbb{E}( \mathbb{E} (\left|f(\boldsymbol{X})\right| | \boldsymbol{X}_{-\mathcal{X}(m)}) )\\
 & \le 2  \mathbb{E}\left|f(\boldsymbol{X}) \right| \\
 & \le 2  \sqrt{\mathbb{E}[(f(\boldsymbol{X}))^2]}.
    \end{split}
\end{equation}

By \eqref{bias.control.3}, we have concluded the proof of \eqref{bias.control.3.b}.

\subsubsection{Proof of \eqref{bias.control.3.2k.b}}\label{proof.theorem3.24}
\noindent\textit{Proof of \eqref{bias.control.3.2k.b}: } For each $K< s \le 2K$ and every set $J$ with $\texttt{\#}J= 2$, let $\mathcal{L}_{1}(J) = \{l : \widehat{J}_{l} \subset J, 1\le l\le K\}$, $\mathcal{L}_{2s}(J) = \{l : \widehat{J}_{l} \cap J \not= \emptyset, K< l<s \}$, and $\mathcal{L}_{3s}(J) = \{l : \widehat{J}_{l} \cap J = \emptyset, K< l<s \}$. Let us deduce that for each $K<s\le 2K$ and every $J\subset\{1, \dots, M\}$ with $\texttt{\#}J=2$,
\begin{equation}
    \begin{split}\label{boundness.bias.4}
& \mathbb{E}[ f(\boldsymbol{X}) - R_{s-1}^{\star}(\boldsymbol{X})  |\boldsymbol{X}_{\mathcal{X}(J)}] \\
& =  g_{J}(\boldsymbol{X}) - \sum_{l \in \mathcal{L}_{1}(J)} \mathbb{E}[  f(\boldsymbol{X}) -  R_{l-1}^{\star}(\boldsymbol{X})  | \boldsymbol{X}_{\mathcal{X}(\widehat{J}_{l})}]   \\
&\qquad - \mathbb{E}\{  \sum_{ l\in \mathcal{L}_{2s}(J) }   \mathbb{E}[  f(\boldsymbol{X}) -  R_{l-1}^{\star}(\boldsymbol{X})  | \boldsymbol{X}_{\mathcal{X}(\widehat{J}_{l})}] |\boldsymbol{X}_{\mathcal{X}(J)}\} \\
&\qquad - \mathbb{E}\{  \sum_{ l\in \mathcal{L}_{3s}(J) }   \mathbb{E}[  f(\boldsymbol{X}) -  R_{l-1}^{\star}(\boldsymbol{X})  | \boldsymbol{X}_{\mathcal{X}(\widehat{J}_{l})}] |\boldsymbol{X}_{\mathcal{X}(J)}\} \\
&\qquad + \mathbb{E}\{ \mathbb{E}[f(\boldsymbol{X}) | \boldsymbol{X}_{-\mathcal{X}(J)} ]  - \sum_{l\le K, l \not\in \mathcal{L}_{1}(J) }   \mathbb{E}[  f(\boldsymbol{X}) -  R_{l-1}^{\star}(\boldsymbol{X})  | \boldsymbol{X}_{\mathcal{X}(\widehat{J}_{l})}] |\boldsymbol{X}_{\mathcal{X}(J)}\},
\end{split}
\end{equation}
where we use  the definition of $g_{J}(\boldsymbol{X})$ in Section~\ref{Sec3qq}, and the expression
$R_{s}^{\star}(\boldsymbol{X}) = \sum_{l=1}^s \mathbb{E}[  f(\boldsymbol{X}) -  R_{l-1}^{\star}(\boldsymbol{X})  | \boldsymbol{X}_{\mathcal{X}(\widehat{J}_{l})}]$, which follows from a recursive application of \eqref{mp.3}.

In light of \eqref{boundness.bias.4}, for each $s>K$ and $J\subset\{1, \dots, M\}$ with $\texttt{\#}J=2$, we define $F_{s}(J)$ such that
\begin{equation}
    \begin{split}\label{boundness.bias.6}
           &\mathbb{E}[  f(\boldsymbol{X}) -  R_{s-1}^{\star}(\boldsymbol{X})  | \boldsymbol{X}_{\mathcal{X}(J) }] \\
           & = F_{s}(J) + g_{J}(\boldsymbol{X}) - \sum_{l \in \mathcal{L}_{1}(J) } \mathbb{E}[  f(\boldsymbol{X}) -  R_{l-1}^{\star}(\boldsymbol{X})  | \boldsymbol{X}_{\mathcal{X}(\widehat{J}_{l})}].
    \end{split}
\end{equation}

To deal with the third term on the RHS of \eqref{boundness.bias.6}, we deduce by the definition of $\mathcal{L}_{1}(J)$ that for each $J\subset\{1, \dots, M\}$ with $\texttt{\#}J = 2$,
\begin{equation}
    \begin{split}\label{boundness.bias.7}
        &\sum_{l \in \mathcal{L}_{1}(J) } \mathbb{E}[  f(\boldsymbol{X}) -  R_{l-1}^{\star}(\boldsymbol{X})  | \boldsymbol{X}_{\mathcal{X}(\widehat{J}_{l})}] \\
        &= \left(\sum_{l \in \mathcal{L}_{1}(J)}g_{\widehat{J}_{l}}(\boldsymbol{X})\right) +  \sum_{l \in \mathcal{L}_{1}(J)} \mathbb{E}[  \mathbb{E}[f(\boldsymbol{X})| \boldsymbol{X}_{-\mathcal{X}(\widehat{J}_{l}) } ] -  R_{l-1}^{\star}(\boldsymbol{X})  | \boldsymbol{X}_{\mathcal{X}(\widehat{J}_{l})}] ,
    \end{split}
\end{equation}
and that almost surely
\begin{equation}
    \begin{split}\label{boundness.bias.5}
        &\left|\sum_{l \in \mathcal{L}_{1}(J)} \mathbb{E}[  \mathbb{E}[f(\boldsymbol{X})| \boldsymbol{X}_{-\mathcal{X}(\widehat{J}_{l}) } ] -  R_{l-1}^{\star}(\boldsymbol{X})  | \boldsymbol{X}_{\mathcal{X}(\widehat{J}_{l})}] \right|\\
        & \le 2\delta_{0} \times \frac{\mathbb{E}|f(\boldsymbol{X})| + \mathbb{E}|R_{l-1}^{\star}(\boldsymbol{X})| }{ p_{\min} } \\
        & \le   \frac{6\delta_{0}\sqrt{\mathbb{E}[(f(\boldsymbol{X}))^2]}}{p_{\min}},
    \end{split}
\end{equation}
where we use Lemma~\ref{lemma.5}, the result that $\mathbb{E}\{\mathbb{E}[  \mathbb{E}[f(\boldsymbol{X})| \boldsymbol{X}_{-\mathcal{X}(\widehat{J}_{l}) } ] -  R_{l-1}^{\star}(\boldsymbol{X})  | \boldsymbol{X}_{\mathcal{X}(\widehat{J}_{l})}]\} = 0$ for $l \in \mathcal{L}_{1}(J)$, Jensen's inequality, Lemma~\ref{lemma.2}, 
the definition of $ p_{\min}$ in Condition~\ref{signal.strength.1}, and that $\texttt{\#}\mathcal{L}_{1}(J) \le 2$ due to the assumption that  Algorithm~\ref{algorithm1} makes the initial $2K$ updates  with  $\widehat{J}_{l} \not= \widehat{J}_{k}$ for each $\{l, k\}\subset \{1, \dots, 2K\}$. In addition, with the required conditions for \eqref{consistency.1}, Condition~\ref{heredity.1}, and the definition of $S_{1}^{*}$, we further have that on the event $\cap_{l=1}^6 E_{4l}$, 
\begin{equation}
    \label{boundness.bias.8}
    \sum_{l \in \mathcal{L}_{1}(J)}g_{\widehat{J}_{l}}(\boldsymbol{X}) = \sum_{m \in J}g_{m}(\boldsymbol{X}),
\end{equation}
where we note that $g_m(\boldsymbol{X}) = 0$ if $\textnormal{Var}(g_m(\boldsymbol{X})) = 0$.

Under the required conditions for \eqref{boundness.bias.6}--\eqref{boundness.bias.7} and \eqref{boundness.bias.8}, we have that
\begin{equation}
    \begin{split}\label{boundness.bias.6.b}
           & \mathbb{E}\left(\left\{\mathbb{E}[  f(\boldsymbol{X}) -  R_{s-1}^{\star}(\boldsymbol{X})  | \boldsymbol{X}_{\mathcal{X}(J) }] \right\}^2\right)  - \mathbb{E}([g_{J}(\boldsymbol{X}) -\sum_{m \in J}g_{m}(\boldsymbol{X})]^2)\\
           & = \mathbb{E}\bigg([F_{s}(J)]^2  + \left[\sum_{l \in \mathcal{L}_{1}(J)} \mathbb{E}[  \mathbb{E}[f(\boldsymbol{X})| \boldsymbol{X}_{-\mathcal{X}(\widehat{J}_{l}) } ] -  R_{l-1}^{\star}(\boldsymbol{X})  | \boldsymbol{X}_{\mathcal{X}(\widehat{J}_{l})}] \right]^2\\
           &  \qquad -2\left[\sum_{l \in \mathcal{L}_{1}(J)} \mathbb{E}[  \mathbb{E}[f(\boldsymbol{X})| \boldsymbol{X}_{-\mathcal{X}(\widehat{J}_{l}) } ] -  R_{l-1}^{\star}(\boldsymbol{X})  | \boldsymbol{X}_{\mathcal{X}(\widehat{J}_{l})}] \right]  \\
           & \qquad\qquad \times\left[F_{s}(J) + g_{J}(\boldsymbol{X}) -\sum_{m \in J}g_{m}(\boldsymbol{X})\right]\\
           & \qquad + 2[F_{s}(J)] \times [g_{J}(\boldsymbol{X}) -\sum_{m \in J}g_{m}(\boldsymbol{X})]\bigg).
    \end{split}
\end{equation}
Based on \eqref{boundness.bias.6.b}, our goal is to derive a result similar to \eqref{bias.control.3} for the case with $K<s \le 2K$ by using the identity $\mathbb{E}\left([g_{J}(\boldsymbol{X}) -\sum_{m \in J}g_{m}(\boldsymbol{X})]^2\right) = \textnormal{Var}\left(g_{J}(\boldsymbol{X}) -\sum_{m \in J}g_{m}(\boldsymbol{X})\right)$, \eqref{boundness.bias.5}, and \eqref{bias.control.6}, along with some upper bound on $|F_s(J)|$. To this end, we  obtain an upper bound for $|F_s(J)|$ as follows.

Recall that $F_s(J)$ is defined in \eqref{boundness.bias.6}, which is the sum of the last three terms on the RHS of \eqref{boundness.bias.4}. Let us establish an upper bound for each of these three terms, and begin with the term involving $\mathcal{L}_{3s}(J)$. By the definition of $\mathcal{L}_{3s}(J)$, Lemma~\ref{lemma.5}, Jensen's inequality, Lemma~\ref{lemma.2}, it holds almost surely that for each $K < s \le 2K$ and each $J\subset\{1, \dots ,M\}$ with $\texttt{\#}J = 2$,
\begin{equation}
    \begin{split}\label{boundness.bias.2}
    &  \left| \sum_{l \in  \mathcal{L}_{3s}(J)} \mathbb{E}\{    \mathbb{E}[  f(\boldsymbol{X}) -  R_{l-1}^{\star}(\boldsymbol{X})  | \boldsymbol{X}_{\mathcal{X}(\widehat{J}_{l})}] |\boldsymbol{X}_{\mathcal{X}(J)}\} \right|\\
        & = \sum_{l \in  \mathcal{L}_{3s}(J)} \Big|\mathbb{E}\{    \mathbb{E}[  f(\boldsymbol{X}) -  R_{l-1}^{\star}(\boldsymbol{X})  | \boldsymbol{X}_{\mathcal{X}(\widehat{J}_{l})}] |\boldsymbol{X}_{\mathcal{X}(J)}\} \\
        &\qquad - \mathbb{E} \left\{    \mathbb{E}[  f(\boldsymbol{X}) -  R_{l-1}^{\star}(\boldsymbol{X})  | \boldsymbol{X}_{\mathcal{X}(\widehat{J}_{l})}] \right\} \Big|\\ 
&\le \sum_{l \in  \mathcal{L}_{3s}(J)} \frac{3 \delta_{0}  \mathbb{E} \left|    \mathbb{E}[  f(\boldsymbol{X}) -  R_{l-1}^{\star}(\boldsymbol{X})  | \boldsymbol{X}_{\mathcal{X}(\widehat{J}_{l})}] \right| }{p_{\min}}\\
        &\le  \sum_{l \in  \mathcal{L}_{3s}(J)}\frac{3 \delta_{0}  \mathbb{E}|f(\boldsymbol{X}) -  R_{l-1}^{\star}(\boldsymbol{X})| }{p_{\min}}\\
        & \le \sum_{l \in  \mathcal{L}_{3s}(J)} \frac{9 \delta_{0}  \sqrt{\mathbb{E}([f(\boldsymbol{X})]^2) } }{p_{\min}},
    \end{split}
\end{equation}
where the first equality follows from Lemma~\ref{lemma.2}, law of total expectation, and the assumption that $\mathbb{E}(f(\boldsymbol{X})) = 0$.

Next, we now proceed to establish an upper bound for the last term on the RHS of \eqref{boundness.bias.4}. By Lemma~\ref{lemma.5}, Jensen's inequality, and Lemma~\ref{lemma.2}, it holds almost surely that for each $J\subset\{1, \dots ,M\}$ with $\texttt{\#}J = 2$,
\begin{equation}
    \begin{split}\label{boundness.bias.3}
        &\Big| \mathbb{E}\{ \mathbb{E}[f(\boldsymbol{X}) | \boldsymbol{X}_{-\mathcal{X}(J)} ]  - \sum_{l\le K, l \not\in \mathcal{L}_{1}(J) }   \mathbb{E}[  f(\boldsymbol{X}) -  R_{l-1}^{\star}(\boldsymbol{X})  | \boldsymbol{X}_{\mathcal{X}(\widehat{J}_{l})}] |\boldsymbol{X}_{\mathcal{X}(J)}\} \Big| \\
        & \le \frac{3\delta_{0} }{p_{\min}} \times \left[\mathbb{E}|f(\boldsymbol{X})| + \mathbb{E}\left|\sum_{l\le K, l \not\in \mathcal{L}_{1}(J) }   \mathbb{E}[  f(\boldsymbol{X}) -  R_{l-1}^{\star}(\boldsymbol{X})  | \boldsymbol{X}_{\mathcal{X}(\widehat{J}_{l})}]\right| \right]\\
        & \le  \frac{3\delta_{0} }{p_{\min}} \times \left[\mathbb{E}|f(\boldsymbol{X})| + \mathbb{E}\left| R_{K}^{\star}(\boldsymbol{X}) \right| +  \mathbb{E}\left|\sum_{l \in \mathcal{L}_{1}(J) }   \mathbb{E}[  f(\boldsymbol{X}) -  R_{l-1}^{\star}(\boldsymbol{X})  | \boldsymbol{X}_{\mathcal{X}(\widehat{J}_{l})}]\right|\right] \\
        & \le  \frac{3\delta_{0} }{p_{\min}} \times \left[ \mathbb{E}|f(\boldsymbol{X})| + 2\sqrt{\mathbb{E}[(f(\boldsymbol{X}))^2]} +  2\mathbb{E}|f(\boldsymbol{X})| + 4\sqrt{\mathbb{E}[(f(\boldsymbol{X}))^2]}\right]\\
        & \le \frac{27\delta_{0} \sqrt{\mathbb{E}[(f(\boldsymbol{X}))^2]}}{p_{\min}},
    \end{split}
\end{equation}
where the first inequality is due to an application of Lemma~\ref{lemma.5}, Jensen's inequality, Lemma~\ref{lemma.2}, and the assumption that $\mathbb{E}(f(\boldsymbol{X})) = 0$; the second inequality follows from a recursive application of \eqref{mp.3}, the definition of $\mathcal{L}_{1}(J)$, and Jensen's inequality; and that the third inequality is due to the definition of $\mathcal{L}_{1}(J)$, Lemma~\ref{lemma.2}, and the assumption that Algorithm~\ref{algorithm1} makes the initial $2K$ updates with $\widehat{J}_{l} \not = \widehat{J}_{k}$ for each $1\le l< k\le 2K$.

Next, the upper bound of the term $\mathbb{E}\{  \sum_{ l\in \mathcal{L}_{2s}(J) }   \mathbb{E}[  f(\boldsymbol{X}) -  R_{l-1}^{\star}(\boldsymbol{X})  | \boldsymbol{X}_{\mathcal{X}(\widehat{J}_{l})}] |\boldsymbol{X}_{\mathcal{X}(J)}\}$ on the RHS of \eqref{boundness.bias.4} is established as follows. By \eqref{boundness.bias.6}--\eqref{boundness.bias.5}, and Lemma~\ref{lemma.4} along with the definition of $\mathcal{L}_{2}(J)$,
\begin{equation}
    \begin{split}\label{boundness.bias.10}
        & \left|\mathbb{E}\{  \sum_{ l\in \mathcal{L}_{2s}(J) }   \mathbb{E}[  f(\boldsymbol{X}) -  R_{l-1}^{\star}(\boldsymbol{X})  | \boldsymbol{X}_{\mathcal{X}(\widehat{J}_{l})}] |\boldsymbol{X}_{\mathcal{X}(J)}\} \right|\\
        & = \left|\sum_{ l\in \mathcal{L}_{2s}(J) } \mathbb{E}\{   
        F_{l}(\widehat{J}_{l}) + g_{\widehat{J}_{l}}(\boldsymbol{X}) - \sum_{q \in \mathcal{L}_{1}(\widehat{J}_{l}) } \mathbb{E}[  f(\boldsymbol{X}) -  R_{q-1}^{\star}(\boldsymbol{X})  | \boldsymbol{X}_{\mathcal{X}(\widehat{J}_{q})}]
        |\boldsymbol{X}_{\mathcal{X}(J)}\} \right|\\
        & \le \left(\sum_{ l\in \mathcal{L}_{2s}(J) } \bar{F}_{l}(\widehat{J}_{l})  \right) + \sum_{ l\in \mathcal{L}_{2s}(J) } \left|  \mathbb{E}\left[  g_{\widehat{J}_{l}}(\boldsymbol{X}) -  \Big(\sum_{q \in \mathcal{L}_{1}(\widehat{J}_{l}) } g_{\widehat{J}_{q}}(\boldsymbol{X})\Big)  \Big| \boldsymbol{X}_{\mathcal{X}(J)} \right]\right|\\ 
        & \qquad + \sum_{ l\in \mathcal{L}_{2s}(J) }\left|\sum_{q \in \mathcal{L}_{1}(\widehat{J}_{l}) } \mathbb{E}\{ \mathbb{E}[  \mathbb{E}[f(\boldsymbol{X}) | \boldsymbol{X}_{-\mathcal{X}(\widehat{J}_{q})}]-  R_{q-1}^{\star}(\boldsymbol{X})  | \boldsymbol{X}_{\mathcal{X}(\widehat{J}_{q})}]
        |\boldsymbol{X}_{\mathcal{X}(J)}\} \right| \\
        & \le \left(\sum_{ l\in \mathcal{L}_{2s}(J) } \bar{F}_{l}(\widehat{J}_{l})  \right)
        + \texttt{\#}\mathcal{L}_{2s}(J) \times \left( \frac{8\delta_{0}\sqrt{\mathbb{E}[(f(\boldsymbol{X}))^2]}}{p_{\min}} + \frac{6 \delta_{0}  \sqrt{\mathbb{E}[(f(\boldsymbol{X}))^2]}}{ p_{\min} }   \right),
    \end{split}
\end{equation}
where for each $J\subset\{1, \dots, M\}$ and $K<l\le 2K$,
$$\bar{F}_{l}(J) \coloneqq \inf\{ c > 0: |F_{l}(J)|\le c \textnormal{ almost surely}\}.$$

By Lemma~\ref{lemma.4}, \eqref{boundness.bias.2}--\eqref{boundness.bias.10},  the definition of $\bar{F}_{l}(J)$ above, and that $F_{s}(J)$ given in \eqref{boundness.bias.6} is defined to be the sum of the last three terms on the RHS of \eqref{boundness.bias.4}, it holds that for each $K<s\le K + \texttt{\#}S_{2}^*$,
\begin{equation}
    \begin{split}\label{boundness.bias.9}
        \bar{F}_{s}(J) & \le \left(\sum_{ l\in \mathcal{L}_{2s}(J) } \bar{F}_{l}(\widehat{J}_{l})  \right) + \texttt{\#}\mathcal{L}_{2s}(J) \times \left( \frac{6\delta_{0}\sqrt{\mathbb{E}[(f(\boldsymbol{X}))^2]}}{p_{\min}} + \frac{8 \delta_{0}  \sqrt{\mathbb{E}[(f(\boldsymbol{X}))^2]}}{ p_{\min} }   \right) \\
        & \qquad+ \left(\sum_{ l\in \mathcal{L}_{3s}(J) } \frac{9 \delta_{0}  \sqrt{\mathbb{E}([f(\boldsymbol{X})]^2) } }{p_{\min}} \right) +
        \frac{27\delta_{0} \sqrt{\mathbb{E}[(f(\boldsymbol{X}))^2]}}{p_{\min}} \\
        & \le \left(\sum_{ l>K}^{s-1} \bar{F}_{l}(\widehat{J}_{l})  \right) + (23(s-K-1) + 27) \times  \frac{\delta_{0} \sqrt{\mathbb{E}[(f(\boldsymbol{X}))^2]}}{p_{\min}}\\
        & \le (23(\texttt{\#}S_{2}^*)^2 + 27\texttt{\#}S_{2}^*) \times  \frac{\delta_{0} \sqrt{\mathbb{E}[(f(\boldsymbol{X}))^2]}}{p_{\min}}\\
        & \le 50(\texttt{\#}S_{2}^*)^2 \times  \frac{\delta_{0} \sqrt{\mathbb{E}[(f(\boldsymbol{X}))^2]}}{p_{\min}},
    \end{split}
\end{equation}
where the second inequality follows from Jensen's inequality, $p_{\min}\le 1$, and the fact that $\texttt{\#}\mathcal{L}_{2s}(J) \vee \texttt{\#}\mathcal{L}_{3s}(J) \le s- K-1$. The third inequality stems from a recursive application of the second inequality. Recall that we have defined that $\sum_{l> K}^{K}1 = 0$.

By the identity $\mathbb{E}\left([g_{J}(\boldsymbol{X}) -\sum_{m \in J}g_{m}(\boldsymbol{X})]^2\right) = \textnormal{Var}\left(g_{J}(\boldsymbol{X}) -\sum_{m \in J}g_{m}(\boldsymbol{X})\right)$, \eqref{boundness.bias.6.b}, \eqref{boundness.bias.9}, \eqref{boundness.bias.5}, and \eqref{bias.control.6}, it holds that on the event $\cap_{l=1}^6 E_{4l}$, for each $K< s\le K + \texttt{\#}S_{2}^*$, each $J\subset \{1, \dots, M\}$ with $\texttt{\#}J=2$,
\begin{equation}
    \begin{split}\label{bias.control.3.2k}
    & \left|\mathbb{E}[(\mathbb{E}(f(\boldsymbol{X}) - R_{s-1}^{\star}(\boldsymbol{X})|\boldsymbol{X}_{\mathcal{X}(J)}))^2  ] - \textnormal{Var}\left[g_{J}(\boldsymbol{X}) - \left(\sum_{j\in J} g_{\{j\}}(\boldsymbol{X}) \right)\right] \right| \\
        & \le \left(\frac{6\delta_{0}\sqrt{\mathbb{E}[(f(\boldsymbol{X}))^2]}}{p_{\min}}\right)^2 + \mathbb{E}[(F_{s}(J))^2 ]\\
           &  \qquad +\left(\frac{12\delta_{0}\sqrt{\mathbb{E}[(f(\boldsymbol{X}))^2]}}{p_{\min}}\right)  \times\left(\mathbb{E}\left|F_{s}(J)\right| + 6\sqrt{\mathbb{E} ([(f(\boldsymbol{X}))^2]} \right) \\
           & \qquad + 12 \sqrt{\mathbb{E} ([(f(\boldsymbol{X}))^2]} \times \mathbb{E}|F_{s}(J)|\\
           & \le \mathbb{E}[(f(\boldsymbol{X}))^2]\left( \frac{36\delta_{0}^2}{p_{\min}^2} + \frac{ 2500 \delta_{0}^2 (\texttt{\#}S_{2}^*)^4}{p_{\min}^2} + \frac{ 600 \delta_{0}^2  (\texttt{\#}S_{2}^*)^2}{p_{\min}^2}  + \frac{72 \delta_{0}  }{p_{\min}} + \frac{ 600 \delta_{0}  (\texttt{\#}S_{2}^*)^2}{p_{\min}^2}\right)\\
           & \le 3808 \delta_{0} {p_{\min}^{-2}} \mathbb{E}[(f(\boldsymbol{X}))^2] (\texttt{\#}S_{2}^*)^4,
    \end{split}
\end{equation}
where we also use $p_{\min}\le 1$ and $\delta_0\le 1$ in the last inequality.

By \eqref{bias.control.3.2k}, we have finished the proof of \eqref{bias.control.3.2k.b}.


\subsection{Proof of Corollary~\ref{corollary.1}}\label{proof.corollary1}

    \noindent\textit{Proof of Corollary~\ref{corollary.1}: }  Let us begin with proving the assertion that $\delta_{0} = 0$ if and only if feature groups $\boldsymbol{X}_{\mathcal{X}(1)}, \dots, \boldsymbol{X}_{\mathcal{X}(M)}$ are independent. Recall that $\delta_{0}$ is defined in Condition~\ref{signal.strength.1}.

When $\delta_{0} = 0$ and $\texttt{\#}\mathcal{X}(m) > 1$, it holds that
for every $\vv{c}_{1} \in \{0, 1\}^{\texttt{\#}\mathcal{X}(m)}$ with $\{\boldsymbol{X}_{\mathcal{X}(m)} = \vv{c}_{1}\} = \{X_{i} = 1\}$ for some $i\in \mathcal{X}(m)$, and every $\vv{c}_{2} \in \{0, 1\}^{p - \texttt{\#}\mathcal{X}(m)}$,
\begin{equation}
    \begin{split}\label{delta_o_1}
        & \mathbb{P}(\boldsymbol{X}_{\mathcal{X}(m)} = \vv{c}_{1}, \boldsymbol{X}_{-\mathcal{X}(m)} = \vv{c}_{2} ) \\
        & = \mathbb{P}(\boldsymbol{X}_{\mathcal{X}(m)} = \vv{c}_{1}| \boldsymbol{X}_{-\mathcal{X}(m)} = \vv{c}_{2} ) \times \mathbb{P}(\boldsymbol{X}_{-\mathcal{X}(m)} = \vv{c}_{2} )   \\
        & = \mathbb{P}(X_{i} = 1 | \boldsymbol{X}_{-\mathcal{X}(m)} = \vv{c}_{2} ) \times \mathbb{P}(\boldsymbol{X}_{-\mathcal{X}(m)} = \vv{c}_{2} )\\
        & = \mathbb{P}(X_{i} = 1 ) \times \mathbb{P}(\boldsymbol{X}_{-\mathcal{X}(m)} = \vv{c}_{2} )\\
        & = \mathbb{P}(\boldsymbol{X}_{\mathcal{X}(m)} = \vv{c}_{1}) \times \mathbb{P}(\boldsymbol{X}_{-\mathcal{X}(m)} = \vv{c}_{2} ), 
    \end{split}
\end{equation}
where the third equality follows from that $\delta_{0}=0$. Recall that $\mathbb{E}(h(\boldsymbol{X})|\boldsymbol{X} = \vv{c}) = 0$ when $\mathbb{P}(\boldsymbol{X} = \vv{c}) = 0$ for every $\vv{c} \in \{0, 1\}^{p}$ and every measurable function $h:\mathbb{R}^p\mapsto\mathbb{R}$. The result of \eqref{delta_o_1} and the assumption that $\boldsymbol{X}_{\mathcal{X}(m)}$ with $\texttt{\#}\mathcal{X}(m) > 1$ consists of one-hot indicators (see Condition~\ref{model.1}) conclude the independence between $\boldsymbol{X}_{\mathcal{X}(m)}$ and $\boldsymbol{X}_{-\mathcal{X}(m)}$ when $\texttt{\#}\mathcal{X}(m) > 1$.

For the case with $\texttt{\#}\mathcal{X}(m) = 1$ and $i\in \mathcal{X}(m)$, we deduce that for every $\vv{c}_{2} \in \{0, 1\}^{p - 1}$,
\begin{equation}
    \begin{split}\label{delta_o_2}
        & \mathbb{P}(X_{i}=1, \boldsymbol{X}_{-\mathcal{X}(m)} = \vv{c}_{2} ) \\
        & = \mathbb{P}(X_{i}=1 | \boldsymbol{X}_{-\mathcal{X}(m)} = \vv{c}_{2} ) \times \mathbb{P}(\boldsymbol{X}_{-\mathcal{X}(m)} = \vv{c}_{2} )   \\
        & = \mathbb{P}(X_{i} = 1 ) \times \mathbb{P}(\boldsymbol{X}_{-\mathcal{X}(m)} = \vv{c}_{2} )  ,  
    \end{split}
\end{equation}
where the last equality is due to $\delta_{0} = 0$. Additionally,
\begin{equation}
    \begin{split}\label{delta_o_3}
        & \mathbb{P}(X_{i}=0, \boldsymbol{X}_{-\mathcal{X}(m)} = \vv{c}_{2} ) \\
        & = \mathbb{P}(X_{i}=0 | \boldsymbol{X}_{-\mathcal{X}(m)} = \vv{c}_{2} ) \times \mathbb{P}(\boldsymbol{X}_{-\mathcal{X}(m)} = \vv{c}_{2} )   \\
        & = [1-\mathbb{P}(X_{i}=1 | \boldsymbol{X}_{-\mathcal{X}(m)} = \vv{c}_{2} )] \times \mathbb{P}(\boldsymbol{X}_{-\mathcal{X}(m)} = \vv{c}_{2} )   \\
        & = [1-\mathbb{P}(X_{i}=1 )] \times \mathbb{P}(\boldsymbol{X}_{-\mathcal{X}(m)} = \vv{c}_{2} )   \\
        & = \mathbb{P}(X_{i}=0 ) \times \mathbb{P}(\boldsymbol{X}_{-\mathcal{X}(m)} = \vv{c}_{2} ) ,
    \end{split}
\end{equation}
where the third equality is due to $\delta_{0} = 0$. The results of \eqref{delta_o_2}--\eqref{delta_o_3} conclude the independence between $\boldsymbol{X}_{\mathcal{X}(m)}$ and $\boldsymbol{X}_{-\mathcal{X}(m)}$ when $\texttt{\#}\mathcal{X}(m) = 1$.

An application of above arguments to each feature group shows that feature groups $\boldsymbol{X}_{\mathcal{X}(1)}, \dots, \boldsymbol{X}_{\mathcal{X}(M)}$ are independent when $\delta_{0} = 0$. On the other hand, it is obvious that $\delta_{0} = 0$ if feature groups  are independent. Therefore, we have finished the proof that $\delta_{0} = 0$ if and only if feature groups $\boldsymbol{X}_{\mathcal{X}(1)}, \dots, \boldsymbol{X}_{\mathcal{X}(M)}$ are independent.

Now, we proceed to deal with the main assertion of Corollary~\ref{corollary.1}, and begin by clarifying notation usage: When $J= \{m\}$, we write that
$h_{m}(\boldsymbol{X}_{\mathcal{X}(m)}) = h_{\{m\}}(\boldsymbol{X}_{\mathcal{X}(\{m)\}}) = h_{J}(\boldsymbol{X}_{\mathcal{X}(J)})$. Let us give an auxiliary result of \eqref{upper.iota.7} below, which is useful for our proof. By the assumption of independent features and the  assumption on $h_{J}(\boldsymbol{X}_{\mathcal{X}(J)})$'s, it holds that for each $m\in S_{1}$, each $J\in S_{2}$, every $H_{1}\subset\{1, \dots, M\}$ with $\texttt{\#} (H_{1}\cap J) = 1$, and every $H_{2}\subset\{1, \dots, M\}$ with $m\not\in H_{2}$,
\begin{equation}
    \begin{split}\label{upper.iota.7}
\mathbb{E}(h_{J}(\boldsymbol{X}_{\mathcal{X}(J)})|\boldsymbol{X}_{\mathcal{X}(H_{1})}) & = \mathbb{E}(h_{J}(\boldsymbol{X}_{\mathcal{X}(J)})|\boldsymbol{X}_{\mathcal{X}(H_{1}\cap J)})\\
& = 0,\\
\mathbb{E}(h_{m}(\boldsymbol{X}_{\mathcal{X}(m)})|\boldsymbol{X}_{\mathcal{X}(H_{2})}) & = \mathbb{E}(h_{m}(\boldsymbol{X}_{\mathcal{X}(m)}))\\
& = 0.
    \end{split}
\end{equation}

Now, we show that, based on our assumption, for any nonrepeated sequence of sets $\widehat{J}_{l}$'s with $\texttt{\#}\widehat{J}_{l} = 1$ for $1\le l\le K$ and $\texttt{\#}\widehat{J}_{l} = 2$ for $K< l\le 2K$, it holds that for each $0<s\le 2K$,
\begin{equation}
    \label{upper.iota.8}
    R_{s}^{\star} (\boldsymbol{X}) = \sum_{l=1}^s h_{\widehat{J}_{l}}(\boldsymbol{X}_{\mathcal{X}(\widehat{J}_{l})})+ \sum_{m \in (\cup_{k=K+1}^{s} \widehat{J}_{k}) \backslash (\cup_{k=1}^{K} \widehat{J}_{k} ) } h_{m}(\boldsymbol{X}_{\mathcal{X}(m)}),
\end{equation} 
where we define that 
\begin{equation}
    \label{upper.iota.10}
    h_{J} (\boldsymbol{X}_{\mathcal{X}(J)}) = 0 \textnormal{ if } J\not\in \{\{m\}: m\in S_{1} \} \cup S_{2},
\end{equation} 
and recall that summation over an empty set is defined to be zero, and that each set has distinct elements. It is important to note that \eqref{upper.iota.8} does not depend on the result of consistent feature selection.

Let us prove \eqref{upper.iota.8}. Suppose it is true  that 
\begin{equation}
    \begin{split}\label{upper.iota.12}
        R_{s-1}^{\star} (\boldsymbol{X}) & = \sum_{l=1}^{s-1} h_{\widehat{J}_{l}}(\boldsymbol{X}_{\mathcal{X}(\widehat{J}_{l})})+ \sum_{m \in (\cup_{k=K+1}^{s-1} \widehat{J}_{k} ) \backslash (\cup_{k=1}^{K} \widehat{J}_{k}) } h_{m}(\boldsymbol{X}_{\mathcal{X}(m)})
    \end{split}
\end{equation}
which in combination with \eqref{upper.iota.7} and the assumption that each element in the sequence $\{\widehat{J}_{l}\}$ is unique leads to that 
\begin{equation}\label{upper.iota.11}
    \mathbb{E}(R_{s-1}^{\star} (\boldsymbol{X}) | \boldsymbol{X}_{\mathcal{X}(\widehat{J}_{s})} ) = \sum_{m\in \widehat{J}_{s} \cap (\cup_{l=1}^{s-1} \widehat{J}_{l} ) } h_{m}(\boldsymbol{X}_{\mathcal{X}(m)}).
\end{equation}
In addition, by the assumption on $f(\boldsymbol{X})$ and \eqref{upper.iota.7}, it holds that
\begin{equation}\label{upper.iota.13}
\begin{split}
    \mathbb{E}(f (\boldsymbol{X}) | \boldsymbol{X}_{\mathcal{X}(J)} ) &= h_{J}(\boldsymbol{X}_{\mathcal{X}(J)}) + \sum_{m\in J} h_{m}(\boldsymbol{X}_{\mathcal{X}(m)}) ,\qquad \textnormal{ for each } J \textnormal{ with } \texttt{\#}J = 2,\\
    \mathbb{E}(f (\boldsymbol{X}) | \boldsymbol{X}_{\mathcal{X}(J)} ) &= h_{J}(\boldsymbol{X}_{\mathcal{X}(J)}), \qquad \qquad\qquad\qquad\qquad\textnormal{ for each } J \textnormal{ with } \texttt{\#}J = 1.
    \end{split}
\end{equation}

By \eqref{upper.iota.12}--\eqref{upper.iota.13}, the assumption that each element in the sequence $\{\widehat{J}_{l}\}$ is unique, the definition that each set  has distinct elements (and hence each element in $\cup J_{l}$ is unique), and \eqref{upper.iota.7}, it holds that for $\widehat{J}_{s}$ with $\texttt{\#}\widehat{J}_{s} = 2$,
    \begin{equation*}
    \begin{split}
        R_{s}^{\star}(\boldsymbol{X}) & = R_{s-1}^{\star} (\boldsymbol{X}) +   \mathbb{E}( f(\boldsymbol{X}) - R_{s-1}^{\star}(\boldsymbol{X}) | \boldsymbol{X}_{\mathcal{X}(\widehat{J}_{s})})
        \\
        & = R_{s-1}^{\star} (\boldsymbol{X})  + \left(h_{\widehat{J}_{s}} (\boldsymbol{X}_{\mathcal{X}(\widehat{J}_{s})}) + \sum_{m \in \widehat{J}_{s}} h_{m} (\boldsymbol{X}_{\mathcal{X}(m)})\right) - \mathbb{E}(R_{s-1}^{\star} (\boldsymbol{X}) | \boldsymbol{X}_{\mathcal{X}(\widehat{J}_{s})} )\\
        & = \sum_{l=1}^{s-1} h_{\widehat{J}_{l}}(\boldsymbol{X}_{\mathcal{X}(\widehat{J}_{l})})+ \sum_{m \in (\cup_{k=K+1}^{s-1} \widehat{J}_{k} ) \backslash (\cup_{k=1}^{K} \widehat{J}_{k}) } h_{m}(\boldsymbol{X}_{\mathcal{X}(m)})\\
        &\qquad + \left(h_{\widehat{J}_{s}} (\boldsymbol{X}_{\mathcal{X}(\widehat{J}_{s})}) + \sum_{m \in \widehat{J}_{s}} h_{m} (\boldsymbol{X}_{\mathcal{X}(m)})\right)
        - \sum_{m\in \widehat{J}_{s} \cap (\cup_{l=1}^{s-1} \widehat{J}_{l} ) } h_{m}(\boldsymbol{X}_{\mathcal{X}(m)}) \\
        & = \sum_{l=1}^s h_{\widehat{J}_{l}}(\boldsymbol{X}_{\mathcal{X}(\widehat{J}_{l})})+ \sum_{m \in (\cup_{k=K+1}^{s} \widehat{J}_{k}) \backslash (\cup_{k=1}^{K} \widehat{J}_{k} ) } h_{m}(\boldsymbol{X}_{\mathcal{X}(m)}),
    \end{split}    
\end{equation*} 
which concludes the desired result by induction in this scenario. The case with $\texttt{\#}\widehat{J}_{s} = 1$ can be proven similarly, and hence we omit the detail. We have finished the proof of the desired result \eqref{upper.iota.8}.

By \eqref{upper.iota.11}--\eqref{upper.iota.13}, it holds that for each $s>0$,
\begin{equation*}
    \begin{split}
    &|\mathbb{E}( f(\boldsymbol{X}) - R_{s-1}^{\star}(\boldsymbol{X}) | \boldsymbol{X}_{\mathcal{X}(\widehat{J}_{s})})|   \\
    &\le  \left( \max_{\vv{c}\in\{0, 1\}^{\texttt{\#} \mathcal{X}(\widehat{J}_{s})}} |h_{\widehat{J}_{s}}(\vv{c})| + \sum_{m\in \widehat{J}_{s}} \max_{\vv{c}\in\{0, 1\}^{\texttt{\#} \mathcal{X}(m)}} |h_{m}(\vv{c})| \right)
    \end{split}
\end{equation*}
which in combination with the definition of $\iota_{l}$'s in \eqref{iota.1} leads to that for each $l\ge 1$,
\begin{equation}
    \begin{split}\label{upper.iota.1}
        \iota_{l} \le  \sum_{J \in \{\{m\}: m\in S_{1} \} \cup S_{2}} \left( \max_{\vv{c}\in\{0, 1\}^{\texttt{\#} \mathcal{X}(J)}} |h_{J}(\vv{c})| + \sum_{m\in J} \max_{\vv{c}\in\{0, 1\}^{\texttt{\#} \mathcal{X}(m)}} |h_{m}(\vv{c})| \right).
    \end{split}
\end{equation}

Now, we turn to establish upper bounds for terms on the RHS of \eqref{upper.iota.1}. By our assumption,  for each $J \in \{\{m\}: m\in S_{1} \} \cup S_{2}$, and every $\vv{v}$ with $\mathbb{P}(\boldsymbol{X}_{\mathcal{X}(J)} = \vv{v}) > 0$,
\begin{equation}
    \begin{split}\label{upper.iota.6}
         w_{n} & = \textnormal{Var}(h_{J}(\boldsymbol{X}_{\mathcal{X}(J)})) \\
         & = \mathbb{E}[(h_{J}(\boldsymbol{X}_{\mathcal{X}(J)}))^2]\\
        & = \sum_{\vv{c} \in \{0, 1\}^{\texttt{\#}\mathcal{X}(J)}} \mathbb{P}(\boldsymbol{X}_{\mathcal{X}(J)} = \vv{c}) \times(h_{J}(\vv{c}))^2\\
        & \ge  p_{\min}\times(h_{J}(\vv{v}))^2,
    \end{split}
\end{equation}
where $p_{\min}$ is defined in Condition~\ref{signal.strength.1}, and that the last inequality is due to
\begin{equation*}
    \min\{ \mathbb{P}(\boldsymbol{X}_{\mathcal{X}(J)} = \vv{c}) > 0:  \vv{c} \in \{0, 1\}^{\texttt{\#}\mathcal{X}(J)},\texttt{\#}J\le 2 \} \ge p_{\min}.
\end{equation*}

The result of \eqref{upper.iota.6} and the definition in \eqref{upper.iota.10} conclude that for each $J\subset\{1, \dots, M\}$ with $\texttt{\#}J\le 2$,
\begin{equation}\label{upper.iota.2}
    |h_{J}(\boldsymbol{X}_{\mathcal{X}(J)})| \le \sqrt{w_{n} p_{\min}^{-1}} \qquad \textnormal{almost surely.}
\end{equation}

By \eqref{upper.iota.1}, \eqref{upper.iota.2}, the definition that $w_{n} = 16L_{1} n^{-\frac{1}{2}} \log{(n)} n^{ \frac{1}{q_{1}}}\sqrt{2K}$, the fact that $\texttt{\#}\widehat{J}_{l}\le 2$ for each $1\le l\le 2K$, and the assumption that $\texttt{\#}S_{1} \vee \texttt{\#}S_{2}  \le K \le (\log{n})^{-1}n^{\frac{1}{5} - \frac{2}{5q_{1}}}$, it holds that 
\begin{equation}\label{upper.iota.3}
    \begin{split}
        \max_{0<l\le 2K}\iota_{l} \le  3\sqrt{w_{n} p_{\min}^{-1}} \times 2K = o(1).
    \end{split}
\end{equation}

By \eqref{upper.iota.3} and some simple calculations, we have that for all large $n$,
\begin{equation}
    \label{upper.iota.4}
    o(1) = w_{n}  \ge 2d_{n},
\end{equation} 
 where we recall that
$$d_{n} = L_{1}  n^{-\frac{1}{2}} \log{(n)} (n^{ \frac{1}{q_{1}}}\sqrt{2K} + \sqrt{\sum_{l=1}^{2K} \iota_{l-1}^2} ) (\iota_{2K-1}^2 + \iota_{2K-1} + 3).$$

By the definition of $f(\boldsymbol{X})$ and similar arguments for \eqref{upper.iota.3}, we have that for all large $n$,
\begin{equation}\label{upper.iota.5}
    \max_{\vv{c} \in \{0, 1\}^{p}}|f(\vv{c})| \le \sum_{J \in \{\{m\}: m\in S_{1} \} \cup S_{2}} \max_{\vv{c}\in\{0, 1\}^{\texttt{\#} \mathcal{X}(J)}} |h_{J}(\vv{c})| \le M_{f}.
\end{equation}

The results of \eqref{upper.iota.4}--\eqref{upper.iota.5} conclude the second assertion of Corollary~\ref{corollary.1}.

Lastly, we proceed to show the last assertion of Corollary~\ref{corollary.1}. The other regularity conditions are given as follows. Consider a modified Algorithm~\ref{algorithm1} such that $\widehat{J}_{l} \not= \widehat{J}_{k}$ for each $\{l, k\}\subset \{1, \dots, 2K\}$. Assume that we are given an  i.i.d. sample $(\boldsymbol{X}_{1}, Y_{1}), \dots, (\boldsymbol{X}_{n}, Y_{n}), (\boldsymbol{X}, Y)$ satisfying Condition~\ref{model.1}. Assume that $p_{\min}\ge p_{01} \vee \frac{2\log{(n)}}{\sqrt{n}}$, and that $\iota_{-1} = \iota_{0}  = 0$. Assume $M\ge K \ge \texttt{\#}S_{1}^* \vee \texttt{\#}S_{2}^*$, , and that $M = O(n^{K_{01}})$ for some arbitrary constant $K_{01}>0$. 

Then, by Theorem~\ref{theorem3}, it holds that $\mathbb{P}(\{S_{1} \not\subset \cup_{l=1}^{K} \widehat{J}_{l}\}\cup ( S_{2} \not\subset \{\widehat{J}_{K+1}, \dots, \widehat{J}_{K+\texttt{\#}S_{2}}\} ) ) = o(1)$. By this result and \eqref{upper.iota.4}--\eqref{upper.iota.5}, we have finished the proof of Corollary~\ref{corollary.1}.

\subsection{Proof of Theorem~\ref{theorem2}}\label{SecB.4}

\noindent\textit{Proof of Theorem~\ref{theorem2}: } The sum of all sample split scores after the $(K + \texttt{\#}S_{2}^*)$th round is given by
\begin{equation}
\begin{split}\label{sample.score.2K.1}
   & n^{-1}\left(\sum_{i=1}^n (Y_{i} - \widehat{U}_{K + \texttt{\#}S_{2}^*}(\boldsymbol{X}_{i}))^2 \right) - n^{-1}\sum_{i=1}^n \left(Y_{i} - \sum_{k=1}^K T_{k}(\boldsymbol{X}_{i}) \right)^2 \\
   & = \Delta + D_{2, K + \texttt{\#}S_{2}^*} + D_{1,K + \texttt{\#}S_{2}^*} + \widehat{\Gamma},
 \end{split}
\end{equation}  
where $\widehat{U}_{s}$'s are the sample predictive functions at the $s$th round defined in \eqref{predictor.trees.b.3}, and that 
\begin{equation*}
    \begin{split} 
        D_{1,K + \texttt{\#}S_{2}^*} & = n^{-1}\sum_{i=1}^n (Y_{i} - \widehat{U}_{K + \texttt{\#}S_{2}^*}(\boldsymbol{X}_{i}))^2 - n^{-1}\sum_{i=1}^n (f(\boldsymbol{X}_{i}) - R_{K + \texttt{\#}S_{2}^*}^{\star}(\boldsymbol{X}_{i}))^2  - n^{-1}\sum_{i=1}^n \varepsilon_{i}^2,\\
        D_{2,K + \texttt{\#}S_{2}^*} & = n^{-1}\sum_{i=1}^n (f(\boldsymbol{X}_{i}) - R_{K + \texttt{\#}S_{2}^*}^{\star}(\boldsymbol{X}_{i}))^2 - \mathbb{E}(f(\boldsymbol{X}) - R_{K + \texttt{\#}S_{2}^*}^{\star}(\boldsymbol{X}) )^2.
    \end{split}
\end{equation*}
Here, we use the result $\mathbb{E} (R_{K + \texttt{\#}S_{2}^*}^{\star}(\boldsymbol{X})) = 0$ from Lemma~\ref{lemma.2} and the assumption $\mathbb{E}(f(\boldsymbol{X})) = 0$ to deduce that $\mathbb{E}(f(\boldsymbol{X}) - R_{K + \texttt{\#}S_{2}^*}^{\star}(\boldsymbol{X}))^2 = \textnormal{Var}(f(\boldsymbol{X}) - R_{K + \texttt{\#}S_{2}^*}^{\star}(\boldsymbol{X})) =\Delta$.

    We now establish the upper bounds for $|D_{2, K+\texttt{\#}S_{2}^*}|$  and $|D_{1,K+\texttt{\#}S_{2}^*}|$ by the results derived in the proof of Theorem~\ref{theorem3}. By \eqref{particular.1}  of Theorem~\ref{theorem3}, the definition \eqref{predictor.trees.b.3} in Theorem~\ref{theorem3}, and the assumed technical conditions, it holds that on the event $\cap_{l=1}^6 E_{4l}$ defined in \eqref{event.2} of Theorem~\ref{theorem3},
\begin{equation}
    \begin{split}\label{xmdi.8}
        |D_{2, K+\texttt{\#}S_{2}^*}| + |D_{1,K+\texttt{\#}S_{2}^*}| \le d_{n}, 
    \end{split}
\end{equation}
where $d_{n} = L_{1}  n^{-\frac{1}{2}} \log{(n)} (n^{ \frac{1}{q_{1}}}\sqrt{2K} + \sqrt{\sum_{l=1}^{2K} \iota_{l-1}^2} ) (\iota_{2K-1}^2 + \iota_{2K-1} + 3)$, with technical parameters given in Theorem~\ref{theorem3}.

The results of \eqref{sample.score.2K.1}--\eqref{xmdi.8} are useful for analyzing our group feature importance measures. Consider $\textnormal{XMDI}_{ij}^{\dagger}$ as an importance measure identical to $\textnormal{XMDI}_{ij}$, except that $\textnormal{XMDI}_{ij}^{\dagger}$ calculates importance only up to the first $K + \texttt{\#}S_{2}^*$ updates, with the remaining split scores ignored. That is, $\textnormal{XMDI}_{J}^{\dagger} = 0$ if $J\not \in \{\widehat{J}_{1}, \dots, \widehat{J}_{K+\texttt{\#}S_{2}^*} \}$, where we use the notation that  $\textnormal{XMDI}_{\{m\}}^{\dagger} = \textnormal{XMDI}_{mm}^{\dagger}$ and $\textnormal{XMDI}_{\{l,k\}}^{\dagger} = \textnormal{XMDI}_{lk}^{\dagger}$.

By \eqref{sample.score.2K.1}--\eqref{xmdi.8} and the definition of XMDI$_{ij}^{\dagger}$'s, it holds that on the event $\cap_{l=1}^6 E_{4l}$,
    \begin{equation}\label{xmdi.4}
    \begin{split}
\sum_{1\le i< j\le M}|\textnormal{XMDI}_{ij} - \textnormal{XMDI}_{ij}^{\dagger}| &\le \Delta + d_{n} + \widehat{\Gamma}.
\end{split}
\end{equation}
In addition, we observe that the estimation of additive effects remains consistent across both definitions of feature importance measures, due to the definition of $\textnormal{XMDI}_{mm}$'s. Specifically, for each $1\le m\le M$,
    \begin{equation}\label{xmdi.3}
        \textnormal{XMDI}_{mm} = \textnormal{XMDI}_{mm}^{\dagger}.
    \end{equation}

Let us define some notation for further analysis: 
\begin{equation*}
    \begin{split}
        \widehat{\Delta}_{\widehat{J}_{s}} & =  \mathbb{E}(f(\boldsymbol{X}) - R_{s-1}^{\star}(\boldsymbol{X}) )^2 - \mathbb{E}(f(\boldsymbol{X}) - R_{s}^{\star}(\boldsymbol{X}) )^2 \textnormal{ for each } 0< s\le K+ \texttt{\#}S_{2}^*,\\
        \widehat{\Delta}_{J} & =  0 \textnormal{ if } J\not \in \{\widehat{J}_{1}, \dots, \widehat{J}_{K+\texttt{\#}S_{2}^*} \},
    \end{split}
\end{equation*} 
where we recall that $R_{s}^{\star}(\boldsymbol{X}) = R_{s-1}^{\star} (\boldsymbol{X})+   \mathbb{E}( f(\boldsymbol{X}) - R_{s-1}^{\star}(\boldsymbol{X}) | \boldsymbol{X}_{\mathcal{X}(\widehat{J}_{s})})$ is defined in \eqref{mp.3}. Also, we use the notation $\widehat{\Delta}_{m} = \widehat{\Delta}_{\{m\}}$ and $\widehat{\Delta}_{kl} = \widehat{\Delta}_{lk}  =\widehat{\Delta}_{\{l,k\}}$ for each $1\le m \le M$ and every $1\le l<k\le M$.

 By Lemma~\ref{lemma.3}, \eqref{equivalence.recur.4.c}, and the assumed conditions in Theorem~\ref{theorem2}, on the event $\cap_{l=1}^6 E_{4l}$, it holds that for each $0<l\le K + \texttt{\#}S_{2}^*$,
\begin{equation}
    \begin{split}\label{xmdi.7}
        |\textnormal{XMDI}_{\widehat{J}_{l}}^{\dagger} - \widehat{\Delta}_{\widehat{J}_{l}} | & \le d_{n}.
    \end{split}
\end{equation}
In addition, by the definitions of $\textnormal{XMDI}_{J}^{\dagger}$'s and $\widehat{\Delta}_{J}$'s, it holds that for every $J\subset\{1, \dots ,M\}$ such that $\texttt{\#}J \le 2$ and $J\not \in \{\widehat{J}_{1}, \dots, \widehat{J}_{K+\texttt{\#}S_{2}^*} \}$,
\begin{equation}
    \label{xmdi.9}
    \textnormal{XMDI}_{J}^{\dagger} = \widehat{\Delta}_{J} = 0.
\end{equation}

By \eqref{consistency.1} and \eqref{consistency.2} and the assumed regularity conditions, it holds that on the event $\cap_{l=1}^6 E_{4l}$,
\begin{equation*}
\begin{split}
S_{1}^{\star} & \subset\cup_{1\le l\le K}\widehat{J}_{l},\\
    S_{2}^* & \subset \{\widehat{J}_{K+1}, \dots, \widehat{J}_{K+\texttt{\#}S_{2}^*}\}.
\end{split}    
\end{equation*} 
By this sure screening result, the definition of significant components in $S_{1}^*$ and $S_{2}^*$, \eqref{xmdi.9}, \eqref{bias.control.3.b}, and \eqref{bias.control.3.2k.b}, it holds that on the event $\cap_{l=1}^6 E_{4l}$, for each $m\in \{1, \dots, M\}$, each $\{l, k\}\in \{\widehat{J}_{K+1}, \dots, \widehat{J}_{K+\texttt{\#}S_{2}^*}\}$, and each $\{i, j\}\subset \{1, \dots, M\}$ with $\{i, j\}\not \in \{\widehat{J}_{K+1}, \dots, \widehat{J}_{K+\texttt{\#}S_{2}^*}\}$,
\begin{equation}
    \begin{split}\label{xmdi.6}
    | \widehat{\Delta}_{m} - \textnormal{Var}(g_{m}(\boldsymbol{X}))| & \le \frac{21\delta_{0}}{(p_{\min})^2}  \mathbb{E}[(f(\boldsymbol{X}))^2], \\
        |\widehat{\Delta}_{lk} - \textnormal{Var}(g_{lk}(\boldsymbol{X} ) - g_{l}(\boldsymbol{X}) - g_{k}(\boldsymbol{X})  ) | & \le  3808 \delta_{0} {p_{\min}^{-2}} \mathbb{E}[(f(\boldsymbol{X}))^2] (\texttt{\#}S_{2}^*)^4,\\
        |\widehat{\Delta}_{ij} - \textnormal{Var}(g_{ij}(\boldsymbol{X} ) - g_{i}(\boldsymbol{X}) - g_{j}(\boldsymbol{X})  ) | & = 0.
    \end{split}
\end{equation}
Note that $\textnormal{Var}(g_{J}(\boldsymbol{X} ) ) = 0$, implying $g_{J}(\boldsymbol{X}) = 0$ almost surely, for every $J\subset\{1, \dots, M\}$ with $\texttt{\#}J = 2$ and $J\not\in S_{2}^*$.

By \eqref{xmdi.3}--\eqref{xmdi.7} and \eqref{xmdi.6}, it holds that on the event $\cap_{l=1}^6 E_{4l}$, for  each $m \in \{1, \dots, M\}$,
\begin{equation}
    \begin{split}\label{gxmdi.add.1}
        & |\textnormal{XMDI}_{mm} - \textnormal{Var}(g_{m}(\boldsymbol{X}))| \\     
        & \le  |\textnormal{XMDI}_{mm} - \textnormal{XMDI}_{mm}^{\dagger}| + |\textnormal{XMDI}_{mm}^{\dagger} - \widehat{\Delta}_{m} | +  | \widehat{\Delta}_{m} - \textnormal{Var}(g_{m}(\boldsymbol{X}))|  \\
        & \le  d_{n} + \frac{21\delta_{0}}{(p_{\min})^2}  \mathbb{E}[(f(\boldsymbol{X}))^2].
    \end{split}
\end{equation}

By \eqref{xmdi.4}, \eqref{xmdi.7},  and \eqref{xmdi.6}, it holds that on the event $\cap_{l=1}^6 E_{4l}$, for each $l\in \{1, \dots ,M\}$,
\begin{equation}
    \begin{split}\label{inter.2}
   & \sum_{k\not= l} |\textnormal{XMDI}_{lk} - \textnormal{Var}(g_{lk}(\boldsymbol{X} ) - g_{l}(\boldsymbol{X}) - g_{k}(\boldsymbol{X}) ) | \\
    & \le \sum_{k\not= l, 1\le k\le M}   \bigg( |\textnormal{XMDI}_{lk} - \textnormal{XMDI}_{lk}^{\dagger}| + |\textnormal{XMDI}_{lk}^{\dagger} - \widehat{\Delta}_{lk}| \\
    &\qquad + |\widehat{\Delta}_{lk} - \textnormal{Var}(g_{lk}(\boldsymbol{X} ) - g_{l}(\boldsymbol{X}) - g_{k}(\boldsymbol{X}) )| \bigg)\\
    & \le (\Delta + d_{n}+ \widehat{\Gamma}) + \texttt{\#}S_{2}^* d_{n} + 3808 \delta_{0} {p_{\min}^{-2}} \mathbb{E}[(f(\boldsymbol{X}))^2] (\texttt{\#}S_{2}^*)^5.
    \end{split}
\end{equation}

By \eqref{gxmdi.add.1}--\eqref{inter.2}, Lemma~\ref{lemma.event.1}, and the assumed regularity conditions, we conclude the desired results of Theorem~\ref{theorem2}.



\renewcommand{\thesubsection}{C.\arabic{subsection}}
\section{Auxiliary Lemmas and their proofs}

\subsection{Lemma~\ref{lemma.2} and its proof}\label{SecC.1}

\begin{lemma}
    \label{lemma.2}
    Assume that $\mathbb{E} (f(\boldsymbol{X}))=0$. For every sequence of sets $J_{l}$'s, and functions $R_{l}$'s defined recursively  such that $R_{0}(\boldsymbol{X}) = 0$ and that for each $s>0$,
$$R_{s}(\boldsymbol{X}) = R_{s-1}(\boldsymbol{X}) +  \mathbb{E} (f(\boldsymbol{X}) - R_{s-1}(\boldsymbol{X}) | \boldsymbol{X}_{\mathcal{X}(J_s)} ), $$
 it holds that 
\begin{equation}
    \begin{split}\label{dependence.8}
    \mathbb{E} (R_{s}(\boldsymbol{X})) & = 0,\\
        \mathbb{E}|R_{s}(\boldsymbol{X})| 
        & \le 2\sqrt{\mathbb{E}[(f(\boldsymbol{X}))^2]} .
    \end{split}
\end{equation}
\end{lemma}

\noindent\textit{Proof of Lemma~\ref{lemma.2}: } Since $\mathbb{E} (f(\boldsymbol{X}))=0$, we deduce that
\begin{equation*}
    \begin{split}
        \mathbb{E}(R_{s} (\boldsymbol{X})) & = \mathbb{E}(R_{s-1} (\boldsymbol{X}) + \mathbb{E} ( f(\boldsymbol{X})- R_{s-1} (\boldsymbol{X}) | \boldsymbol{X}_{\mathcal{X}(J_{s})}) ) \\
        & = \mathbb{E}(R_{s-1} (\boldsymbol{X}) - \mathbb{E} (  R_{s-1} (\boldsymbol{X}) | \boldsymbol{X}_{\mathcal{X}(J_{s})}) ) + \mathbb{E} ( f(\boldsymbol{X})) \\
        & = \mathbb{E}(R_{s-1} (\boldsymbol{X})) - \mathbb{E} (  R_{s-1} (\boldsymbol{X})) \\
        & = 0
    \end{split}
\end{equation*} 
for each $s>0$, which concludes the first assertion of Lemma~\ref{lemma.2}.

Next, we proceed to show the second assertion of Lemma~\ref{lemma.2}. By assumption, we deduce have that for each $s>0$,
\begin{equation}
    \begin{split}\label{dependence.7}
& \mathbb{E} [ (f(\boldsymbol{X}) - R_{s}(\boldsymbol{X}))^2] \\ 
& = \mathbb{E} \{[ f(\boldsymbol{X}) - R_{s-1}(\boldsymbol{X}) -  \mathbb{E} (f(\boldsymbol{X}) - R_{s-1}(\boldsymbol{X}) | \boldsymbol{X}_{\mathcal{X}(J_s)} ) ]^2\} \\
& =  \mathbb{E} [ (f(\boldsymbol{X}) - R_{s-1}(\boldsymbol{X}))^2]  - \mathbb{E}\{ [\mathbb{E} (f(\boldsymbol{X}) - R_{s-1}(\boldsymbol{X}) | \boldsymbol{X}_{\mathcal{X}(J_s)}) ]^2\} \\
& \le \mathbb{E} [ (f(\boldsymbol{X}) - R_{s-1}(\boldsymbol{X}))^2]\\
&\le \mathbb{E}[(f(\boldsymbol{X}))^2],
    \end{split}
\end{equation}
where  the last inequality follows from a recursive application of the  results before the last inequality, and the definition that $R_{0}(\boldsymbol{X}) = 0$.

By \eqref{dependence.7} and Jensen's inequality,
\begin{equation}
    \begin{split}
        \mathbb{E}|R_{s}(\boldsymbol{X})| & \le \mathbb{E}|f(\boldsymbol{X}) - R_{s}(\boldsymbol{X})| + \mathbb{E}|f(\boldsymbol{X})| \\        
        & \le 2\sqrt{\mathbb{E}[(f(\boldsymbol{X}))^2]} .
    \end{split}
\end{equation}


\subsection{Lemma~\ref{lemma.1} and its proof}

\begin{lemma}\label{lemma.1}
For each $l>0$ and set $J$ with $\texttt{\#}J\le 2$,
\begin{equation}\label{dependence.5.b}
    |\mathbb{E}(f(\boldsymbol{X}) - R_{l-1}^{\star}(\boldsymbol{X}) | \boldsymbol{X}_{\mathcal{X}(J)})| \le 3\sqrt{\mathbb{E}[(f(\boldsymbol{X}))^2]}  \times p_{\min}^{-1} \qquad \textnormal{almost surely,}
\end{equation} 
   where $p_{\min}>0$ is defined in Condition~\ref{signal.strength.1}. Additionally, for each $s>0$,
\begin{equation}\label{dependence.3.b}
    \iota_{s} \le s\times 3\sqrt{\mathbb{E}[(f(\boldsymbol{X}))^2]}  \times p_{\min}^{-1} ,
\end{equation}
where $\iota_{l}$'s are defined in \eqref{iota.1}.
\end{lemma}

\noindent\textit{Proof of Lemma~\ref{lemma.1}: }
 By the definition of $p_{\min}$ in Condition~\ref{signal.strength.1}, it holds that 
\begin{equation}\label{dependence.1.b}
    \min\{ \mathbb{P}(\boldsymbol{X}_{\mathcal{X}(J)} = \vv{c}) > 0:  \vv{c} \in \{0, 1\}^{\texttt{\#}\mathcal{X}(J)},\texttt{\#}J\le 2 \} \ge p_{\min}.
\end{equation}

By the law of total expectation, we have that for every $J\subset\{1, \dots, p\}$ with $\texttt{\#}J\le 2$,
\begin{equation}
    \begin{split} \label{dependence.2.b}
         \mathbb{E}(|f(\boldsymbol{X})|) & = \mathbb{E}[\mathbb{E}(|f(\boldsymbol{X})| | \boldsymbol{X}_{\mathcal{X}(J)})] \\
        & = \sum_{\vv{c} \in \{0, 1\}^{\texttt{\#}\mathcal{X}(J)}} \mathbb{P}(\boldsymbol{X}_{\mathcal{X}(J)} = \vv{c}) \times  \mathbb{E}(|f(\boldsymbol{X})| | \boldsymbol{X}_{\mathcal{X}(J)} = \vv{c}).
    \end{split}
\end{equation}
By  \eqref{dependence.1.b}--\eqref{dependence.2.b} and Jensen's inequality, for every $J\subset\{1, \dots, p\}$ with $\texttt{\#}J\le 2$, it holds that almost surely
$$\mathbb{E}(|f(\boldsymbol{X})| | \boldsymbol{X}_{\mathcal{X}(J)})\le (\mathbb{E}|f(\boldsymbol{X})|) \times (p_{\min})^{-1} \le  \sqrt{\mathbb{E}[(f(\boldsymbol{X}))^2]} \times p_{\min}^{-1}.$$
By a similar calculation and the result of \eqref{dependence.8} from Lemma~\ref{lemma.2}, we have that for each $l\ge 0$ and every set $J$ with $\texttt{\#}J \le 2$,
\begin{equation}
    \begin{split}
        \mathbb{E}(|R_{l}^{\star}(\boldsymbol{X})| | \boldsymbol{X}_{\mathcal{X}(J)}) & \le   2\sqrt{\mathbb{E}[(f(\boldsymbol{X}))^2]}  \times p_{\min}^{-1}.
    \end{split}
\end{equation}
Therefore, it holds that almost surely,
\begin{equation*}
    |\mathbb{E}(f(\boldsymbol{X}) - R_{l-1}^{\star}(\boldsymbol{X}) | \boldsymbol{X}_{\mathcal{X}(J)})| \le 3\sqrt{\mathbb{E}[(f(\boldsymbol{X}))^2]}  \times p_{\min}^{-1}
\end{equation*} 
for each $l>0$ and every set $J$ with $\texttt{\#}J\le 2$, which in combination with the definition of $\iota_{s}$ in \eqref{iota.1} concludes the two desired results of Lemma~\ref{lemma.1}.

\subsection{Lemma~\ref{lemma.3} and its proof}

\begin{lemma}
    \label{lemma.3}
    For each $s>0$ and every set $J\subset \{1, \dots, M\}$,
    \begin{equation}
        \begin{split}
        &\mathbb{E} (f(\boldsymbol{X}) - R_{s-1}^{\star}(\boldsymbol{X}))^2  - \mathbb{E} (f(\boldsymbol{X}) - R_{s, J}^{\star}(\boldsymbol{X}))^2 \\
        & =\mathbb{E}[(\mathbb{E}(f(\boldsymbol{X}) - R_{s-1}^{\star}(\boldsymbol{X})|\boldsymbol{X}_{\mathcal{X}(J)})  )^2]\\
        &=  \textnormal{Var}[\mathbb{E}( f(\boldsymbol{X}) - R_{s-1}^{\star}(\boldsymbol{X})|\boldsymbol{X}_{\mathcal{X}(J)}  )].
        \end{split}
    \end{equation}
\end{lemma}

   \noindent \textit{Proof of Lemma~\ref{lemma.3}: } 
    By the definitions of $R_{l}^{\star}$'s in \eqref{mp.3} and $R_{s, J}^{\star}$ in \eqref{predictor.trees.b.3}, the law of iterated expectation,
\begin{equation*}
\begin{split}\label{bias.4}
 & \mathbb{E} (f(\boldsymbol{X}) - R_{s, J}^{\star}(\boldsymbol{X}))^2  \\
 & = \mathbb{E} (f(\boldsymbol{X}) - R_{s-1}^{\star}(\boldsymbol{X}) -   \mathbb{E} (f(\boldsymbol{X}) - R_{s-1}^{\star}(\boldsymbol{X})| \boldsymbol{X}_{\mathcal{X}(J)}) )^2  \\
 & = \mathbb{E} (f(\boldsymbol{X}) - R_{s-1}^{\star}(\boldsymbol{X}))^2 \\
 &\qquad - 2 \mathbb{E}[ (f(\boldsymbol{X}) - R_{s-1}^{\star}(\boldsymbol{X})) \times  \mathbb{E}(f(\boldsymbol{X}) - R_{s-1}^{\star}(\boldsymbol{X})|\boldsymbol{X}_{\mathcal{X}(J)})  ]\\
 &\qquad + \mathbb{E}[(\mathbb{E}(f(\boldsymbol{X}) - R_{s-1}^{\star}(\boldsymbol{X})|\boldsymbol{X}_{\mathcal{X}(J)})  )^2]\\
 & = \mathbb{E} (f(\boldsymbol{X}) - R_{s-1}^{\star}(\boldsymbol{X}))^2 - \mathbb{E}[(\mathbb{E}(f(\boldsymbol{X}) - R_{s-1}^{\star}(\boldsymbol{X})|\boldsymbol{X}_{\mathcal{X}(J)})  )^2].
\end{split}
\end{equation*}

Lastly, since we have assumed $\mathbb{E}(f (\boldsymbol{X})) = 0$, and that
\begin{equation*}
    \begin{split}
        \mathbb{E}(R_{s}^{\star} (\boldsymbol{X})) & = \mathbb{E}(R_{s-1}^{\star} (\boldsymbol{X}) + \mathbb{E} ( f(\boldsymbol{X})- R_{s-1}^{\star} (\boldsymbol{X}) | \boldsymbol{X}_{\mathcal{X}(\widehat{J}_{s})}) ) \\
        & = \mathbb{E}(R_{s-1}^{\star} (\boldsymbol{X}) - \mathbb{E} (  R_{s-1}^{\star} (\boldsymbol{X}) | \boldsymbol{X}_{\mathcal{X}(\widehat{J}_{s})}) ) + \mathbb{E} ( f(\boldsymbol{X})) \\
        & = \mathbb{E}(R_{s-1}^{\star} (\boldsymbol{X})) - \mathbb{E} (  R_{s-1}^{\star} (\boldsymbol{X})) \\
        & = 0
    \end{split}
\end{equation*} 
for each $s>0$, we deduce that 
$$\mathbb{E}[(\mathbb{E}(f(\boldsymbol{X}) - R_{s-1}^{\star}(\boldsymbol{X})|\boldsymbol{X}_{\mathcal{X}(J)})  )^2] = \textnormal{Var}[\mathbb{E}( f(\boldsymbol{X}) - R_{s-1}^{\star}(\boldsymbol{X})|\boldsymbol{X}_{\mathcal{X}(J)}  )] ,$$
which conclude the proof of Lemma~\ref{lemma.3}.

\subsection{Lemma~\ref{lemma.4} and its proof}

\begin{lemma}\label{lemma.4}
    For every $\{j, l, k\}\subset \{1, \dots, M\}$, it holds almost surely that
    \begin{equation*}
        |\mathbb{E}\{  g_{lk}(\boldsymbol{X}) -g_{l}(\boldsymbol{X}) -g_{k}(\boldsymbol{X})|\boldsymbol{X}_{\mathcal{X}(lj)}\} | \le \frac{8 \delta_{0}  \sqrt{\mathbb{E}[(f(\boldsymbol{X}))^2]}}{ p_{\min} }  ,
    \end{equation*}
    where $g_{J}(\boldsymbol{X})$ for each $J\subset \{1, \dots, M\}$ is defined in Section~\ref{Sec3qq}, and $p_{\min}$ is defined in Condition~\ref{signal.strength.1}.
\end{lemma}

    \noindent\textit{Proof of Lemma~\ref{lemma.4}: }
Recall that $g_{lk}(\boldsymbol{X}) = \mathbb{E}(f(\boldsymbol{X})  - \mathbb{E} (f(\boldsymbol{X})| \boldsymbol{X}_{-\mathcal{X}(lk)}) |\boldsymbol{X}_{\mathcal{X}(lk)})$, and that $\mathcal{X}(jl) = \mathcal{X}(j) \cup \mathcal{X}(l)$. For every $\{j, l, k\}\subset \{1, \dots, M\}$, we deduce that 
\begin{equation}
    \begin{split}\label{dependence.21}
        & \mathbb{E}\{  g_{lk}(\boldsymbol{X}) -g_{l}(\boldsymbol{X}) -g_{k}(\boldsymbol{X})|\boldsymbol{X}_{\mathcal{X}(lj)}\} \\ 
        & = \mathbb{E}\Big\{ \mathbb{E}\left[f(\boldsymbol{X}) - \mathbb{E} (f(\boldsymbol{X})| \boldsymbol{X}_{-\mathcal{X}(lk)}) |\boldsymbol{X}_{\mathcal{X}(lk)}\right]
        \\
        & \qquad - \mathbb{E}\left[f(\boldsymbol{X}) - \mathbb{E} (f(\boldsymbol{X})| \boldsymbol{X}_{-\mathcal{X}(l)}) | \boldsymbol{X}_{\mathcal{X}(l)}\right]
        |\boldsymbol{X}_{\mathcal{X}(lj)} \Big\} - \mathbb{E}\{g_{k}(\boldsymbol{X})|\boldsymbol{X}_{\mathcal{X}(lj)}\} \\
        & = \mathbb{E}\left\{ \mathbb{E}(f(\boldsymbol{X}) |\boldsymbol{X}_{\mathcal{X}(lk)}) - \mathbb{E}(f(\boldsymbol{X}) | \boldsymbol{X}_{\mathcal{X}(l)}) |\boldsymbol{X}_{\mathcal{X}(lj)} \right\} \\
        &\qquad + \mathbb{E}\left\{ \mathbb{E} [\mathbb{E} (f(\boldsymbol{X})| \boldsymbol{X}_{-\mathcal{X}(l)}) | \boldsymbol{X}_{\mathcal{X}(l)} ]  - \mathbb{E} [ \mathbb{E} (f(\boldsymbol{X})| \boldsymbol{X}_{-\mathcal{X}(lk)}) | \boldsymbol{X}_{\mathcal{X}(lk)} ] |\boldsymbol{X}_{\mathcal{X}(lj)} \right\} \\
        & \qquad - \mathbb{E}\{g_{k}(\boldsymbol{X})|\boldsymbol{X}_{\mathcal{X}(lj)}\}.
    \end{split}
\end{equation}

Let us deal with the first term on the RHS of \eqref{dependence.21}. By Lemma~\ref{lemma.5} (where $j$ corresponds to $l$ in \eqref{dependence.18.c}) and Jensen's inequality,
\begin{equation}
    \begin{split}\label{dependence.20}
        & | \mathbb{E}\left\{ \mathbb{E}(f(\boldsymbol{X}) |\boldsymbol{X}_{\mathcal{X}(lk)})  |\boldsymbol{X}_{\mathcal{X}(lj)} \right\} - \mathbb{E}\left\{ \mathbb{E}(f(\boldsymbol{X}) |\boldsymbol{X}_{\mathcal{X}(lk)})  |\boldsymbol{X}_{\mathcal{X}(l)} \right\}| \\
        & \le   \frac{2\delta_{0}  \mathbb{E}|\mathbb{E}(f(\boldsymbol{X}) |\boldsymbol{X}_{\mathcal{X}(lk)})| }{p_{\min} }\\
        & \le \frac{2\delta_{0}  \mathbb{E}|f(\boldsymbol{X})| }{p_{\min}}.
    \end{split}
\end{equation}
In addition, by the law of iterated expectation,
\begin{equation}
    \begin{split}\label{dependence.24}
        & \mathbb{E}\left\{ \mathbb{E}(f(\boldsymbol{X}) |\boldsymbol{X}_{\mathcal{X}(lk)})  |\boldsymbol{X}_{\mathcal{X}(l)} \right\} =  \mathbb{E}(f(\boldsymbol{X}) |\boldsymbol{X}_{\mathcal{X}(l)}) ,\\
        & \mathbb{E}\left\{ \mathbb{E}(f(\boldsymbol{X}) | \boldsymbol{X}_{\mathcal{X}(l)}) |\boldsymbol{X}_{\mathcal{X}(lj)} \right\}= \mathbb{E}(f(\boldsymbol{X}) | \boldsymbol{X}_{\mathcal{X}(l)}).
    \end{split}
\end{equation}
Hence, by \eqref{dependence.20}--\eqref{dependence.24},
\begin{equation}\label{dependence.25}
\begin{split}
    & \left| \mathbb{E}\left\{ \mathbb{E}(f(\boldsymbol{X}) |\boldsymbol{X}_{\mathcal{X}(lk)}) - \mathbb{E}(f(\boldsymbol{X}) | \boldsymbol{X}_{\mathcal{X}(l)}) |\boldsymbol{X}_{\mathcal{X}(lj)} \right\} \right| \\
& = \bigg| \mathbb{E}\left\{ \mathbb{E}(f(\boldsymbol{X}) |\boldsymbol{X}_{\mathcal{X}(lk)})|\boldsymbol{X}_{\mathcal{X}(lj)} \right\} - \mathbb{E}\left\{ \mathbb{E}(f(\boldsymbol{X}) |\boldsymbol{X}_{\mathcal{X}(lk)})|\boldsymbol{X}_{\mathcal{X}(l)} \right\} \\
& \qquad + \mathbb{E}\left\{ \mathbb{E}(f(\boldsymbol{X}) |\boldsymbol{X}_{\mathcal{X}(lk)})|\boldsymbol{X}_{\mathcal{X}(l)} \right\} - \mathbb{E}\left\{ \mathbb{E}(f(\boldsymbol{X}) | \boldsymbol{X}_{\mathcal{X}(l)}) |\boldsymbol{X}_{\mathcal{X}(lj)} \right\} \bigg| \\    
    &\le \frac{2\delta_{0}  \mathbb{E}|f(\boldsymbol{X})| }{p_{\min}}.
    \end{split}
\end{equation}

Now, let us deal with the second term  on the RHS of \eqref{dependence.21}.  By Lemma~\ref{lemma.5} and Jensen's inequality,
\begin{equation*}
    \begin{split}
        \mathbb{E}\left\{  \mathbb{E} (f(\boldsymbol{X})| \boldsymbol{X}_{-\mathcal{X}(lk)}) |\boldsymbol{X}_{\mathcal{X}(lk)} \right\} & \le \frac{3\delta_{0}\mathbb{E}|f(\boldsymbol{X})|}{ p_{\min} },\\
        \mathbb{E}\left\{  \mathbb{E} (f(\boldsymbol{X})| \boldsymbol{X}_{-\mathcal{X}(l)}) |\boldsymbol{X}_{\mathcal{X}(l)} \right\} & \le  \frac{\delta_{0}\mathbb{E}|f(\boldsymbol{X})|}{ p_{\min} },
    \end{split}
\end{equation*}
implying that almost surely
\begin{equation}
    \label{dependence.27}
    \begin{split}        
     & \mathbb{E}\left\{ \mathbb{E} [\mathbb{E} (f(\boldsymbol{X})| \boldsymbol{X}_{-\mathcal{X}(l)}) | \boldsymbol{X}_{\mathcal{X}(l)} ]  - \mathbb{E} [ \mathbb{E} (f(\boldsymbol{X})| \boldsymbol{X}_{-\mathcal{X}(lk)}) | \boldsymbol{X}_{\mathcal{X}(lk)} ] |\boldsymbol{X}_{\mathcal{X}(lj)} \right\} \\
     & \le \frac{4 \delta_{0}\mathbb{E}|f(\boldsymbol{X})|}{ p_{\min} }.
     \end{split}
\end{equation}

For the third term on the RHS of \eqref{dependence.21}, by Lemma~\ref{lemma.5} and \eqref{bias.control.6}, we deduce that
\begin{equation}\label{dependence.28}
    |\mathbb{E}\{g_{k}(\boldsymbol{X})|\boldsymbol{X}_{\mathcal{X}(lj)}\}| \le  \frac{\delta_{0}\mathbb{E}|g_{k}(\boldsymbol{X})|}{ p_{\min} } \le \frac{2 \delta_{0}  \sqrt{\mathbb{E}[(f(\boldsymbol{X}))^2]}}{ p_{\min} }   .
\end{equation}

By \eqref{dependence.21}, \eqref{dependence.25}--\eqref{dependence.28}, and Jensen's inequality, it holds almost surely that 
\begin{equation}
    \begin{split}\label{boundness.bias.1}
        & \left| \mathbb{E}\{  g_{lk}(\boldsymbol{X}) -g_{l}(\boldsymbol{X}) -g_{k}(\boldsymbol{X})|\boldsymbol{X}_{\mathcal{X}(lj)}\} \right| \\
        & \le \frac{2\delta_{0}  \mathbb{E}|f(\boldsymbol{X})| }{p_{\min}} + \frac{4 \delta_{0}\mathbb{E}|f(\boldsymbol{X})|}{ p_{\min} } + \frac{2 \delta_{0}  \sqrt{\mathbb{E}[(f(\boldsymbol{X}))^2]}}{ p_{\min} }   \\
        & \le \frac{8 \delta_{0}  \sqrt{\mathbb{E}[(f(\boldsymbol{X}))^2]}}{ p_{\min} }  .
    \end{split}
\end{equation}

\subsection{Lemma~\ref{lemma.event.1} and its proof}\label{SecC.5}

Let $E_{41}, \dots, E_{46}$ be defined as in \eqref{event.2}.
\begin{lemma}
    \label{lemma.event.1} 
    Assume Condition~\ref{model.1}, that $(\boldsymbol{X}_{1}, Y_{1}), \dots, (\boldsymbol{X}_{n}, Y_{n}), (\boldsymbol{X}, Y)$ are i.i.d., and that $\mathbb{E}(\varepsilon \times \boldsymbol{1}\{|\varepsilon| \le n^{\frac{1}{q_{1}}}\} |\boldsymbol{X}) = 0$ for each $n>0$, in which $q_{1}$ is an arbitrary constant with $q_{0} > q_{1} > 0$.
  Assume $M = O(n^{K_{01}})$ for an arbitrary constant $K_{01}>0$, $t_{1}=t_2 = \log{(n)}n^{\frac{1}{2} + \frac{1}{q_{1}}}$, $t_3 = t_4 = t_{6}= \log{(n)}\sqrt{n}$, and $t_{5} = 2n\mathbb{E}\varepsilon^2$. Then, it holds that $\mathbb{P}((\cap_{l=1}^6 E_{4l})^c) = o(1)$, where $A^c$ denotes the complementary event of event $A$.
\end{lemma}

\noindent\textit{Proof of Lemma~\ref{lemma.event.1}: }
Define $\epsilon_{i} = \varepsilon_{i}\times \boldsymbol{1}\{|\varepsilon_{i}| \le n^{\frac{1}{q_{1}}}\}$, and that $E_{4} = \cap_{i=1}^n\{ \epsilon_{i} = \varepsilon_{i} \} $. By assumption,
\begin{equation}
    \label{model.error.2}
    \mathbb{E}(\epsilon_{i}|\boldsymbol{X}_{i}) = 0,
\end{equation}
and that 
\begin{equation}
    \begin{split}\label{error.distribution.1}
        \mathbb{P}(\cup_{i=1}^n\{ \epsilon_{i} \not = \varepsilon_{i} \} ) & \le \sum_{i=1}^n \mathbb{P}(|\varepsilon_{i}| \ge n^{\frac{1}{q_{1} }} )  \\
        & = \sum_{i=1}^n \mathbb{P}(|\varepsilon_{i}|^{q_{0}}  \ge n^{\frac{q_{0}}{q_{1} }} ) \\
        & \le n^{1-\frac{q_{0}}{q_{1}}} \times \mathbb{E} |\varepsilon|^{q_{0}}.
    \end{split}
\end{equation}

The above results of \eqref{model.error.2} and \eqref{error.distribution.1} are required for establishing the upper bound of $\mathbb{P}((\cap_{l=1}^6 E_{4l})^c)$ as follows. Let us deduce that
\begin{equation}
    \begin{split}
        \label{lemma.event.2}
        & \mathbb{P}((\cap_{l=1}^6 E_{4l})^c) \\
        & \le \mathbb{P}(E_{4}^c) + \mathbb{P}(E_{4}\cap (\cap_{l=1}^6 E_{4l})^c)\\
        & \le \mathbb{P}(E_{4}^c) + \mathbb{P}\left(\left\{\left|\sum_{i=1}^n \epsilon_{i} f(\boldsymbol{X}_{i})\right| > t_{1} \right\}\right)\\
        & \qquad + \mathbb{P}\left(  \cup_{\texttt{\#}J\le 2, \vv{c}\in \{0, 1\}^{\texttt{\#}\mathcal{X}(J)}}\left\{ \left| \sum_{i=1}^n \epsilon_{i} \boldsymbol{1}_{\boldsymbol{X}_{i \mathcal{X}(J)} = \vv{c}} \right| >  t_{2 }\right\} \right) + \mathbb{P}(E_{43}^c) + \mathbb{P}(E_{44}^c) \\
        & \qquad + \mathbb{P}\left(\left|\sum_{i=1}^n \epsilon_{i}^2\right| > t_{5}  \right) + \mathbb{P}(E_{46}^c).
    \end{split}
\end{equation}

An upper bound on the probability $ \mathbb{P}(E_{4}^c)$ has been established in \eqref{error.distribution.1}. Additional probability upper bounds \eqref{additiona.probability.1}--\eqref{error.distribution.2} are derived as follows. By Hoeffding's inequality, that 
$$- n^{q_{1}^{-1}} \max_{\vv{c} \in \{0, 1\}^{p}} |f(\vv{c})|\le \epsilon_{i}f(\boldsymbol{X}_{i}) \le n^{q_{1}^{-1}} \max_{\vv{c} \in \{0, 1\}^{p}} |f(\vv{c})|,$$ 
and that $\mathbb{E} (\epsilon_{i} f(\boldsymbol{X}_{i})) = 0$ due to \eqref{model.error.2}, it holds that for each $t_{1}>0$,
\begin{equation}
    \begin{split}\label{additiona.probability.1}
         \mathbb{P}\left( \left\{\left|\sum_{i=1}^n \epsilon_{i} f(\boldsymbol{X}_{i})\right| \ge t_{1} \right\}\right)  \le \exp\left( \frac{-2t_{1}^2}{4 n^{1+\frac{2}{q_{1}}} \max_{\vv{c} \in \{0, 1\}^{p}} |f(\vv{c})|^2 }\right).
    \end{split}
\end{equation}

Let us move on to the next probability upper bound. By Hoeffding's inequality, that $- n^{q_{1}^{-1}} \le \epsilon_{i} \boldsymbol{1}_{\boldsymbol{X}_{i \mathcal{X}(J)} = \vv{c}} \le n^{q_{1}^{-1}} $ for every set $J$ and $\vv{c} \in \{0, 1\}^{\texttt{\#}\mathcal{X}(J)}$, and that $\mathbb{E} (\epsilon_{i} \boldsymbol{1}_{\boldsymbol{X}_{i \mathcal{X}(J)} = \vv{c}} ) = 0$ due to \eqref{model.error.2}, it holds that for each $t_{2}>0$,
\begin{equation}
    \begin{split}
        & \mathbb{P}\left(  \cup_{\texttt{\#}J\le 2, \vv{c}\in \{0, 1\}^{\texttt{\#}\mathcal{X}(J)}}\left\{ \left| \sum_{i=1}^n \epsilon_{i} \boldsymbol{1}_{\boldsymbol{X}_{i \mathcal{X}(J)} = \vv{c}} \right| >  t_{2 }\right\} \right) \\
        & \le \sum_{\texttt{\#}J\le 2, \vv{c}\in \{0, 1\}^{\texttt{\#}\mathcal{X}(J)}} \mathbb{P}\left(  \left\{ \left| \sum_{i=1}^n \epsilon_{i} \boldsymbol{1}_{\boldsymbol{X}_{i \mathcal{X}(J)} = \vv{c}} \right| >  t_{2 }\right\} \right) \\
        & \le 2^{2M_{\mathcal{X}}} \times M^2\times \exp\left( \frac{-2t_{2}^2}{4 n^{1+\frac{2}{q_{1}}}  }\right).
    \end{split}
\end{equation}

Let us move on to the next probability upper bound. By Hoeffding's inequality, that $- \max_{\vv{c} \in \{0, 1\}^{p}} |f(\vv{c})| \le f(\boldsymbol{X}_{i}) \boldsymbol{1}_{\boldsymbol{X}_{i\mathcal{X}(J)} = \vv{c}} \le \max_{\vv{c} \in \{0, 1\}^{p}} |f(\vv{c})|$ for every set $J$ and $\vv{c} \in \{0, 1\}^{\texttt{\#}\mathcal{X}(J)}$, it holds that for each $t_{3}>0$,
\begin{equation}
    \begin{split}
        & \mathbb{P}\left( \cup_{\texttt{\#}J\le 2, \vv{c}\in \{0, 1\}^{\texttt{\#}\mathcal{X}(J)}} \left\{ |\sum_{i=1}^n \left(f(\boldsymbol{X}_{i}) \boldsymbol{1}_{\boldsymbol{X}_{i\mathcal{X}(J)} = \vv{c}} -\mathbb{E}(f(\boldsymbol{X}) \boldsymbol{1}_{\boldsymbol{X}_{\mathcal{X}(J)} = \vv{c}} )\right) | > t_{3}\right\}\right)\\
        & \le \sum_{\texttt{\#}J\le 2, \vv{c}\in \{0, 1\}^{\texttt{\#}\mathcal{X}(J)} }\mathbb{P}\left( \left|\sum_{i=1}^n \left(f(\boldsymbol{X}_{i}) \boldsymbol{1}_{\boldsymbol{X}_{i\mathcal{X}(J)} = \vv{c}} -\mathbb{E}(f(\boldsymbol{X}) \boldsymbol{1}_{\boldsymbol{X}_{\mathcal{X}(J)} = \vv{c}} )\right) \right| > t_{3}\right)\\
        & \le 2^{2M_{\mathcal{X}}} \times M^2\times \exp\left( \frac{-2t_{3}^2}{4 n \max_{\vv{c} \in \{0, 1\}^{p}} |f(\vv{c})|^2 }  \right).
    \end{split}
\end{equation}

Let us move on to the next probability upper bound. By Hoeffding's inequality, that $-1\le \boldsymbol{1}_{\boldsymbol{X}_{i\mathcal{X}(J)} = \vv{c}} -\mathbb{P}(\boldsymbol{X}_{\mathcal{X}(J)} = \vv{c} ) \le 1$ for every set $J$ and $\vv{c} \in \{0, 1\}^{\texttt{\#}\mathcal{X}(J)}$, it holds that for each $t_{4}>0$,
\begin{equation}
    \begin{split}
        & \mathbb{P}\left( \cup_{\texttt{\#}J\le 4, \vv{c}\in \{0, 1\}^{\texttt{\#}\mathcal{X}(J)}} \left\{ |\sum_{i=1}^n \left( \boldsymbol{1}_{\boldsymbol{X}_{i\mathcal{X}(J)} = \vv{c}} -\mathbb{P}(\boldsymbol{X}_{\mathcal{X}(J)} = \vv{c} )\right) | > t_{4}\right\} \right) \\
        & \le \sum_{\texttt{\#}J\le 4, \vv{c}\in \{0, 1\}^{\texttt{\#}\mathcal{X}(J)}} \mathbb{P}\left( \left|\sum_{i=1}^n \left( \boldsymbol{1}_{\boldsymbol{X}_{i\mathcal{X}(J)} = \vv{c}} -\mathbb{P}(\boldsymbol{X}_{\mathcal{X}(J)} = \vv{c} )\right) \right| > t_{4}\right)\\
        & \le 2^{4M_{\mathcal{X}}} \times  M^4 \times \exp\left( \frac{-2t_{4}^2}{ 4n  }  \right).
    \end{split}
\end{equation}

Let us move on to the next probability upper bound. By Hoeffding's inequality, that $-n^{2q_{1}^{-1}} \le \epsilon_{i}^2 - \mathbb{E}(\epsilon_{i})^2 \le n^{2q_{1}^{-1}}$, it holds that for each $t_{5}>0$,
\begin{equation}
    \begin{split}
        & \mathbb{P}\left(\left\{\left|\sum_{i=1}^n (\epsilon_{i}^2 -\mathbb{E}(\epsilon_{1}^2) + \mathbb{E}(\epsilon_{1}^2))\right| > t_{5} \right\} \right) \\
        & \le \mathbb{P}\left(\left\{\left|\sum_{i=1}^n (\epsilon_{i}^2 -\mathbb{E}(\epsilon_{1}^2) \right| > t_{5} - n \mathbb{E}(\epsilon_{1}^2)) \right\} \right) \\
        & \le \exp\left( \frac{-2(t_{5} - n \mathbb{E}(\epsilon_{1}^2))^2}{ 4n^{1+ 4q_{1}^{-1}}  }  \right).
    \end{split}
\end{equation}
In addition, simple calculations show that 
\begin{equation}
    \mathbb{E}\epsilon_{1}^2 = \mathbb{E}\left[ \varepsilon_{1}^2\boldsymbol{1}\{|\varepsilon_{1}|\le n^{q_{1}^{-1}} \} \right] \le \mathbb{E}\varepsilon_{1}^2 = \mathbb{E}\varepsilon^2.
\end{equation}

Let us move on to the next probability upper bound. By Hoeffding's inequality and that $-\max_{\vv{c} \in \mathbb{R}^p} |f(\vv{c})|^2\le (f(\boldsymbol{X}))^2 - \mathbb{E}(f(\boldsymbol{X}))^2 \le \max_{\vv{c} \in \mathbb{R}^p} |f(\vv{c})|^2$ almost surely, it holds that for each $t_{6}> 0$,
\begin{equation}
    \begin{split}\label{error.distribution.2}
        \mathbb{P}\left(\left| \sum_{i=1}^n [(f(\boldsymbol{X}_{i}))^2 - \mathbb{E}(f(\boldsymbol{X}))^2 ] \right| > t_{6} \right)  \le \exp\left(\frac{ -2t_{6}^2 }{4n \times \max_{\vv{c} \in \mathbb{R}^p} |f(\vv{c})|^4  } \right).
    \end{split}
\end{equation}

By \eqref{error.distribution.1}--\eqref{error.distribution.2}, the assumed regularity conditions, and the choice of $t_{1}, t_{2}, t_{3}, t_{4}, t_{5}$, and $t_{6}$, we conclude the desired result of Lemma~\ref{lemma.event.1} that the probability $\mathbb{P}((\cap_{l=1}^6 E_{4l})^c)$ becomes asymptotically negligible as the number of samples increases.

\subsection{Lemma~\ref{lemma.5} and its proof}\label{SecC.6}

\begin{lemma}
    \label{lemma.5}
    For each $m\in \{1, \dots, M\}$ and every measurable function $q:\mathbb{R}^{p- \texttt{\#}\mathcal{X}(m)}\mapsto\mathbb{R}$,
\begin{equation}\label{dependence.9}
|\mathbb{E}(q(\boldsymbol{X}_{-\mathcal{X}(m)})) - \mathbb{E}(q(\boldsymbol{X}_{-\mathcal{X}(m)}) | \boldsymbol{X}_{\mathcal{X}(m)}) | \le \frac{\delta_{0} \mathbb{E}|q(\boldsymbol{X}_{-\mathcal{X}(m)})|}{ \min_{1\le m\le M, l\in \mathcal{X}(m), a \in \{0, 1\}} \mathbb{P}(X_{l}=a) } ,
\end{equation}
where $\delta_0$ is given in Condition~\ref{signal.strength.1}, and the result holds almost surely. In addition, for every $\{l, k\}\subset\{1, \dots, M\}$ and every measurable function $q:\mathbb{R}^{p- \texttt{\#}\mathcal{X}(l)}\mapsto\mathbb{R}$,
\begin{equation}
    \begin{split}\label{dependence.18.c}
        & |\mathbb{E}(q( \boldsymbol{X}_{-\mathcal{X}(l)}) |\boldsymbol{X}_{\mathcal{X}(k)}, \boldsymbol{X}_{\mathcal{X}(l)} ) - \mathbb{E}(q( \boldsymbol{X}_{-\mathcal{X}(l)})| \boldsymbol{X}_{\mathcal{X}(k)} ) |   \le \frac{2\delta_{0}  \mathbb{E}|q(\boldsymbol{X}_{-\mathcal{X}(l)})| }{p_{\min}},
    \end{split}
\end{equation}
where $p_{\min}$ is given in Condition~\ref{signal.strength.1}, and the result holds almost surely. Moreover, for every $\{l, k\}\subset\{1, \dots, M\}$ and every measurable function $q:\mathbb{R}^{p- \texttt{\#}\mathcal{X}(lk)}\mapsto\mathbb{R}$, it holds almost surely that 
\begin{equation}
    \begin{split}\label{dependence.19}
    & |\mathbb{E}(q( \boldsymbol{X}_{-\mathcal{X}(kl)}) |\boldsymbol{X}_{\mathcal{X}(k)}, \boldsymbol{X}_{\mathcal{X}(l)}) - \mathbb{E}(q( \boldsymbol{X}_{-\mathcal{X}(kl)}) ) |  \le \frac{3\delta_{0}  \mathbb{E}|q(\boldsymbol{X}_{-\mathcal{X}(kl)})| }{p_{\min} }.
    \end{split}
\end{equation}

\end{lemma}

\noindent\textit{Proof of Lemma~\ref{lemma.5}: }
For the reader's convenience, let us reiterate the definition of $\delta_{0}$ in Condition~\ref{signal.strength.1} as follows.
$$\delta_{0} =  \max_{1\le m\le M}\inf \left\{\delta\ge 0 : \mathbb{P} \left( \max_{i\in\mathcal{X}(m)}| \mathbb{P}(X_{i} = 1|\boldsymbol{X}_{-\mathcal{X}(m)})  - \mathbb{P}(X_{i} = 1) | \le \delta \right) = 1 \right\}. $$

For each $1\le m\le M$, each $l\in\mathcal{X}(m)$, and each $\vv{c} \in \{0,1\}^{p-\texttt{\#} \mathcal{X}(m)}$, let $\varepsilon_{\vv{c}, l}\in\mathbb{R}$  be a real constant such that 
$\mathbb{P}(X_{l} = 1 | \boldsymbol{X}_{-\mathcal{X}(m)} = \vv{c}) = \mathbb{P}( X_{l} = 1 ) + \varepsilon_{\vv{c}, l}$. By the definition of $\delta_{0}$, it holds that for each $\vv{c} \in \{0,1\}^{p-\texttt{\#} \mathcal{X}(m)}$ and every $l\in\mathcal{X}(m)$,
\begin{equation}
    \label{dependence.1}
    |\varepsilon_{\vv{c}, l}| \le \delta_{0}.
\end{equation}

Since features are binary with $\sum_{j\in \mathcal{X}(m)} \boldsymbol{1}_{X_{j} = 1} = 1$ when $\texttt{\#}\mathcal{X}(m) > 1$, on $\{X_{l} = 1\}$ for some $l\in \mathcal{X}(m)$, 
\begin{equation}
    \begin{split}\label{dependence.2}
        &\mathbb{E}(q(\boldsymbol{X}_{-\mathcal{X}(m)})) - \mathbb{E}(q(\boldsymbol{X}_{-\mathcal{X}(m)}) | \boldsymbol{X}_{\mathcal{X}(m)}) \\
        & = \mathbb{E}(q(\boldsymbol{X}_{-\mathcal{X}(m)})) - \frac{ \mathbb{E}(q(\boldsymbol{X}_{-\mathcal{X}(m)}) \boldsymbol{1}_{ X_{l}=1} ) }{\mathbb{P}(X_{l}=1)} \\
        & = \mathbb{E}(q(\boldsymbol{X}_{-\mathcal{X}(m)})) - \frac{ \sum_{\vv{c} \in \{0, 1\}^{p-\texttt{\#}\mathcal{X}(m)}} q(\vv{c}) \times\mathbb{E}( \boldsymbol{1}_{\boldsymbol{X}_{-\mathcal{X}(m)} = \vv{c}}\boldsymbol{1}_{ X_{l}=1} ) }{\mathbb{P}(X_{l}=1)}\\
        & = \mathbb{E}(q(\boldsymbol{X}_{-\mathcal{X}(m)})) - \sum_{\vv{c} \in \{0, 1\}^{p-\texttt{\#}\mathcal{X}(m)}} \frac{  \mathbb{P}(\boldsymbol{X}_{-\mathcal{X}(m)} = \vv{c})\times q(\vv{c}) \times\mathbb{E}( \boldsymbol{1}_{\boldsymbol{X}_{-\mathcal{X}(m)} = \vv{c}}\boldsymbol{1}_{ X_{l}=1} ) }{\mathbb{P}(\boldsymbol{X}_{-\mathcal{X}(m)} = \vv{c})\times \mathbb{P}(X_{l}=1)}\\
        & = \mathbb{E}(q(\boldsymbol{X}_{-\mathcal{X}(m)})) - \sum_{\vv{c} \in \{0, 1\}^{p-\texttt{\#}\mathcal{X}(m)}} \frac{  \mathbb{P}(\boldsymbol{X}_{-\mathcal{X}(m)} = \vv{c})\times q(\vv{c}) \times(\mathbb{P}(X_{l}=1) + \varepsilon_{\vv{c}, l} ) }{\mathbb{P}(X_{l}=1)} \\
        & =  - \sum_{\vv{c} \in \{0, 1\}^{p-\texttt{\#}\mathcal{X}(m)}} \frac{  \mathbb{P}(\boldsymbol{X}_{-\mathcal{X}(m)} = \vv{c})\times q(\vv{c}) \times \varepsilon_{\vv{c}, l} }{\mathbb{P}(X_{l}=1)}.
    \end{split}
\end{equation}

By \eqref{dependence.1} and \eqref{dependence.2}, it holds that on $\{X_{l} = 1\}$,
\begin{equation}
\begin{split}\label{dependence.4}
    & |\mathbb{E}(q(\boldsymbol{X}_{-\mathcal{X}(m)})) - \mathbb{E}(q(\boldsymbol{X}_{-\mathcal{X}(m)}) | \boldsymbol{X}_{\mathcal{X}(m)})| \\
    & \le \sum_{\vv{c} \in \{0, 1\}^{p-\texttt{\#}\mathcal{X}(m)}} \frac{  \mathbb{P}(\boldsymbol{X}_{-\mathcal{X}(m)} = \vv{c})\times |q(\vv{c})| \times \delta_{0} }{\mathbb{P}(X_{l}=1)} \\
    & \le \delta_{0} \times \frac{\mathbb{E}|q(\boldsymbol{X}_{-\mathcal{X}(m)})| }{\mathbb{P}(X_{l}=1)}.
    \end{split}
\end{equation}

By similar arguments for \eqref{dependence.4}, it holds almost surely that for each measurable $q:\mathbb{R}^{p-\texttt{\#}\mathcal{X}(m)} \mapsto\mathbb{R}$  and $\texttt{\#}\mathcal{X}(m) > 1$,
\begin{equation}\label{dependence.6}
|\mathbb{E}(q(\boldsymbol{X}_{-\mathcal{X}(m)})) - \mathbb{E}(q(\boldsymbol{X}_{-\mathcal{X}(m)}) | \boldsymbol{X}_{\mathcal{X}(m)}) | \le \delta_{0} \times \frac{\mathbb{E}|q(\boldsymbol{X}_{-\mathcal{X}(m)})|}{ \min_{l\in \mathcal{X}(m)} \mathbb{P}(X_{l}=1) } .
\end{equation}
By similar arguments, it holds  almost surely that for $\texttt{\#}\mathcal{X}(m) =1$,
\begin{equation*}
|\mathbb{E}(q(\boldsymbol{X}_{-\mathcal{X}(m)})) - \mathbb{E}(q(\boldsymbol{X}_{-\mathcal{X}(m)}) | \boldsymbol{X}_{\mathcal{X}(m)}) | \le \delta_{0} \times \frac{\mathbb{E}|q(\boldsymbol{X}_{-\mathcal{X}(m)})|}{ \mathbb{P}(X_{l}=1)\wedge \mathbb{P}(X_{l}=0) } .
\end{equation*}

Therefore, we deduce almost surely that for each $m\in \{1, \dots, M\}$,
\begin{equation*}
|\mathbb{E}(q(\boldsymbol{X}_{-\mathcal{X}(m)})) - \mathbb{E}(q(\boldsymbol{X}_{-\mathcal{X}(m)}) | \boldsymbol{X}_{\mathcal{X}(m)}) | \le \delta_{0} \times \frac{\mathbb{E}|q(\boldsymbol{X}_{-\mathcal{X}(m)})|}{ \min_{1\le m\le M, l\in \mathcal{X}(m), a \in \{0, 1\}} \mathbb{P}(X_{l}=a) } ,
\end{equation*}
which concludes the proof of the first assertion of Lemma~\ref{lemma.5}.

We proceed to deal with the cases involving two groups. For every $\{l, k\}\subset \{1, \dots, M\}$, $\vv{c}_{l} \in \{0, 1\}^{\texttt{\#} \mathcal{X}(l)}$, and $\vv{c}_{k} \in \{0, 1\}^{\texttt{\#} \mathcal{X}(k)}$, it holds that
\begin{equation}
    \begin{split}\label{dependence.11}
        &|\mathbb{P}( \boldsymbol{X}_{\mathcal{X}(k)} = \vv{c}_{k}, \boldsymbol{X}_{\mathcal{X}(l)} = \vv{c}_{l})  - \mathbb{P} (\boldsymbol{X}_{\mathcal{X}(l)}  = \vv{c}_{l}) \mathbb{P} (\boldsymbol{X}_{\mathcal{X}(k)}  = \vv{c}_{k})| \\
        & = | \eta_{1}| \times \mathbb{P} (\boldsymbol{X}_{\mathcal{X}(k)}  = \vv{c}_{k}) \\
        & \le \delta_{0}\times \mathbb{P} (\boldsymbol{X}_{\mathcal{X}(k)}  = \vv{c}_{k}),
    \end{split}
\end{equation}
where $\eta_{1} = \mathbb{P}(\boldsymbol{X}_{\mathcal{X}(l)} = \vv{c}_{l} | \boldsymbol{X}_{\mathcal{X}(k)}  = \vv{c}_{k}) - \mathbb{P} (\boldsymbol{X}_{\mathcal{X}(l)}  = \vv{c}_{l})$, and the inequality is due to the definition of $\delta_{0}$. Recall that we have defined the probability conditional  on an event with zero probability to be zero.

By \eqref{dependence.11} and similar arguments for \eqref{dependence.4}, it holds that for every $\{l, k\} \subset\{1, \dots, M\}$, every pair of one-hot vectors $\vv{c}_{k}\in\mathbb{R}^{\texttt{\#} \mathcal{X}(k)}$ and $\vv{c}_{l}\in\mathbb{R}^{\texttt{\#} \mathcal{X}(l)}$ with  $\mathbb{P}(\boldsymbol{X}_{\mathcal{X}(k)}=\vv{c}_{k}, \boldsymbol{X}_{\mathcal{X}(l)}=\vv{c}_{l}) > 0$, and every measurable function $q:\mathbb{R}^{p- \texttt{\#}\mathcal{X}(l)}\mapsto\mathbb{R}$,
\begin{equation}
    \begin{split}\label{dependence.18.b}
    & |\mathbb{E}(q( \boldsymbol{X}_{-\mathcal{X}(l)}) |\boldsymbol{X}_{\mathcal{X}(k)}=\vv{c}_{k}, \boldsymbol{X}_{\mathcal{X}(l)}=\vv{c}_{l}) - \mathbb{E}(q( \boldsymbol{X}_{-\mathcal{X}(l)})| \boldsymbol{X}_{\mathcal{X}(k)}=\vv{c}_{k} ) | 
        \\
        & =  \left( \mathbb{E}(q( \boldsymbol{X}_{-\mathcal{X}(l)}) \boldsymbol{1}_{ \boldsymbol{X}_{\mathcal{X}(k)}=\vv{c}_{k}} \boldsymbol{1}_{ \boldsymbol{X}_{\mathcal{X}(l)}=\vv{c}_{l}} ) - \mathbb{E}(q( \boldsymbol{X}_{-\mathcal{X}(l)}) \boldsymbol{1}_{\boldsymbol{X}_{\mathcal{X}(k)}=\vv{c}_{k}}) \mathbb{P}( \boldsymbol{X}_{\mathcal{X}(l)}=\vv{c}_{l})\right)  \\
        & \qquad\qquad \times \frac{1}{\mathbb{P}( \boldsymbol{X}_{\mathcal{X}(k)}=\vv{c}_{k}, \boldsymbol{X}_{\mathcal{X}(l)}=\vv{c}_{l})} \\
        & \quad + 
\left( \frac{\mathbb{P}( \boldsymbol{X}_{\mathcal{X}(l)}=\vv{c}_{l})  \mathbb{P}( \boldsymbol{X}_{\mathcal{X}(k)}=\vv{c}_{k}) - \mathbb{P}( \boldsymbol{X}_{\mathcal{X}(k)}=\vv{c}_{k}, \boldsymbol{X}_{\mathcal{X}(l)}=\vv{c}_{l})}{\mathbb{P}( \boldsymbol{X}_{\mathcal{X}(k)}=\vv{c}_{k}, \boldsymbol{X}_{\mathcal{X}(l)}=\vv{c}_{l}) \mathbb{P}( \boldsymbol{X}_{\mathcal{X}(k)}=\vv{c}_{k}) } \right) \\
&\qquad\qquad \times        
        \mathbb{E}(q( \boldsymbol{X}_{-\mathcal{X}(l)}) \boldsymbol{1}_{\boldsymbol{X}_{\mathcal{X}(k)}=\vv{c}_{k}})\\
        & \le  \left| \mathbb{E}(q( \boldsymbol{X}_{-\mathcal{X}(l)}) \boldsymbol{1}_{ \boldsymbol{X}_{\mathcal{X}(k)}=\vv{c}_{k}} \boldsymbol{1}_{ \boldsymbol{X}_{\mathcal{X}(l)}=\vv{c}_{l}} ) - \mathbb{E}(q( \boldsymbol{X}_{-\mathcal{X}(l)}) \boldsymbol{1}_{\boldsymbol{X}_{\mathcal{X}(k)}=\vv{c}_{k}}) \mathbb{P}( \boldsymbol{X}_{\mathcal{X}(l)}=\vv{c}_{l})\right|  \\
        & \qquad\qquad \times \frac{1}{\mathbb{P}( \boldsymbol{X}_{\mathcal{X}(k)}=\vv{c}_{k}, \boldsymbol{X}_{\mathcal{X}(l)}=\vv{c}_{l})} \\
        & \quad + 
\left| \frac{\mathbb{P}( \boldsymbol{X}_{\mathcal{X}(l)}=\vv{c}_{l})  \mathbb{P}( \boldsymbol{X}_{\mathcal{X}(k)}=\vv{c}_{k}) - \mathbb{P}( \boldsymbol{X}_{\mathcal{X}(k)}=\vv{c}_{k}, \boldsymbol{X}_{\mathcal{X}(l)}=\vv{c}_{l})}{\mathbb{P}( \boldsymbol{X}_{\mathcal{X}(k)}=\vv{c}_{k}, \boldsymbol{X}_{\mathcal{X}(l)}=\vv{c}_{l}) \mathbb{P}( \boldsymbol{X}_{\mathcal{X}(k)}=\vv{c}_{k}) } \right| \\
&\qquad\qquad \times        
        \mathbb{E}|q( \boldsymbol{X}_{-\mathcal{X}(l)}) \boldsymbol{1}_{\boldsymbol{X}_{\mathcal{X}(k)}=\vv{c}_{k}}|\\
        & \le \frac{ 2\delta_{0}\mathbb{E}|q(\boldsymbol{X}_{-\mathcal{X}(l)}) \boldsymbol{1}_{\boldsymbol{X}_{\mathcal{X}(k)}=\vv{c}_{k}}|}{\mathbb{P}( \boldsymbol{X}_{\mathcal{X}(k)}=\vv{c}_{k},  \boldsymbol{X}_{\mathcal{X}(l)}=\vv{c}_{l})} \\
        & \le \max_{\ i\in \mathcal{X}(k), j\in \mathcal{X}(l), (a, b)\in \{0, 1\}^2 }\frac{2\delta_{0}  \mathbb{E}|q(\boldsymbol{X}_{-\mathcal{X}(l)})| }{\mathbb{P}( X_{i}=a,  X_{j}=b)},
    \end{split}
\end{equation}
which, in conjunction with the definition of $p_{\min}$ in Condition~\ref{signal.strength.1}, implies almost surely that for every $\{l, k\} \subset\{1, \dots, M\}$,
\begin{equation*}
    \begin{split}
        & |\mathbb{E}(q( \boldsymbol{X}_{-\mathcal{X}(l)}) |\boldsymbol{X}_{\mathcal{X}(k)}, \boldsymbol{X}_{\mathcal{X}(l)} ) - \mathbb{E}(q( \boldsymbol{X}_{-\mathcal{X}(l)})| \boldsymbol{X}_{\mathcal{X}(k)} ) |  \\
        &\le  \max_{\ i\in \mathcal{X}(k), j\in \mathcal{X}(l), (a, b)\in \{0, 1\}^2 } \frac{2\delta_{0}  \mathbb{E}|q(\boldsymbol{X}_{-\mathcal{X}(l)})| }{\mathbb{P}( X_{i}=a,  X_{j}=b)}\\
        & \le \frac{2\delta_{0}  \mathbb{E}|q(\boldsymbol{X}_{-\mathcal{X}(l)})| }{p_{\min}}.
    \end{split}
\end{equation*}
This result concludes the proof of the second assertion of Lemma~\ref{lemma.5}.


In addition, by \eqref{dependence.4} and \eqref{dependence.11}, for every $1\le k<  l\le M$, every pair of one-hot vectors $\vv{c}_{k}\in\mathbb{R}^{\texttt{\#} \mathcal{X}(k)}$ and $\vv{c}_{l}\in\mathbb{R}^{\texttt{\#} \mathcal{X}(l)}$ with $\mathbb{P}(\boldsymbol{X}_{\mathcal{X}(k)}=\vv{c}_{k}, \boldsymbol{X}_{\mathcal{X}(l)}=\vv{c}_{l}) > 0$, and every measurable function $q:\mathbb{R}^{p- \texttt{\#}\mathcal{X}(lk)}\mapsto\mathbb{R}$,
\begin{equation}
    \begin{split}\label{dependence.18}
    & |\mathbb{E}(q( \boldsymbol{X}_{-\mathcal{X}(kl)}) |\boldsymbol{X}_{\mathcal{X}(k)}=\vv{c}_{k}, \boldsymbol{X}_{\mathcal{X}(l)}=\vv{c}_{l}) - \mathbb{E}(q( \boldsymbol{X}_{-\mathcal{X}(kl)}) ) | 
        \\
        & =  \Big|\frac{1}{\mathbb{P}( \boldsymbol{X}_{\mathcal{X}(k)}=\vv{c}_{k}, \boldsymbol{X}_{\mathcal{X}(l)}=\vv{c}_{l})} \times \\
        & \qquad \Big[ \mathbb{E}(q( \boldsymbol{X}_{-\mathcal{X}(kl)}) \boldsymbol{1}_{ \boldsymbol{X}_{\mathcal{X}(k)}=\vv{c}_{k}} \boldsymbol{1}_{ \boldsymbol{X}_{\mathcal{X}(l)}=\vv{c}_{l}} ) \\
        & \qquad\qquad - \mathbb{E}(q( \boldsymbol{X}_{-\mathcal{X}(kl)}) \boldsymbol{1}_{\boldsymbol{X}_{\mathcal{X}(k)}=\vv{c}_{k}}) \mathbb{P}( \boldsymbol{X}_{\mathcal{X}(l)}=\vv{c}_{l}) \\
        & \quad + \mathbb{E}(q( \boldsymbol{X}_{-\mathcal{X}(kl)}) \boldsymbol{1}_{\boldsymbol{X}_{\mathcal{X}(k)}=\vv{c}_{k}}) \mathbb{P}( \boldsymbol{X}_{\mathcal{X}(l)}=\vv{c}_{l}) \\
        & \qquad \qquad - \mathbb{E}(q( \boldsymbol{X}_{-\mathcal{X}(kl)})) \mathbb{P}( \boldsymbol{X}_{\mathcal{X}(k)}=\vv{c}_{k}) \mathbb{P}( \boldsymbol{X}_{\mathcal{X}(l)}=\vv{c}_{l}) \\
        & \quad + \mathbb{E}(q( \boldsymbol{X}_{-\mathcal{X}(kl)})) \mathbb{P}( \boldsymbol{X}_{\mathcal{X}(k)}=\vv{c}_{k}) \mathbb{P}( \boldsymbol{X}_{\mathcal{X}(l)}=\vv{c}_{l})  \\
        &\qquad\qquad -  \mathbb{E}(q( \boldsymbol{X}_{-\mathcal{X}(kl)})) \mathbb{P}( \boldsymbol{X}_{\mathcal{X}(k)}=\vv{c}_{k}, \boldsymbol{X}_{\mathcal{X}(l)}=\vv{c}_{l}) \\
        & \quad + \mathbb{E}(q( \boldsymbol{X}_{-\mathcal{X}(kl)})) \mathbb{P}( \boldsymbol{X}_{\mathcal{X}(k)}=\vv{c}_{k}, \boldsymbol{X}_{\mathcal{X}(l)}=\vv{c}_{l}) \Big] -  \mathbb{E}(q( \boldsymbol{X}_{-\mathcal{X}(kl)}) ) \Big| \\
        & \le \frac{1}{\mathbb{P}( \boldsymbol{X}_{\mathcal{X}(k)}=\vv{c}_{k},  \boldsymbol{X}_{\mathcal{X}(l)}=\vv{c}_{l})} \times  \Big[ \delta_{0}\mathbb{E}|q(\boldsymbol{X}_{-\mathcal{X}(kl)}) \boldsymbol{1}_{\boldsymbol{X}_{\mathcal{X}(k)}=\vv{c}_{k}}| \\
        & \qquad + \delta_{0} \mathbb{E}|q( \boldsymbol{X}_{-\mathcal{X}(kl)}) | \mathbb{P}(\boldsymbol{X}_{\mathcal{X}(l)}=\vv{c}_{l}) + \mathbb{E}|q( \boldsymbol{X}_{-\mathcal{X}(kl)}) |   \delta_{0}\mathbb{P}(\boldsymbol{X}_{\mathcal{X}(k)}=\vv{c}_{k})\Big] \\
        & \le \max_{\ i\in \mathcal{X}(k), j\in \mathcal{X}(l), (a, b)\in \{0, 1\}^2 }\frac{3\delta_{0}  \mathbb{E}|q(\boldsymbol{X}_{-\mathcal{X}(kl)})| }{\mathbb{P}( X_{i}=a,  X_{j}=b)}\\
        & \le \frac{3\delta_{0}  \mathbb{E}|q(\boldsymbol{X}_{-\mathcal{X}(kl)})| }{p_{\min} }.
    \end{split}
\end{equation}

By similar arguments for \eqref{dependence.18}, we derive that for every $1\le k< l \le M$, it holds that  almost surely,
\begin{equation*}
    \begin{split}
    & |\mathbb{E}(q( \boldsymbol{X}_{-\mathcal{X}(kl)}) |\boldsymbol{X}_{\mathcal{X}(k)}, \boldsymbol{X}_{\mathcal{X}(l)}) - \mathbb{E}(q( \boldsymbol{X}_{-\mathcal{X}(kl)}) ) | \\
    & \le  \max_{\ i\in \mathcal{X}(k), j\in \mathcal{X}(l), (a, b)\in \{0, 1\}^2 }\frac{3\delta_{0}  \mathbb{E}|q(\boldsymbol{X}_{-\mathcal{X}(kl)})| }{\mathbb{P}( X_{i}=a,  X_{j}=b)}\\
    & \le \frac{3\delta_{0}  \mathbb{E}|q(\boldsymbol{X}_{-\mathcal{X}(kl)})| }{p_{\min} },
    \end{split}
\end{equation*}
which concludes the proof of the third assertion of Lemma~\ref{lemma.5}, and hence finishes the proof of Lemma~\ref{lemma.5}.


\end{document}